%% file: main.tex
\documentclass[twocolumn]{aastex62}
\usepackage{amsmath,amstext}
\usepackage[T1]{fontenc}
\usepackage{apjfonts} 
\usepackage[figure,figure*]{hypcap}
\usepackage{braket}

\shorttitle{RRLs, OH, and NH$_3$ in Nearby Galaxies}
\shortauthors{Eisner et al.}

\begin{document}

\title{A Spectral Analysis of the Centimeter Regime of Nearby Galaxies: RRLs, Excited OH, and NH$_3$}

\author{Brian A. Eisner}
\affiliation{Department of Physics \& Astronomy, Macalester College, 1600 Grand Avenue, St Paul, MN 55105, USA}
\affiliation{Summer student at the National Radio Astronomy Observatory, P.O. Box O, 1003 Lopezville Road, Socorro, NM 87801, USA}
\affiliation{Department of Astronomy, University of Virginia, P.O. Box 400325, 530 McCormick Road, Charlottesville, VA 22904, USA}
\author{Juergen Ott}
\affiliation{National Radio Astronomy Observatory, P.O. Box O, 1003 Lopezville Road, Socorro, NM 87801, USA}
\author{David S. Meier}
\affiliation{Department of Physics, New Mexico Institute of Mining and Technology, 801 Leroy Place, Socorro, NM 87801, USA}
\affiliation{National Radio Astronomy Observatory, P.O. Box O, 1003 Lopezville Road, Socorro, NM 87801, USA}
\author{John M. Cannon}
\affiliation{Department of Physics \& Astronomy, Macalester College, 1600 Grand Avenue, St Paul, MN 55105, USA}

\begin{abstract}
Centimeter-wave transitions are important counterparts to the rotational mm-wave transitions usually observed to study gas in star-forming regions. However, given their relative weakness, these transitions have historically been neglected. We present Australia Telescope Compact Array 4cm- and 15mm-band spectral line observations of nine nearby star-forming galaxies in the H75 array configuration. Thirteen different molecular lines are detected across the sample from OH, NH$_3$, H$_2$O, H$_2$CO, and \textit{c}-C$_3$H$_2$, as well as 18 radio recombination lines (RRLs) in NGC~253. Excited OH $^2\Pi_{3/2}$ absorption is detected towards NGC~253 (J=5/2), NGC~4945 (J=9/2), and Circinus (J=9/2); the latter two represent only the third and fourth extragalactic J=9/2 detections. These lines in Circinus suggest rotation temperatures in excess of 2000~K, and thus it is likely that the populations of OH rotational states are not governed by a Boltzmann distribution. Circinus's OH lines are blueshifted from the systemic velocity by \textasciitilde35~km~s$^{-1}$, while NGC~4945's are redshifted by \textasciitilde100~km~s$^{-1}$. NGC~4945's OH absorption likely indicates infall onto the nucleus. The NH$_3$ (1,1) through (6,6) lines in NGC~4945 display a superposition of emission and absorption similar to that seen in other dense gas tracers. Strong (3,3) emission points towards maser activity. The relative NH$_3$ absorption strengths in NGC~4945 show similar anomalies as in previous studies of Arp~220 (weak (1,1) and strong (5,5) absorption). A trend towards higher LTE electron temperatures with increasing RRL frequency is present in NGC~253, likely indicative of stimulated emission within the nuclear region.
\end{abstract}


\input{Intro}
\input{Sources}
\input{Methods}
\input{Results}
\input{RRL}
\input{OH}
\input{NH3}
\input{UnidentifiedLine}
\input{Nondetections}
\input{Conclusions}

\acknowledgments
Acknowledgments.

B.~A.~E. would like to thank the National Radio Astronomy Observatory and the National Science Foundation for providing the Research Experience for Undergraduates program in which he participated. Specifically, he would like to thank Dr. Adam Ginsburg (National Radio Astronomy Observatory) for providing assistance with technical aspects of this work, and Drs. Amy Mioduszewski, Anna Kapinska, and Jim Braatz (National Radio Astronomy Observatory) for coordinating all summer student operations at the National Radio Astronomy Observatory. He would also like to thank Macalester College for providing the support to continue work on the project between the two summers at the NRAO, and specifically to Drs. James Heyman and Tom Varberg for providing valuable feedback on the paper. We would additionally like to thank Dr. Dana Balser (National Radio Astronomy Observatory) for his valuable input on radio recombination line physics.

The National Radio Astronomy Observatory is a facility of the National Science Foundation operated under cooperative agreement by Associated Universities, Inc. The Australia Telescope Compact Array is part of the Australia Telescope National Facility which is funded by the Australian Government for operation as a National Facility managed by CSIRO. This research is based on photographic data obtained using The UK Schmidt Telescope. The UK Schmidt Telescope was operated by the Royal Observatory Edinburgh, with funding from the UK Science and Engineering Research Council, until 1988 June, and thereafter by the Anglo-Australian Observatory. Original plate material is copyright6 (c) of the Royal Observatory Edinburgh and the Anglo-Australian Observatory. The plates were processed into the present compressed digital form with their permission. The Digitized Sky Survey was produced at the Space Telescope Science Institute under US Government grant NAG W-2166. This research has made use of NASA's Astrophysics Data System. This research has made use of the NASA/IPAC Extragalactic Database (NED) which is operated by the Jet Propulsion Laboratory, California Institute of Technology, under contract with the National Aeronautics and Space Administration. This work made use of the IPython package. This research made use of NumPy. This research made use of MatPlotLib, a Python library for publication quality graphics. This research made use of Astropy, a community-developed core Python package for Astronomy. This research made use of SpectralCube, a library for astronomical spectral data cubes. This research made use of PySpecKit, an open-source spectral analysis and plotting package for Python hosted at http://pyspeckit.bitbucket.org. This material is based upon work supported by the National Science Foundation under Grant No. 1358169.

\facilities{ATCA} 
\software{MIRIAD \citep{1995ASPC...77..433S}, CASA \citep{2007ASPC..376..127M}, IPython \citep{10.1109/MCSE.2007.53}, PySpecKit \citep{2011ascl.soft09001G}, Astropy \citep{2018AJ....156..123A}, NumPy \citep{numpy}, MatPlotLib \citep{10.1109/MCSE.2007.55}, SpectralCube, Karma \citep{1996ASPC..101...80G}}

\bibliographystyle{yahapj}

\appendix
\input{NondetectionsAppendix}

\end{document}

%% file: Intro.tex
\section{INTRODUCTION \& MOTIVATION}\label{Ch1}

Understanding the molecular gas of the ISM in star-forming regions allows for characterization of the star formation process on both local and galactic scales \citep{2012ARAaA..50..531K}. These regions are often studied using molecular spectroscopy and astrochemistry, as molecular transitions are able to provide additional information not available from continuum observations such as velocity, excitation, and chemical abundances \citep[e.g.][]{2013AaA...559A..47B,2015ApJ...801...63M,2018AaA...615A.155H}. In Galactic sources, star-forming regions can be probed to sub-AU scales, revealing in exquisite detail the processes underlying star formation \citep[e.g.][]{2013AaA...559A..47B,2015ApJ...808L...3A}. In extragalactic sources this is generally not possible with current technology, and star formation is instead studied on the scale of entire star-forming regions or even the entire galaxy \citep[e.g.][]{2015ApJ...801...25L,2015ApJ...801...63M,2018AaA...615A.155H}.

However, different molecules trace different gas properties. The study of a single molecule cannot comprehensively characterize the properties of the star formation; multiple transitions and molecules are necessary to fully characterize the gas \citep{2005ApJ...618..259M}.

While typically the millimeter wavelengths, in particular the CO $\Delta J$ transitions, are used to trace the column density of the gas \citep[e.g.][]{1988AaA...207....1S,2001ApJ...547..792D,2013ARAaA..51..207B}, the centimeter wavelengths offer relatively untapped potential. In particular, studies of lines arising from mechanisms other than pure rotation are important counterparts to the multitude of rotational surveys present in the literature \citep[e.g.][]{2006ApJS..164..450M,2015ApJ...801...63M,2018AaA...615A.155H}. Centimeter transitions are unlike millimeter in that the majority of the most common transitions are not purely rotational in origin. The most abundant molecules' non-rotational transitions tend to lie at cm-wavelengths, such as the $\Lambda$-doubling transitions of OH \citep[e.g.][]{1967ARAaA...5..183R,1986AaA...155..193H} and the inversion transitions of NH$_3$ \citep[e.g.][]{1983ARAaA..21..239H,1983AaA...122..164W}.

Observing these cm-wave lines can have significant benefits. For example, NH$_3$ has great utility as a temperature probe due to its unique structural properties (a symmetric top with a heavy atom above a plane of light atoms). Its inversion lines' strengths are nearly independent of other ISM properties such as density \citep[e.g.][]{1983ARAaA..21..239H,1983AaA...122..164W}. Similarly, the free radical nature of OH provides it a large number of transitions for a diatomic molecule, equating to more constraints on parameters such as abundance, temperature, and density than is easily attainable with typical molecules. OH also traces different regions than CO, such as AGN disks, and different properties than CO, such as cosmic-ray ionization rates. Other simple molecules such as H$_2$O and H$_2$CO have well-studied cm-wave transitions; H$_2$O at 22~GHz is often a strong maser \citep{1995ApJ...440..619G} and H$_2$CO is a well-known density probe \citep{2013ApJ...766..108M}. Thus, observations of the cm-wave transitions complement those of the rotational mm-wave transitions, and combined the two can comprehensively characterize the ISM in ways that observations of the millimeter wavelengths alone cannot.

In addition to the multitude of molecular lines present at cm-wavelengths, radio recombination lines (RRLs) are also common. Unlike the majority of molecular transitions, RRLs trace the location of ionized gas with intensities related to star-formation rate \citep[e.g.][]{1990ApJ...365..606G,2011ApJ...739L..24K,2016MNRAS.463..252B}, and thus their observations complement those of molecules to paint a more complete picture of star-forming regions. The frequency-spacing of lines decreases as a function of decreasing frequency; between 4~GHz and 50~GHz there are 67 different H(n)$\alpha$ RRLs and even larger numbers of H$\beta$ and other higher-order RRLs. This makes low frequencies productive for RRL studies, as large numbers of lines are observable in a single pointing, allowing for increased sensitivity from the stacking of multiple lines.

Additionally, RRLs can be used for calculations of electron temperatures in these ionized regions. Specifically, the line to free-free-continuum ratio has a one-to-one correspondence with electron temperature, assuming local thermodynamic equilibrium. RRLs are unobscured compared to common optical/UV atomic lines (e.g. H$\alpha$) used to probe star formation rates, increasing the ease of calculations. Moreover, because line and continuum opacities decrease with frequency, electron temperatures measured with different RRLs can probe the non-LTE structure of \ion{H}{2} regions \citep[e.g.][]{2017ApJ...844...73B}.

Historically, searches for lines in the cm-wave regime has been impeded by the lack of wideband correlators. The lines are also typically weaker, requiring higher sensitivities to detect than typical lines at mm-wavelengths. Fortunately, within the last decade, the advent of new correlators on instruments, such as the NSF's Karl G. Jansky Very Large Array (VLA) and the Australia Telescope Compact Array (ATCA), has allowed these centimeter regimes to be surveyed in a nearly-complete sense. However, the only extragalactic cm-wave wideband line survey is the Survey of Water and Ammonia in Nearby Galaxies (SWAN) \citep{2017ApJ...842..124G,2018ApJ...856..134G}. Their sample consists of four star-forming galaxies (NGC~253, NGC~2146, NGC~6946, and IC~342) covering a wide range of physical parameters surveyed over the K- and Ka-bands using the VLA with \textasciitilde1$^{\prime\prime}$ resolution. Within this survey, lines from seven different atomic or molecular species are detected, although the authors focus their analysis on NH$_3$, H$_2$O, and CH$_3$OH.

While it is helpful to study the gas within galaxies in a resolved sense, as in the SWAN, there are advantages of lower resolution as well. In particular, low angular resolution observations allow for the comprehensive determination of global properties of galaxy nuclei and allow for greater surface brightness sensitivity. The SWAN survey also only focuses on the K- and Ka-bands, and no surveys of the lower-frequency bands have been conducted. In addition, SWAN only targets four galaxies, a sample size too small to draw any conclusions about global galaxies properties.

Our purpose with this study is twofold: we target a large number of understudied, likely noteworthy cm-wave transitions in our observations, and also cover large areas of frequency space. Our target objects (Section \ref{Ch2}) represent nine of the most promising objects for observing lines, and our target lines represent the most promising transitions to detect. Using the transitions we detect, we derive physical properties of the gas in the galaxies such as rotation temperatures, and compare our detections between galaxies. Comparison to archival data is also employed to further construct a thorough interpretation of our results, and the origin of the spectral features are investigated. In Sections \ref{Ch2} and \ref{Ch3} we introduce the sources and observations used for this study. In Section \ref{Ch4} we present our detected lines, and interpret them in the following sections. Finally, in Section \ref{Ch6} we outline our main conclusions.

%% file: Sources.tex
\section{SOURCES}\label{Ch2}

\begin{figure*}
  \centering
  \plotone{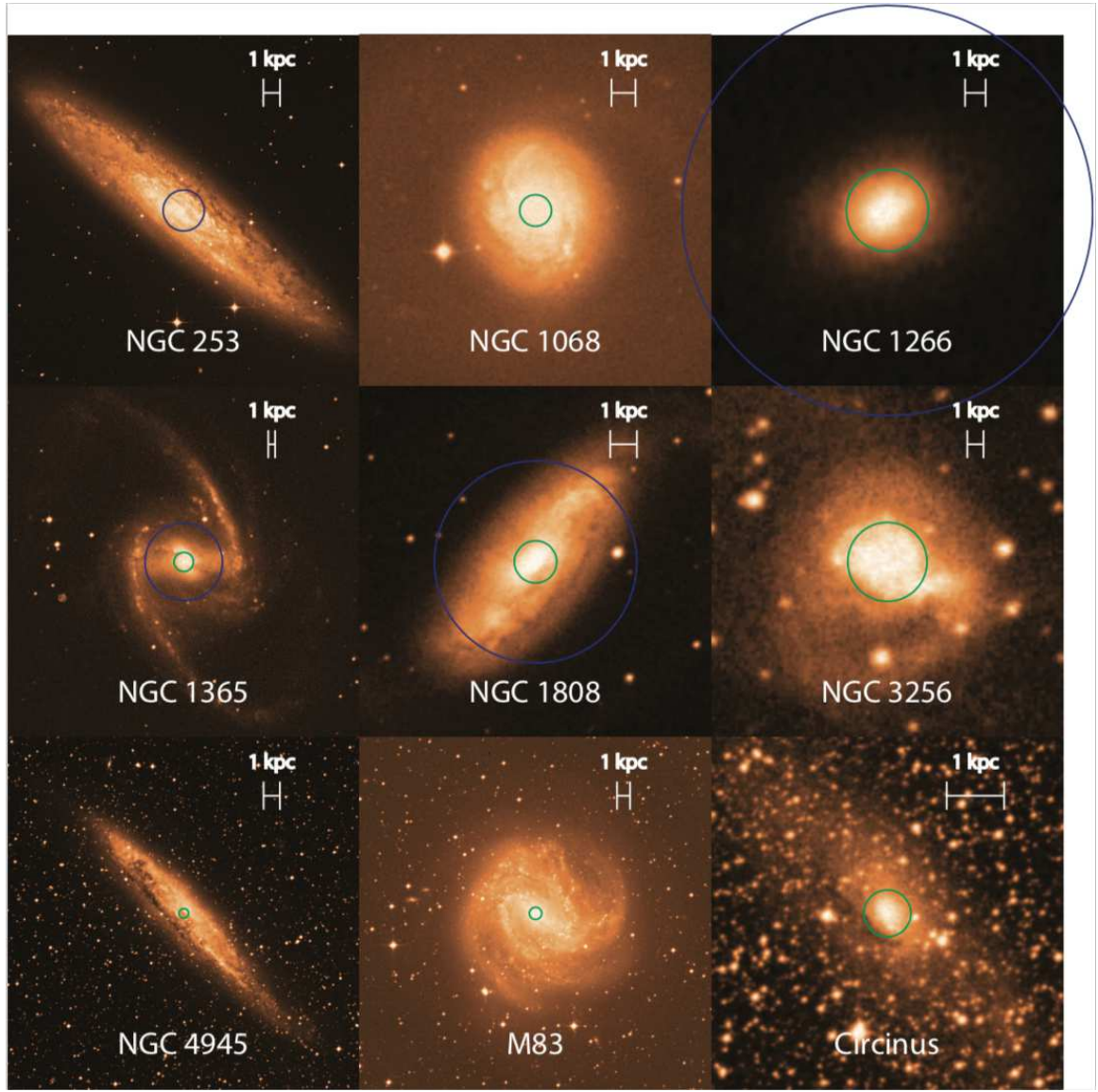}
  \caption{European Southern Observatory Digitized Sky Survey (DSS) images of all galaxies, overlain with our ATCA synthesized beams at the pointing location. Blue circles represent synthesized 4cm beams, and green circles 15mm beams.}
  \label{fig:Galaxies_pointings}
\end{figure*}

Nine nearby galaxies (Table \ref{table:sources}) with varying rates of star formation were chosen as sources. The sample was crafted to contain nearby, diverse sources bright enough and close enough to have a reasonable possibility of detecting spectral lines with the southern-sky ATCA. All objects are located in the southern celestial hemisphere, allowing for comparisons to pre-existing and/or future ALMA data. Optical images of the galaxies can be seen in Figure \ref{fig:Galaxies_pointings}.

\begin{deluxetable*}{lccccclll}
\tablecaption{Global parameters of observed galaxies\label{table:sources}}
\tablehead{
\colhead{Source} & \colhead{RA (J2000)} & \colhead{Dec (J2000)} & \colhead{$d$ (Mpc)} & \colhead{$\upsilon$ (km s$^{-1}$)\tablenotemark{a}} & \colhead{SFR\tablenotemark{b} ($M_\odot$ yr$^{-1}$)} &  \colhead{SFR method\tablenotemark{c}} & \colhead{AGN}
 & \colhead{References}}

\startdata
\object{NGC 253}  & 00$^{h}$ 47$^{m}$ 33$^{s}$.1 & $-$25$^\circ$ 17$^\prime$ 18$^{\prime\prime}$ & 3.56 $\pm$ 0.28 &  236 $\pm$ 2 & 1.73 $\pm$ 0.12        & RL,FF   & None & 1,5,9 \\
\object{NGC 1068} & 02$^{h}$ 42$^{m}$ 40$^{s}$.7 & $-$00$^\circ$ 00$^\prime$ 48$^{\prime\prime}$ & 10.1$\pm$ 3.9   & 1126 $\pm$ 3 & 0.36$^{+0.36}_{-0.18}$ & PAH     & Seyfert 1 & 2,6,10,17 \\
\object{NGC 1266} & 03$^{h}$ 16$^{m}$ 00$^{s}$.7 & $-$02$^\circ$ 25$^\prime$ 38$^{\prime\prime}$ & 29.4 $\pm$ 2.9\tablenotemark{d} & 2157 $\pm$ 5 & 0.87$^{+0.63}_{-0.43}$ & FF & AGN & 3,7,11,18 \\
\object{NGC 1365} & 03$^{h}$ 33$^{m}$ 36$^{s}$.4 & $-$36$^\circ$ 08$^\prime$ 25$^{\prime\prime}$ & 17.8 $\pm$ 1.8  & 1619 $\pm$ 1 & $<$0.05                & PAH     & Seyfert 1.9 & 1,8,10,19 \\
\object{NGC 1808} & 05$^{h}$ 07$^{m}$ 42$^{s}$.3 & $-$37$^\circ$ 30$^\prime$ 47$^{\prime\prime}$ & 9.1 $\pm$ 1.8  &  976 $\pm$ 4 & 0.18$^{+0.18}_{-0.09}$ & RL      & Seyfert 2? & 1,5,12,20 \\
\object{NGC 3256} & 10$^{h}$ 27$^{m}$ 51$^{s}$.3 & $-$43$^\circ$ 54$^\prime$ 13$^{\prime\prime}$ & 37 $\pm$ 14     & 2792 $\pm$ 6 & 21$^{+21}_{-11}$\tablenotemark{e}   & SED     & Seyfert 2\tablenotemark{e} & 4,5,13,21 \\
\object{NGC 4945} & 13$^{h}$ 05$^{m}$ 27$^{s}$.5 & $-$49$^\circ$ 28$^\prime$ 06$^{\prime\prime}$ & 3.72 $\pm$ 0.30 &  558 $\pm$ 3 & 4.42 $\pm$ 0.49        & RL,FF   & Seyfert 2 & 1,5,14,22,23 \\
\object{M83}      & 13$^{h}$ 37$^{m}$ 00$^{s}$.9 & $-$29$^\circ$ 51$^\prime$ 56$^{\prime\prime}$ & 4.66 $\pm$ 0.33 &  514 $\pm$ 2 & 1.6$^{+1.6}_{-0.8}$    & FUV     & None & 1,5,15 \\
\object{Circinus} & 14$^{h}$ 13$^{m}$ 09$^{s}$.9 & $-$65$^\circ$ 20$^\prime$ 21$^{\prime\prime}$ & 4.2 $\pm$ 1.6  &  429 $\pm$ 3 & 3--8                   & MIR,FIR & Seyfert 2 & 2,5,16,19
\enddata

\tablenotetext{a}{All velocities quoted in LSRK frame.}
\tablenotetext{b}{Nuclear SFR only with the exception of M83, where the SFR denotes the entire galaxy.}
\tablenotetext{c}{RL: recombination lines; FF: free-free continuum; PAH: polycyclic aromatic hydrocarbon; MIR: mid-IR; FIR: far-IR; SED: spectral energy distribution; FUV: far-UV.}
\tablenotetext{d}{Calculated from radial velocity and adding a 200~km~s$^{-1}$ error to account for peculiar velocity.}
\tablenotetext{e}{Double-nucleus galaxy with Sy2 activity in southern nucleus only; SFR is sum total of both nuclei.}
\tablecomments{(1) \citet{2013AJ....146...86T}; (2) \citet{2009AJ....138..323T}; (3) \citet{1991rc3..book.....D}; (4) \citet{1988ngc..book.....T}; (5) \citet{2004AJ....128...16K}; (6) \citet{1999ApJS..121..287H}; (7) \citet{2011MNRAS.413..813C}; (8) \citet{1996ApJ...463...60B}; (9) \citet{2015MNRAS.450L..80B}; (10) \citet{2014ApJ...780...86E}; (11) \citet{2015ApJ...798...31A}; (12) \citet{2017AaA...598A..55B}; (13) \citet{2008MNRAS.384..316L}; (14) \citet{2016MNRAS.463..252B}; (15) \citet{2013AJ....146...46K}; (16) \citet{2012MNRAS.425.1934F}; (17) \citet{2009MNRAS.398.1165G}; (18) \citet{2011ApJ...735...88A}; (19) \citet{2012MNRAS.426.1750M}; (20) \citet{2005AaA...442..861J}; (21) \citet{2015ApJ...805..162O}; (22) \citet{1993ApJ...409..155I}; (23) \citet{1994ApJ...429..602M}}
\end{deluxetable*}

The galaxies in our sample range over an order of magnitude in distance, between 3.5 and 40 Mpc. One galaxy (NGC~1266) is lenticular, one (NGC~3256) is an interacting pair of former spirals, while the remaining seven are spirals. Several of the galaxies (NGC~253, NGC~1068, NGC~1266, NGC~1808, and Circinus) are known to host massive molecular outflows from the nucleus \citep{2013Natur.499..450B,2014AaA...567A.125G,2011ApJ...735...88A,2016ApJ...823...68S,1997ApJ...479L.105V}. Many (NGC~253, NGC~1808, NGC~3256, NGC~4945, M83, Circinus) are currently starbursting \citep{2015MNRAS.450L..80B,2017AaA...598A..55B,2008MNRAS.384..316L,1994ApJ...429..602M,2016ApJ...832..142Z}, and NGC~1266 may host a dust-obscured starburst \citep{2015ApJ...798...31A}. NGC~1068 underwent two starbursts in the recent past \citep{2012ApJ...755...87S}, and Circinus's outflow suggests starburst characteristics \citep{2016ApJ...832..142Z}. Seven of the nine galaxies display AGN activity, with the majority of these containing Seyfert nuclei. Nuclear star formation rates range over several orders of magnitude, from $\leq$0.05 to \textasciitilde20~$M_\odot$~yr$^{-1}$.

NGC 253, one of the most-studied of all astronomical objects, is near enough (3.56 Mpc) that individual protoclusters have recently been resolved in the nucleus \citep{2018ApJ...869..126L}. Two large (\textasciitilde100~pc) molecular superbubbles are present in a \textasciitilde500~pc circumnuclear disk \citep{2006ApJ...636..685S}, and multiple smaller bubbles are present in the disk \citep{2005ApJ...629..767O}. NGC~253, along with M82, was the first known source of extragalactic molecular lines, following the detection of the 1665~MHz and 1667~MHz $^2\Pi_{3/2}$ $J=3/2$ OH transitions by \citet{1971ApJ...167L..47W}. Since then, NGC~253 has served as a premier laboratory for extragalactic astrochemistry, with multiple mm-wave line surveys conducted \citep[e.g.][]{2006ApJS..164..450M,2015ApJ...801...63M,2015MNRAS.450L..80B,2017ApJ...835..265W,2017ApJ...849...81A}. Notably, NGC~253 is the only galaxy in our sample for which a dedicated cm-wave survey of spectral lines exists. \citet{2017ApJ...842..124G} surveyed 4~GHz ranges in both the K- and Ka-bands for the SWAN survey using the VLA and detected 17 spectral lines from seven different molecular species.

NGC~1068, the prototypical Seyfert galaxy, is also one of the most well-studied objects. Multiple ALMA line surveys have been conducted \citep[e.g.][]{2014AaA...567A.125G,2015PASJ...67....8N,2016ApJ...823L..12G,2016MNRAS.459.3629I}. The remaining galaxies are all significantly less-studied with regards to astrochemistry, with only NGC~4945 \citep[][]{2018AaA...615A.155H} and NGC~3256 \citep{2014ApJ...797...90S,2018ApJ...855...49H} having dedicated ALMA line surveys in the millimeter.

%% file: Methods.tex
\section{OBSERVATIONS \& DATA REDUCTION}\label{Ch3}

We targeted and observed the nuclear regions of all nine galaxies listed in Table \ref{table:sources} using the 4cm and/or 15mm bands of the Australia Telescope Compact Array (ATCA) in its H75 configuration. As the most compact ATCA layout, the H75 configuration yields angular resolution comparable to large-diameter single-dish measurements but maintains an inherently better baseline stability that significantly improves the ability to detect faint lines. The primary beam of the ATCA ranges from 12$^\prime$ at the start of the 4cm-band to 2$^\prime$ at the end of the 15mm band. The 4cm band spectral range is roughly equivalent to the VLA's C- and X-bands, while the 15mm band is analogous to the Ku- and K-bands.

The majority of observations were taken throughout the APRS semester of 2016 during filler time slots (project code CX297), utilizing the ATCA's remote observing capabilities; one NGC 1365 observation was taken in 2015. The CABB correlator was utilized in its 64M-32k mode, providing 32 total zoom bands with resolutions of 32~kHz (equivalent to \textasciitilde1.5~km~s$^{-1}$ at 6~GHz and \textasciitilde0.4~km~s$^{-1}$ at 23~GHz), each with 2048 channels. The two continuum/spectral bands were centered roughly at rest frequencies 5.7~GHz and 8.6~GHz in the 4cm-band, and at 22.1~GHz and 24.5~GHz in the 15mm-band. The 32 zoom bands were placed at representative frequencies within these bands, many targeting plausible spectral lines. The rest frequencies of these zoom bands are noted in Table \ref{table:obs_freqs}; actual observational frequencies were corrected for the galaxy redshifts listed in Table \ref{table:sources}. We emphasize that using the CABB correlator it is impossible to cover the entire 4cm- or 15mm-band; we have only sampled the band at multiple, but specific, frequency ranges. In galaxies where only one band could be observed due to a lack of observing time, the 15mm band was chosen for observation. As NGC~253 was observed in the K- and Ka-bands using the VLA by \citet{2017ApJ...842..124G}, we did not sample its 15mm band.

\startlongtable
\begin{deluxetable*}{cccc}
\tablecaption{Central frequencies of zoom bands in observations for a target at rest, plus rest frequencies for targeted spectral lines\label{table:obs_freqs}}
\tablehead{
\colhead{Center freq. (MHz)} & \colhead{Bandwidth (MHz)\tablenotemark{a}} & \colhead{Target lines} & \colhead{Line freq. (MHz)\tablenotemark{b}}
}

\startdata
4770 & 58 & H(111)$\alpha$ & 4744.18 \\
     &    & OH $^2\Pi_{1/2}$ J=1/2 F=1 & 4750.66 \\
4850 & 89 & H$_2$CO 1$_{10}$--1$_{11}$ & 4829.66 \\
     &    & H(110)$\alpha$ & 4874.16 \\
4994 & 58 & H(109)$\alpha$ & 5008.92 \\
5154 & 58 & H(108)$\alpha$ & 5148.70 \\
5298 & 89 & H(107)$\alpha$ & 5293.73 \\
5442 & 58 & H(106)$\alpha$ & 5444.26 \\
5602 & 58 & H(105)$\alpha$ & 5600.55 \\
5762 & 58 & H(104)$\alpha$ & 5762.88 \\
5923 & 58 & H(103)$\alpha$ & 5931.54 \\
6019 & 58 & OH $^2\Pi_{3/2}$ J=5/2 F=2 & 6030.75 \\
     &    & OH $^2\Pi_{3/2}$ J=5/2 F=3 & 6035.09 \\
6115 & 58 & H(102)$\alpha$ & 6106.86 \\
6275 & 58 & H(101)$\alpha$ & 6289.14 \\
6467 & 58 & H(100)$\alpha$ & 6478.76 \\
6659 & 58 & CH$_3$OH 5$_{15}$--6$_{06}$ & 6668.52 \\
     &    & H(99)$\alpha$ & 6676.08 \\
7784 & 122 & H(94)$\alpha$ & 7792.87 \\
7832 & 58 & \nodata & \nodata \\
7992 & 58 & \nodata & \nodata \\
8056 & 58 & H(93)$\alpha$ & 8045.60 \\
8120 & 58 & \nodata & \nodata \\
8184 & 58 & \nodata & \nodata \\
8248 & 58 & \nodata & \nodata \\
8601 & 58 & H(91)$\alpha$ & 8584.82 \\
8809 & 89 & \nodata & \nodata \\
9049 & 58 & \nodata & \nodata \\
9113 & 58 & HC$_3$N 1--0 & 9098.12 \\
9177 & 58 & H(89)$\alpha$ & 9173.32 \\
9241 & 58 & \nodata & \nodata \\
9497 & 58 & H(88)$\alpha$ & 9487.82 \\
21203 & 58 & \nodata & \nodata \\
21300 & 58 & \nodata & \nodata \\
21364 & 58 & H(67)$\alpha$ & 21384.79 \\
21588 & 58 & \textit{c}-C$_3$H$_2$ 2$_{20}$--2$_{11}$ & 21587.40 \\
21973 & 58 & HNCO 1$_{01}$--0$_{00}$ & 21981.57 \\
22037 & 58 & \nodata & \nodata \\
22229 & 58 & H$_2$O 6$_{16}$--5$_{23}$ & 22235.08 \\
22326 & 122 & H(66)$\alpha$ & 22364.17 \\
22470 & 89 & \nodata & \nodata \\
22678 & 58 & \nodata & \nodata \\
22839 & 58 & \nodata & \nodata \\
22999 & 58 & \nodata & \nodata \\
23095 & 58 & \nodata & \nodata \\
23716 & 89 & NH$_3$ (1,1) & 23694.50 \\
      &    & NH$_3$ (2,2) & 23722.63 \\
23844 & 89 & OH $^2\Pi_{3/2}$ J=9/2 F=4 & 23817.62 \\
      &    & OH $^2\Pi_{3/2}$ J=9/2 F=5 & 23826.62 \\
      &    & NH$_3$ (3,3) & 23870.13 \\
23957 & 58 & \nodata & \\
24149 & 58 & NH$_3$ (4,4) & 24139.42 \\
24309 & 58 & \nodata & \\
24518 & 89 & NH$_3$ (5,5) & 24532.99 \\
      &    & H(64)$\alpha$ & 24509.90 \\
24662 & 58 & \nodata & \nodata \\
24758 & 58 & \nodata & \nodata \\
24790 & 58 & \nodata & \nodata \\
24918 & 58 & \nodata & \nodata \\
25046 & 58 & NH$_3$ (6,6) & 25056.03 \\
25335 & 58 & \nodata & \nodata \\
25399 & 58 & \nodata & \nodata 
\enddata
\tablenotetext{a}{The outermost 100 channels (\textasciitilde3~MHz) were excised from each spectral window; the bandwidth given is after removal.}
\tablenotetext{b}{\citet{2007AAS...21113211R}}
\end{deluxetable*}

Data flagging and calibration were performed using the MIRIAD software \citep{1995ASPC...77..433S}. Standard procedures were used: erroneous data were flagged and sources were calibrated using flux (PKS 1934-638), bandpass, and gain calibrators. Table \ref{table:reductions} shows the reduction parameters and imaging properties for all sources. The data were then inverted and cleaned, with a cell size of 4$^{\prime\prime}$ for all 15mm cubes and 10$^{\prime\prime}$ for 4cm cubes. To increase the S/N, each spectrum was smoothed over 7 channels in frequency space prior to imaging, for a final resolution of 224~kHz (\textasciitilde11~km~s$^{-1}$ at 6~GHz and \textasciitilde2.9~km~s$^{-1}$ at 23~GHz). Using CASA \citep{2007ASPC..376..127M}, each cube was examined in channel space. Potential line-like features in each cube were noted, and continuum subtraction was performed using line-free and interference-free channels. Due to the large frequency range observed for each source, the synthesized beam size varied significantly from the low-frequency end to the high-frequency end of the spectra. Thus, all cubes within a given band were smoothed to a circular beam size using the long-axis beam diameter of the lowest-frequency channel.

The resulting synthesized beams, superimposed on top of optical images of the galaxies, can be seen in Figure \ref{fig:Galaxies_pointings}. As seen, the pointings vary from completely global (e.g. NGC 1266 4cm-band) to purely nuclear (e.g. M83 15mm-band) among the sample.

\begin{deluxetable*}{llllccccc}
\tablecaption{Reduction parameters for each galaxy\label{table:reductions}}
\tablehead{   
\colhead{Source} & \colhead{Band} & \colhead{Bandpass cal\tablenotemark{a}} & \colhead{Gain cal} & \colhead{Beam (arcsec)} & \colhead{Beam (kpc)} & \colhead{Time (min)} & \colhead{Continuum\tablenotemark{b} (Jy)} & \colhead{RMS (mJy bm$^{-1}$)} \\
}

\startdata
NGC 253  & 4cm  & \nodata  & 0023-263   & 140 & 2.42 & 262 & 1.4   & 3.4 \\
NGC 1068 & 15mm & 1934-638 & 0237+040   & 27  & 1.32 & 128 & 0.4   & 14 \\
NGC 1266 & 4cm  & \nodata  & 0336-109   & 140 & 20.0 & 299 & 0.035 & 0.8 \\
         & 15mm & 1921-293 & 0336-109   & 28  & 3.99 & 602 & 0.012 & 1.1 \\
NGC 1365 & 4cm  & \nodata  & 0332-403   & 132 & 11.4 & 238 & 0.15  & 0.65 \\
         & 15mm & 1921-293 & 0332-403   & 33  & 2.85 & 224 & 0.047 & 3.1 \\
NGC 1808 & 4cm  & \nodata  & 0426-380   & 172 & 7.57 & 221 & 0.18  & 1.3 \\
NGC 3256 & 15mm & 0723-008 & 1039-47    & 27  & 4.90 & 180 & 0.050 & 4.4 \\
NGC 4945 & 15mm & 1921-293 & j1326-5256 & 33  & 0.59 & 241 & 0.96  & 4.8 \\
M83      & 15mm & 0723-008 & 1313-333   & 42  & 0.95 & 179 & 0.047 & 4.2 \\
Circinus & 15mm & 0723-008 & 1532-63    & 40  & 0.82 & 340 & 0.15  & 3.7 
\enddata

\tablenotetext{a}{A dedicated bandpass calibrator was only used for 15mm data. For 4cm data, bandpass calibration was performed using the flux calibrator 1934-638.}
\tablenotetext{b}{An intrinsic 10\% uncertainty is applied for all flux measurements}
\end{deluxetable*}

The post-subtraction cubes were then examined, and for all sources a spectrum was extracted from the central synthesized beam pixel using PySpecKit \citep{2011ascl.soft09001G}. Lines appearing roughly Gaussian were fit using PySpecKit's Gaussian fitting algorithm, while those with a clearly non-Gaussian profile were fit using multiple Gaussians if possible. Lines were then cross-identified with Splatalogue \citep{2007AAS...21113211R} for possible matches. We apply an intrinsic 10\% uncertainty to all flux measurements.

%% file: Results.tex
\section{LINES DETECTED}\label{Ch4}

We detected at least one molecular spectral line in eight of our nine galaxies. No morphological structure is observed in either the continuum or spectral line emission from any of the galaxies within the spatially smoothed cubes, and the two nuclei of the double-nucleus source NGC~3256 (separation 5$^{\prime\prime}$) are observed as a single object. Table \ref{table:lines_detected} shows the parameters of the molecular lines detected within the galaxies; transitions selected for further analysis are shown in boldface. Hydrogen radio recombination lines (RRLs) were only detected in NGC~253; the observed RRL parameters can be seen in Table \ref{table:RRLs_detected}. Transitions observed but not detected are shown in Appendix \ref{A1} for each galaxy.

Throughout the remainder of this paper, we focus our analysis on RRL, OH, and NH$_3$ transitions. All velocities are quoted in the radio LSRK frame.

\startlongtable
\begin{deluxetable*}{llcccccc}
\tablecaption{Molecular lines detected; transitions in boldface are selected for further analysis\label{table:lines_detected}}
\tablehead{
\colhead{Galaxy} & \colhead{Species} & \colhead{Transition} & \colhead{$\upsilon_{obs}$ (km s$^{-1}$)} & \colhead{S$_{\nu,peak}$ (mJy bm$^{-1}$)} & \colhead{$\Delta \upsilon_{FWHM}$ (km s$^{-1}$)} & \colhead{$T_{mb}$ (mK)} & \colhead{$\int T_{mb} d\upsilon$ (K km s$^{-1}$)}
}
\startdata
NGC~253 & \nodata & \textbf{5.737~GHz\tablenotemark{a}} & \nodata & \textbf{6.7~$\pm$~1.2} & \textbf{121~$\pm$~21} & \textbf{12.7~$\pm$~2.2} & \textbf{1.65~$\pm$~0.40} \\
 & H$_2$CO & 1$_{10}$--1$_{11}$ & 228~$\pm$~4 & -14.9~$\pm$~1.6 & 159~$\pm$~9 & -39.7~$\pm$~4.4 & -6.75~$\pm$~0.84 \\
 & \textbf{OH $^2\Pi_{3/2}$} & \textbf{J=5/2 F=2} & 231~$\pm$~9 & -12.3~$\pm$~2.3 & 123~$\pm$~29 & -21.1~$\pm$~3.9 & -2.76~$\pm$~0.83 \\
 & \textbf{OH $^2\Pi_{3/2}$} & \textbf{J=5/2 F=3} & 229~$\pm$~10 & -13.3~$\pm$~2.2 & 146~$\pm$~27 & -22.8~$\pm$~3.7 & -3.55~$\pm$~0.87 \\
NGC~1068 & H$_2$O & 6$_{16}$--5$_{23}$ & 1403~$\pm$~7 & 215~$\pm$~48 & 54~$\pm$~16 & 740~$\pm$~160 & 42~$\pm$~16 \\
 &  &  & 1134~$\pm$~2 & 106~$\pm$~18 & 33~$\pm$~5 & 362~$\pm$~63 & 12.9~$\pm$~3.1 \\
 &  &  & 845~$\pm$~2 & 78~$\pm$~18 & 24~$\pm$~7 & 266~$\pm$~61 & 6.7~$\pm$~2.2 \\
NGC~1266 & NH$_3$ & (3,3) & 2176~$\pm$~2 & 1.19~$\pm$~0.38 & 95~$\pm$~36 & 3.3~$\pm$~1.1 & 0.33~$\pm$~0.17 \\
NGC~1365 & H$_2$CO & 1$_{10}$--1$_{11}$ & 1598~$\pm$~27 & -1.26~$\pm$~0.29 & 244~$\pm$~66 & -3.40~$\pm$~0.78 & -0.88~$\pm$~0.31 \\
 & \textbf{NH$_3$} & \textbf{(1,1)} & 1552~$\pm$~10 & 6.0~$\pm$~1.2 & 115~$\pm$~24 & 12.1~$\pm$~2.4 & 1.48~$\pm$~0.42 \\
 & \textbf{NH$_3$} & \textbf{(3,3)} & 1711~$\pm$~8 & 5.4~$\pm$~1.5 & 61~$\pm$~18 & 10.8~$\pm$~3.0 & 0.70~$\pm$~0.28 \\
 &  &  & 1545~$\pm$~6 & 8.8~$\pm$~1.3 & 112~$\pm$~14 & 17.5~$\pm$~2.5 & 2.08~$\pm$~0.40 \\
NGC~3256 & H$_2$O & 6$_{16}$--5$_{23}$ & 2855~$\pm$~3 & 16.1~$\pm$~2.1 & 69.6~$\pm$~7.7 & 55.6~$\pm$~7.4 & 4.12~$\pm$~0.71 \\
 &  &  & 2789~$\pm$~14 & 7.0~$\pm$~1.2 & 102~$\pm$~45 & 24.2~$\pm$~4.2 & 2.6~$\pm$~1.2 \\
 &  &  & 2699~$\pm$~4 & 5.7~$\pm$~1.5 & 32.0~$\pm$~9.6 & 19.7~$\pm$~5.2 & 0.67~$\pm$~0.27 \\
NGC~4945 & \textit{c}-C$_3$H$_2$ & 2$_{20}$--2$_{11}$ & 564~$\pm$~6 & -17.6~$\pm$~2.0 & 220~$\pm$~13 & -42.5~$\pm$~4.8 & -9.9~$\pm$~1.3 \\
 & H$_2$O & 6$_{16}$--5$_{23}$ & \nodata & \nodata & \nodata & \nodata & \nodata \\
 & \textbf{NH$_3$} & \textbf{(1,1)} & 707~$\pm$~3 & 18.7~$\pm$~3.1 & 47.8~$\pm$~7.5 & 37.6~$\pm$~6.3 & 1.91~$\pm$~0.44 \\
 &  &  & 626~$\pm$~3 & -21.2~$\pm$~3.4 & 39.4~$\pm$~6.3 & -42.6~$\pm$~6.9 & -1.78~$\pm$~0.41 \\
 &  &  & 563~$\pm$~3 & -14.3~$\pm$~3.5 & 28.4~$\pm$~7.5 & -28.7~$\pm$~7.0 & -0.87~$\pm$~0.31 \\
 &  &  & 458~$\pm$~5 & 14.5~$\pm$~2.8 & 76~$\pm$~13 & 29.1~$\pm$~5.6 & 2.35~$\pm$~0.60 \\
 & \textbf{NH$_3$} & \textbf{(2,2)} & 681~$\pm$~3 & -22.3~$\pm$~5.8 & 42.9~$\pm$~9.3 & -45~$\pm$~12 & -2.04~$\pm$~0.69 \\
 &  &  & 596~$\pm$~14 & -16.2~$\pm$~2.3 & 110~$\pm$~69 & -32.4~$\pm$~4.7 & -3.8~$\pm$~2.4 \\
 &  &  & 513~$\pm$~10 & 12.7~$\pm$~7.8 & 58~$\pm$~22 & 25~$\pm$~16 & 1.6~$\pm$~1.1 \\
 & \textbf{OH $^2\Pi_{3/2}$} & \textbf{J=9/2 F=4} & 672~$\pm$~2 & -21.5~$\pm$~2.6 & 52~$\pm$~4 & -42.7~$\pm$~5.2 & -2.36~$\pm$~0.34 \\
 & \textbf{OH $^2\Pi_{3/2}$} & \textbf{J=9/2 F=5} & 658~$\pm$~2 & -20.3~$\pm$~2.6 & 56~$\pm$~6 & -40.3~$\pm$~5.3 & -2.40~$\pm$~0.40 \\
 & \textbf{NH$_3$} & \textbf{(3,3)} & \nodata & \nodata & \nodata & \nodata & \nodata \\
 & \textbf{NH$_3$} & \textbf{(4,4)} & 615~$\pm$~3 & -10.8~$\pm$~3.3 & 18.5~$\pm$~6.4 & -20.9~$\pm$~6.3 & -0.41~$\pm$~0.19 \\
 &  &  & 569~$\pm$~3 & -11.2~$\pm$~2.7 & 28.7~$\pm$~7.6 & -21.7~$\pm$~5.3 & -0.66~$\pm$~0.24 \\
 &  &  & 435~$\pm$~9 & 7.7~$\pm$~1.5 & 103~$\pm$~21 & 14.9~$\pm$~2.9 & 1.63~$\pm$~0.46 \\
 & \textbf{NH$_3$} & \textbf{(5,5)}\tablenotemark{b} & 712~$\pm$~77 & 8.1~$\pm$~3.5 & 160~$\pm$~100 & 15.2~$\pm$~6.5 & 2.6~$\pm$~2.0 \\
 &  &  & 643~$\pm$~11 & -10.2~$\pm$~9.9 & 69~$\pm$~40 & -19~$\pm$~18 & -1.4~$\pm$~1.6 \\
 &  &  & 556~$\pm$~11 & -12.7~$\pm$~3.6 & 84~$\pm$~20 & -23.8~$\pm$~6.8 & -2.11~$\pm$~0.80 \\
 & \textbf{NH$_3$} & \textbf{(6,6)} & 637~$\pm$~10 & -8.5~$\pm$~6.2 & 32~$\pm$~23 & -15~$\pm$~11 & -0.52~$\pm$~0.53 \\
 &  &  & 575~$\pm$~25 & -7.1~$\pm$~2.6 & 85~$\pm$~68 & -12.7~$\pm$~4.7 & -1.1~$\pm$~1.0 \\
 &  &  & 433~$\pm$~13 & 8.0~$\pm$~2.5 & 85~$\pm$~34 & 14.3~$\pm$~4.5 & 1.29~$\pm$~0.66 \\
M83 & \textbf{NH$_3$} & \textbf{(1,1)} & 504~$\pm$~9 & 19.7~$\pm$~7.9 & 46~$\pm$~20 & 24.4~$\pm$~9.7 & 1.18~$\pm$~0.71 \\
 & \textbf{NH$_3$} & \textbf{(2,2)} & 546~$\pm$~8 & 7.8~$\pm$~2.2 & 60~$\pm$~19 & 9.6~$\pm$~2.8 & 0.62~$\pm$~0.26 \\
 & \textbf{NH$_3$} & \textbf{(3,3)} & 510~$\pm$~6 & 12.8~$\pm$~2.0 & 108~$\pm$~14 & 15.7~$\pm$~2.4 & 1.81~$\pm$~0.37 \\
Circinus & H$_2$O & 6$_{16}$--5$_{23}$ & \nodata & \nodata & \nodata & \nodata & \nodata \\
 & \textbf{OH $^2\Pi_{3/2}$} & \textbf{J=9/2 F=4} & 392~$\pm$~4 & -15.8~$\pm$~2.2 & 83~$\pm$~11 & -21.3~$\pm$~3.0 & -1.87~$\pm$~0.36 \\
 & \textbf{OH $^2\Pi_{3/2}$} & \textbf{J=9/2 F=5} & 402~$\pm$~8 & -13.0~$\pm$~1.9 & 120~$\pm$~21 & -17.5~$\pm$~2.6 & -2.25~$\pm$~0.51
\enddata
\tablenotetext{a}{No plausible species were identified; we thus show the observed line frequency instead.}
\tablenotetext{b}{Contamination from the H(64)$\alpha$ line is likely present.}
\tablecomments{Gaussian fit parameters are included for all lines that are not confirmed or likely masers. No lines were detected in NGC~1808 (see note in Table \ref{table:reductions}.}
\end{deluxetable*}

\begin{deluxetable*}{lccccc}
\tablecaption{Radio recombination lines detected within NGC~253\label{table:RRLs_detected}}
\tablehead{
\colhead{Transition} & \colhead{$\upsilon_{obs}$ (km s$^{-1}$)} & \colhead{S$_{\nu,peak}$ (mJy bm$^{-1}$)} & \colhead{$\Delta \upsilon_{FWHM}$ (km s$^{-1}$)} & \colhead{$T_{mb}$ (mK)} & \colhead{$\int T_{mb}d\upsilon$ (K km s$^{-1}$)}
}
\startdata
H(111)$\alpha$ & 186~$\pm$~9 & 10.2~$\pm$~1.3 & 237~$\pm$~22 & 28.3~$\pm$~3.6 & 7.12~$\pm$~1.13 \\
H(110)$\alpha$ & 229~$\pm$~7 & 8.2~$\pm$~1.1 & 159~$\pm$~16 & 21.4~$\pm$~2.7 & 3.64~$\pm$~0.60 \\
H(109)$\alpha$ & 265~$\pm$~9 & 8.4~$\pm$~1.0 & 244~$\pm$~21 & 21.0~$\pm$~2.6 & 5.46~$\pm$~0.83 \\
H(108)$\alpha$\tablenotemark{a} & 196~$\pm$~11 & 5.5~$\pm$~1.1 & 130~$\pm$~26 & 13.1~$\pm$~2.7 & 1.81~$\pm$~0.52 \\
H(107)$\alpha$ & 229~$\pm$~8 & 10.6~$\pm$~1.4 & 182~$\pm$~20 & 23.6~$\pm$~3.2 & 4.56~$\pm$~0.80 \\
H(106)$\alpha$ & 188~$\pm$~8 & 8.0~$\pm$~1.1 & 175~$\pm$~19 & 16.8~$\pm$~2.4 & 3.13~$\pm$~0.56 \\
H(105)$\alpha$ & 237~$\pm$~5 & 12.2~$\pm$~1.3 & 222~$\pm$~15 & 24.3~$\pm$~2.7 & 5.74~$\pm$~0.74 \\
H(104)$\alpha$ & 228~$\pm$~4 & 11.2~$\pm$~1.2 & 182~$\pm$~11 & 21.1~$\pm$~2.3 & 4.08~$\pm$~0.52 \\
H(103)$\alpha$\tablenotemark{a} & 278~$\pm$~14 & 5.4~$\pm$~0.9 & 213~$\pm$~33 & 9.5~$\pm$~1.6 & 2.16~$\pm$~0.50 \\
H(102)$\alpha$ & 161~$\pm$~8 & 10.6~$\pm$~1.4 & 204~$\pm$~20 & 17.7~$\pm$~2.3 & 3.84~$\pm$~0.63 \\
H(101)$\alpha$\tablenotemark{a} & 255~$\pm$~18 & 5.7~$\pm$~1.3 & 183~$\pm$~43 & 9.0~$\pm$~2.1 & 1.75~$\pm$~0.58 \\
H(100)$\alpha$ & 186~$\pm$~7 & 13.4~$\pm$~1.6 & 222~$\pm$~23 & 20.0~$\pm$~2.3 & 4.73~$\pm$~0.73 \\
H(099)$\alpha$ & 214~$\pm$~8 & 12.9~$\pm$~1.5 & 312~$\pm$~19 & 18.0~$\pm$~2.0 & 5.98~$\pm$~0.77 \\
H(094)$\alpha$ & 220~$\pm$~11 & 10.9~$\pm$~1.6 & 216~$\pm$~26 & 11.2~$\pm$~1.7 & 2.57~$\pm$~0.49 \\
H(093)$\alpha$ & 214~$\pm$~6 & 11.6~$\pm$~1.9 & 162~$\pm$~24 & 11.2~$\pm$~1.8 & 1.92~$\pm$~0.42 \\
H(091)$\alpha$ & 188~$\pm$~20 & 8.0~$\pm$~1.4 & 283~$\pm$~48 & 6.8~$\pm$~1.2 & 2.04~$\pm$~0.50 \\
H(089)$\alpha$ & 207~$\pm$~8 & 13.5~$\pm$~2.0 & 149~$\pm$~19 & 10.0~$\pm$~1.5 & 1.58~$\pm$~0.31 \\
H(088)$\alpha$ & 180~$\pm$~16 & 7.6~$\pm$~1.6 & 184~$\pm$~39 & 5.3~$\pm$~1.1 & 1.03~$\pm$~0.31
\enddata
\tablenotetext{a}{The baselines near these transitions suffer from significant ringing effects due to the problems with antenna CA05, and thus the results of the Gaussian fits are not representative of the true line strength. They have been excluded from further calculations.}
\end{deluxetable*}

During the NGC~253 observations, a ringing effect was produced by antenna CA05, resulting in poor-quality data. As the antenna was positioned on the edge of the array at the time of observation, the antenna's data could not be removed without causing a severely-elongated beam, impeding the inversion process, and substantially increasing the noise. As a consequence, a slight baseline ripple is present within the data, most notable at lower frequencies.

The majority of detected lines were observed with roughly Gaussian profiles of no more than two separable components, the few non-Gaussian lines do not have intensity or FWHM values in Table \ref{table:lines_detected}. H$_2$CO has been extensively studied by \citet{2008ApJ...673..832M,2013ApJ...766..108M} and H$_2$O has been subject to numerous studies in each detection in our sample \citep[e.g.][]{2001ApJ...556..694G,2009AaA...502..529S,1997ApJ...481L..23G,2003ApJ...590..162G}, while \textit{c}-C$_3$H$_2$ is only detected from a single transition in a single galaxy. We thus focus our analysis primarily on the RRL, OH, and NH$_3$ transitions.

Discounting the RRLs in NGC~253, we detect the most lines in NGC~4945. In NGC~4945, the lines from different species are strikingly different from each other in morphology. The OH and \textit{c}-C$_3$H$_2$ lines are both detected solely in absorption, but with drastically different FWHMs of \textasciitilde55~km~s$^{-1}$ and \textasciitilde 220~km~s$^{-1}$, respectively. Furthermore, the NH$_3$ transitions bear no resemblance at all to either of the former two lines, with a complicated superposition of emission and absorption. The H$_2$O maser has a double-peaked velocity structure.

%% file: RRL.tex
\section{RADIO RECOMBINATION LINES}
\begin{figure}
  \plotone{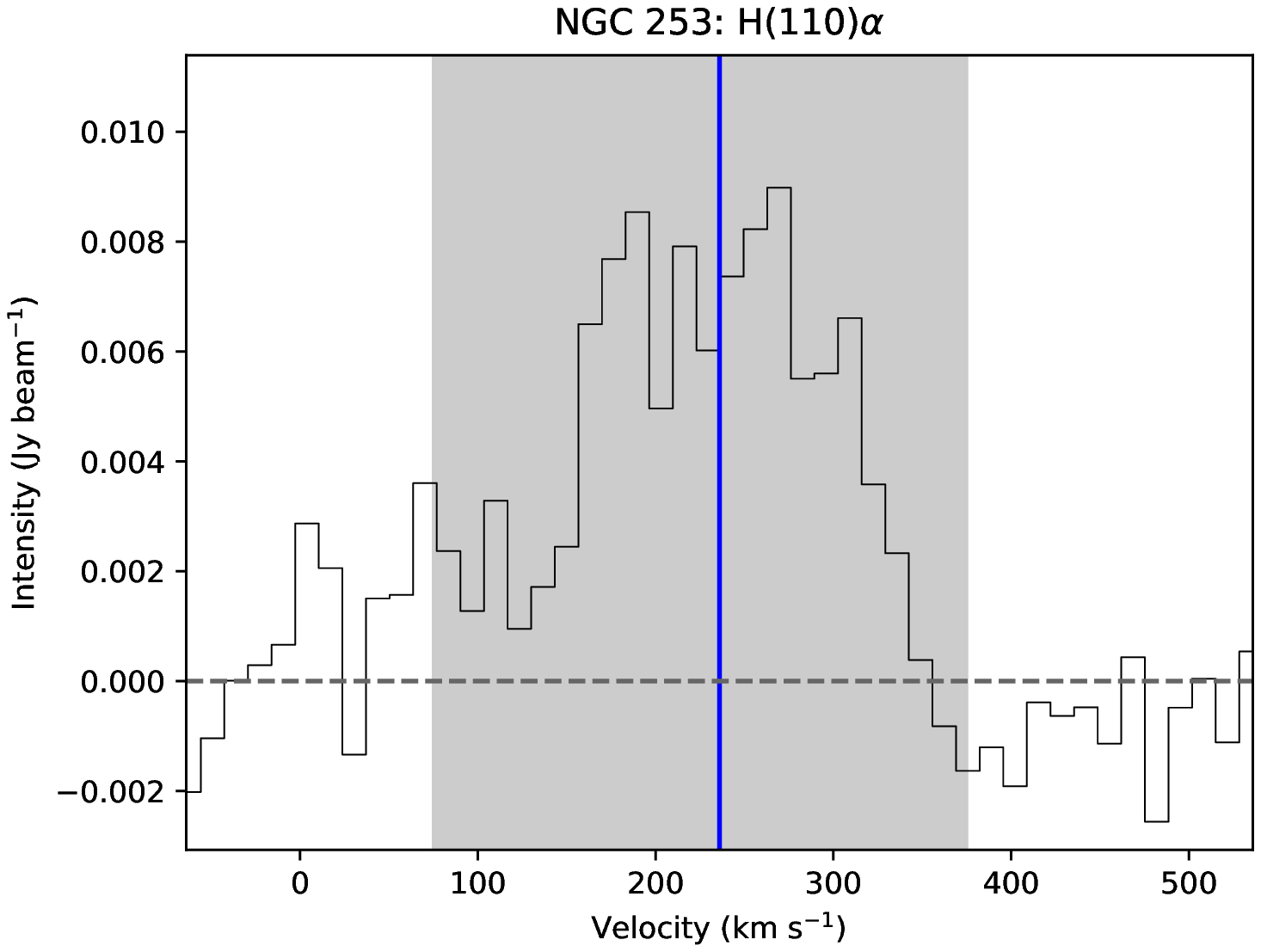}
  \plotone{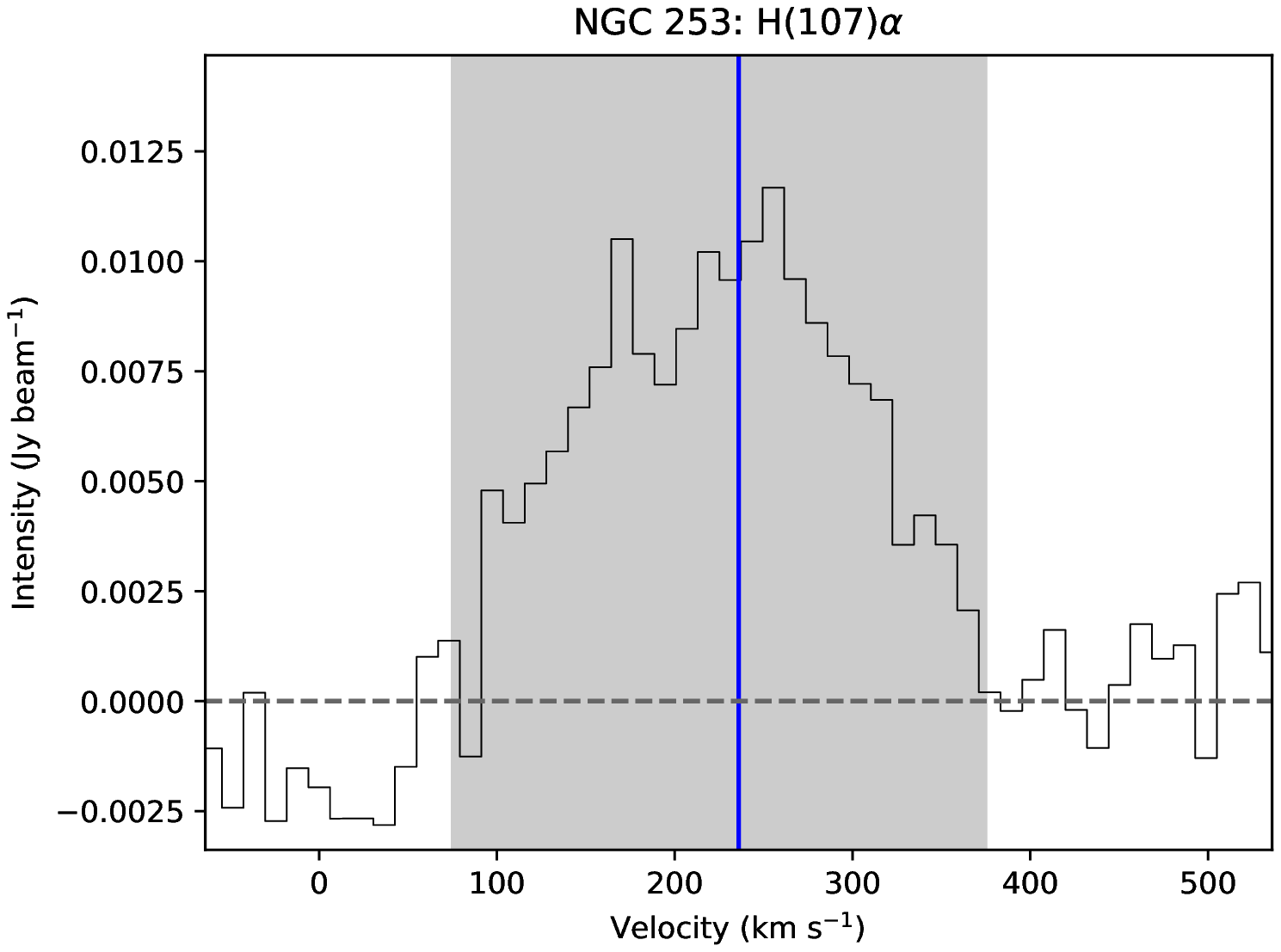}
  \plotone{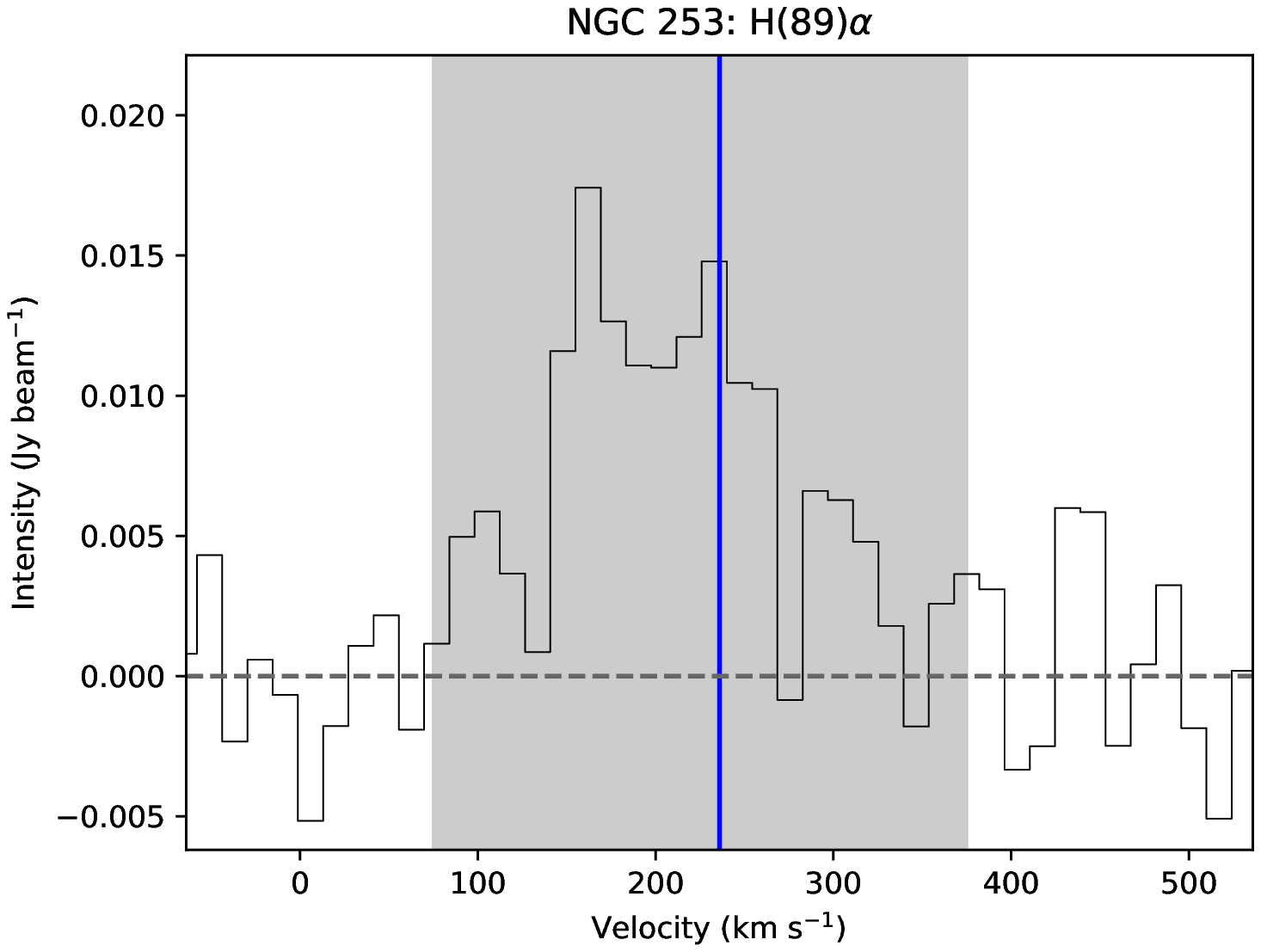}
  \caption{Profiles of three detected RRLs within NGC~253; other H(n)$\alpha$ RRLs not shown have similar morphologies. The blue vertical lines show the systemic velocity \citep{2004AJ....128...16K}, while the shaded region represents the approximate extent of the line.}
  \label{fig:n253_RRL_lines}
\end{figure}

We detect radio recombination lines (RRLs) within a single galaxy: NGC~253. All 18 H$\alpha$ RRLs within our spectral windows were detected, while neither higher-order H$\beta$ transitions nor lines from other elements were detected. All lines display roughly Gaussian profiles (e.g. Figure \ref{fig:n253_RRL_lines}). Three lines --- H(108)$\alpha$, H(103)$\alpha$, and H(101)$\alpha$ --- are corrupted by the antenna CA05 problems, and we exclude these lines from further analysis. The remaining galaxies contain no discernible RRL emission to our thresholds.

To attempt to detect RRLs in the remaining galaxies with 4cm-band data (NGC~1266, NGC~1365, and NGC~1808), we stacked all RRL spectra from each galaxy. The spectra containing the RRL frequencies were smoothed to a common velocity resolution of 15 km s$^{-1}$, centered in velocity, and averaged together. No clear evidence of RRLs was present in any of the galaxies after stacking to upper limits of \textasciitilde0.25~mJy~bm$^{-1}$ in NGC~1266 and NGC~1365, and \textasciitilde0.35~mJy~bm$^{-1}$ in NGC~1808.

\subsection{NGC~253 and Electron Temperature}

The detection of a radio recombination line from a source allows for the calculation of the average electron temperature as long as the emitting \ion{H}{2} region is under LTE. The electron temperature is a kinetic temperature, namely a measure of the electrons' average velocities according to the Maxwell-Boltzmann distribution. The electron temperature assuming LTE, $T_e^*$, is related to the line-to-continuum ratio, via the formula \citep{2000tra..book.....R}
\begin{equation}
\label{eq:e_temp_LTE}
\frac{T_e^*}{\textrm{K}} = \left[7.0\times10^3 \left(\frac{\nu}{\textrm{GHz}}\right)^{1.1} 1.08^{-1} \left(\frac{\Delta \upsilon}{\textrm{km s}^{-1}}\right)^{-1} \left(\frac{T_C}{T_L}\right)\right]^{0.87},
\end{equation}
where $T_L/T_c$ is the line-to-thermal-continuum peak brightness temperature ratio, and $\Delta \upsilon$ is the FWHM of the line. In addition to LTE conditions, this formula assumes that the \ion{H}{2} region is isothermal, the optical depths of the line $\tau_L$ and continuum $\tau_c$ obey the relations $|\tau_L - \tau_C| \ll 1$ and $\tau_C \ll 1$, and the ratio of helium atoms to hydrogen atoms $N_{\textrm{He}}/N_{\textrm{H}} = 0.08$.

The detection of 18 RRLs within NGC~253 (e.g. Figure \ref{fig:n253_RRL_lines}), in theory, allows for the calculation of $T_e$ for each line. However, the process is complicated by the fact that not all of the continuum emission from NGC~253 is from thermal free-free interactions. Synchrotron emission is often the dominant emission mechanism at centimeter wavelengths, with free-free only contributing a small fraction. Thus, the free-free continuum $S_{\nu,ffc}$ must be separately estimated. We utilized the work of \citet{2010ApJ...710.1462W}, who compare their data to past literature to separate the nonthermal and thermal components of NGC~253's spectrum between 1~GHz and 7~GHz. They have a comparable angular resolution to our data and well-sample the frequency range, and thus their data are ideal for our purposes. From their data, we extract the formula:
\begin{equation}
\label{eq:ff_cont}
S_{\nu,ffc} = \frac{1.73}{1.37} \frac{4.58 \times 10^{-6} \nu^3}{e^{4.79 \times 10^{-6} \nu} - 1} \left(1 - e^{-0.27 \nu^{-2.1}}\right),
\end{equation}
where $\nu$ is in GHz and $S_{\nu,ffc}$ in Jy, a fit to the data for their paper. The prefactor $1.73/1.37$ scales their flux density at 5~GHz to our observed flux density. The second factor is simply blackbody continuum emission, while the final factor is the optical depth correction $1 - e^{-\tau}$ evaluated using the optical depths from their paper. To estimate the free-free continuum for the RRLs at frequencies higher than 7~GHz, we assume that the extrapolation of this formula is valid until 10~GHz. Between 4~GHz and 10~GHz, this free-free continuum SED is nearly flat, varying between 0.28~Jy and 0.26~Jy, so the extrapolation does not affect the result significantly. Comparing the continuum slope from \citet{2010ApJ...710.1462W} to those from our data, this free-free continuum fit appears reasonable when applied to our data.

\begin{figure}
  \centering
  \plotone{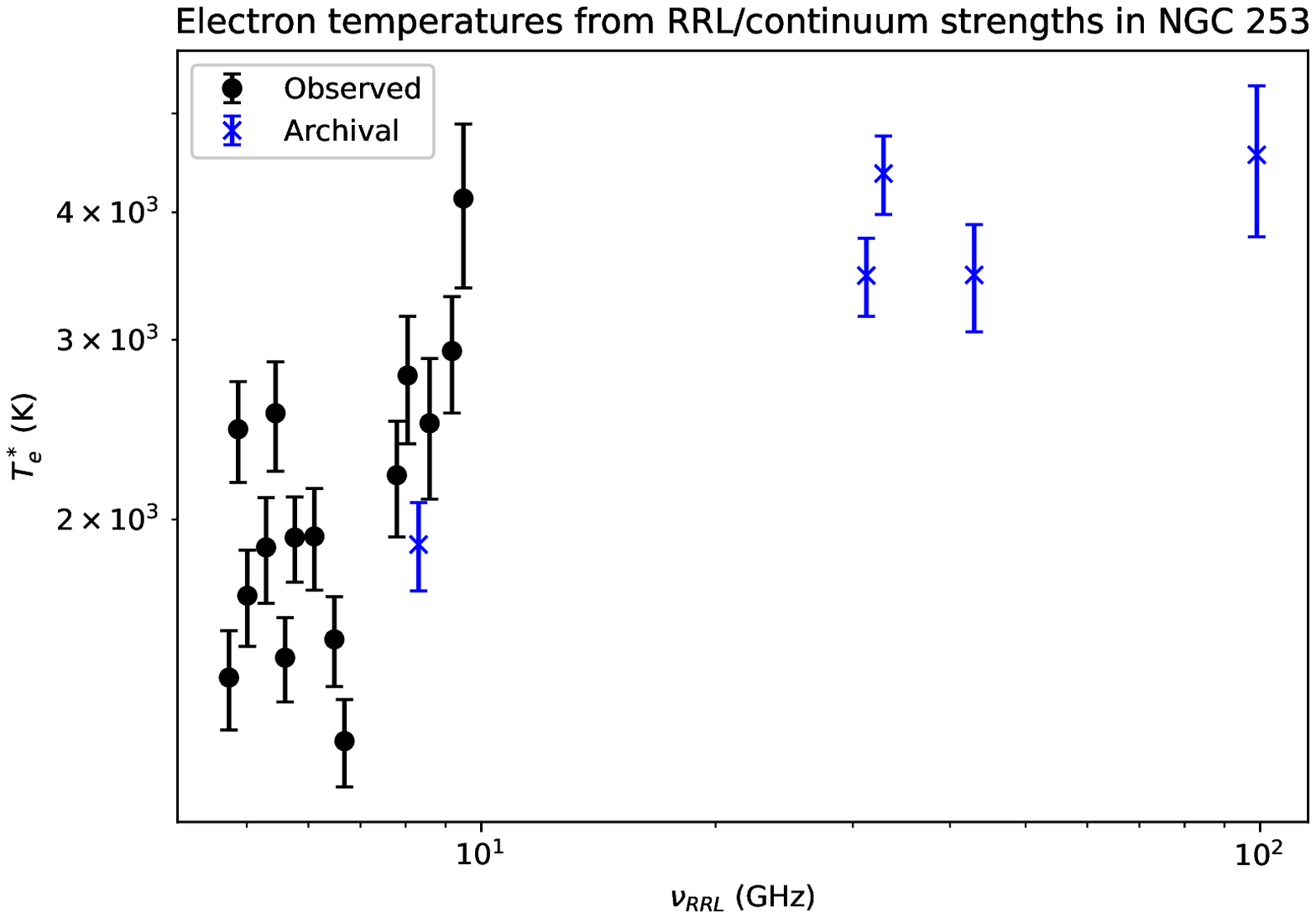}
  \plotone{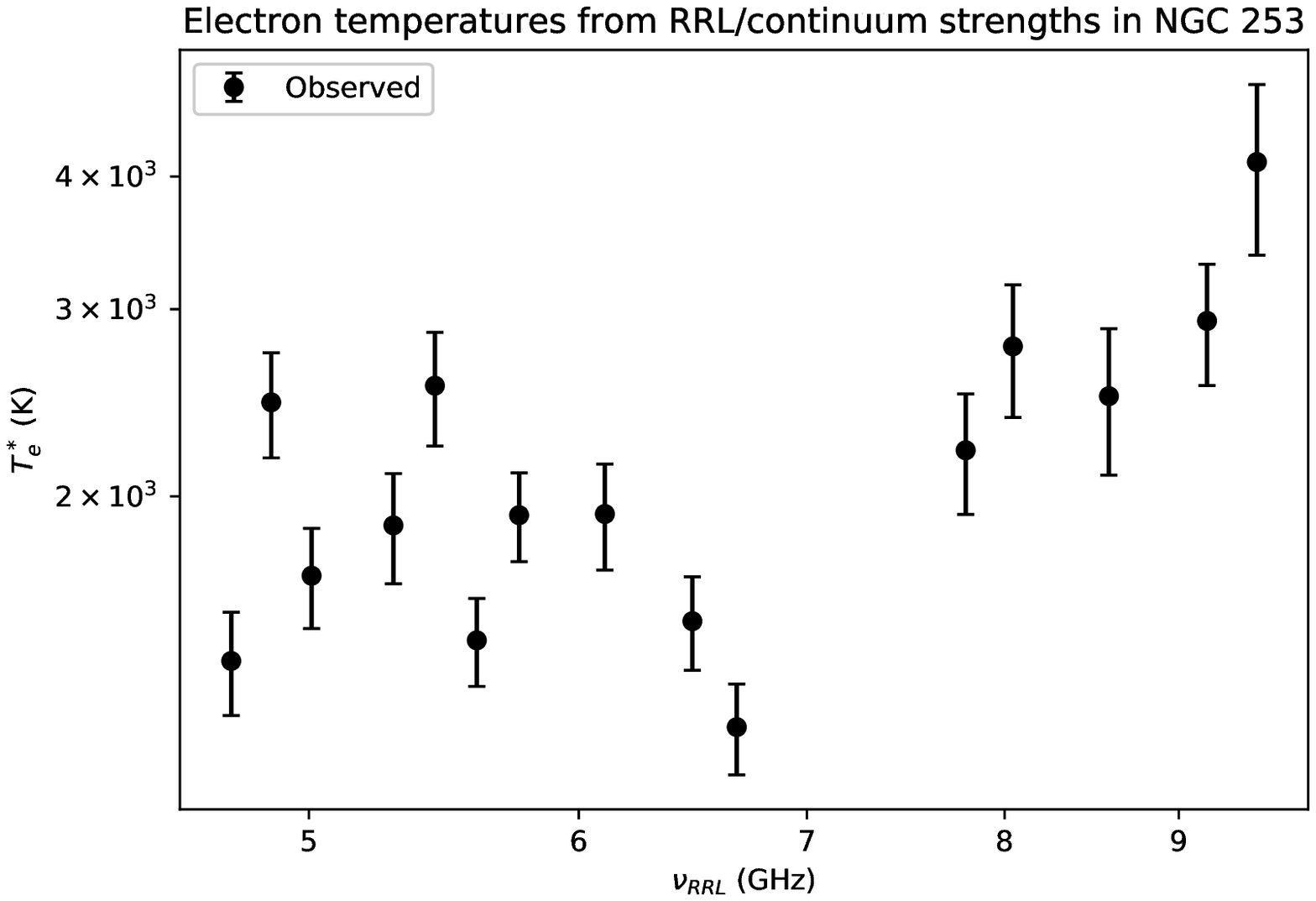}
  \caption{LTE Electron temperatures in NGC~253, calculated from RRL line-to-continuum ratios. On the bottom is only our data, while the top includes archival data from Table \ref{table:n253_rrl_hist}.}
  \label{fig:n253_e_temp}
\end{figure}

Historical RRL observations (Table \ref{table:n253_rrl_hist}) with associated continuum measurements have also been included. Only historical observations and measurements over the entire nuclear region are utilized; we exclude studies that focus on smaller spatial scales. From the literature values of $S_{\nu,l}$, $S_{\nu,ffc}$, and $\Delta \upsilon$, we calculate electron temperatures using Equation \ref{eq:e_temp_LTE}. The results can be seen in Table \ref{table:n253_e_temp} and Figure \ref{fig:n253_e_temp}. The displayed error bars are solely statistical errors from the Gaussian fits. Our derived electron temperatures are close to that of the H(92)$\alpha$ line from \citet{2006ApJ...644..914R}.

\begin{deluxetable}{lcc}
\tablecaption{LTE Electron temperatures calculated from each RRL in NGC~253\label{table:n253_e_temp}}
\tablehead{
\colhead{Transition} & \colhead{$\nu_0$ (GHz)} & \colhead{$T_e^*$ (K)}
}
\startdata
H(111)$\alpha$ & 4.744183 & 1400 $\pm$ 160 \\
H(110)$\alpha$ & 4.874157 & 2450 $\pm$ 280 \\
H(109)$\alpha$ & 5.008922 & 1680 $\pm$ 180 \\
H(107)$\alpha$ & 5.293732 & 1880 $\pm$ 220 \\
H(106)$\alpha$ & 5.444260 & 2540 $\pm$ 310 \\
H(105)$\alpha$ & 5.600550 & 1460 $\pm$ 140 \\
H(104)$\alpha$ & 5.762880 & 1920 $\pm$ 180 \\
H(102)$\alpha$ & 6.106855 & 1930 $\pm$ 220 \\
H(100)$\alpha$ & 6.478759 & 1530 $\pm$ 150 \\
H(99)$\alpha$ & 6.676075 & 1210 $\pm$ 120 \\
H(94)$\alpha$ & 7.792870 & 2210 $\pm$ 290 \\
H(93)$\alpha$ & 8.045602 & 2770 $\pm$ 400 \\
H(91)$\alpha$ & 8.584820 & 2490 $\pm$ 390 \\
H(89)$\alpha$ & 9.173320 & 2930 $\pm$ 380 \\
H(88)$\alpha$ & 9.487820 & 4130 $\pm$ 760
\enddata
\end{deluxetable}

\begin{deluxetable*}{lcccccc}
\tablecaption{Historical unresolved RRL detections and peak intensities in NGC~253\label{table:n253_rrl_hist}, along with calculated LTE electron temperatures}
\tablehead{
\colhead{Transition} & \colhead{$\theta_{FWHM}$ (arcsec)} & \colhead{$\nu_0$ (GHz)} & \colhead{$S_\nu$ (mJy)} & \colhead{$S_{\nu,ffc}$ (mJy)} & \colhead{Reference} & \colhead{$T_e^*$\tablenotemark{a} (K)}
}

\startdata
H(92)$\alpha$ & 1.5 $\times$ 1.0 & 8.309 & 9.0 $\pm$ 0.5 & 150 $\pm$ 15 & 1 & 1890 $\pm$ 188 \\
H(59)$\alpha$ & 1.8 $\times$ 1.6 & 31.223 & 16.3 $\pm$ 0.2 & 140 $\pm$ 15\tablenotemark{b} & 2 & 3468 $\pm$ 304 \\
H(58)$\alpha$ & 1.8 $\times$ 1.5 & 32.852 & 17.5 $\pm$ 0.3 & 140 $\pm$ 15\tablenotemark{b} & 2 & 4366 $\pm$ 386 \\
H(53)$\alpha$ & 1.5 $\times$ 1.0 & 42.952 & 21 $\pm$ 2 & 140 $\pm$ 15 & 1 & 3473 $\pm$ 417 \\
H(40)$\alpha$ & 1.9 $\times$ 1.6 & 99.023 & 38 $\pm$ 5 & 102 $\pm$ 15 & 3 & 4554 $\pm$ 768
\enddata
\tablenotetext{a}{Calculated using Equation \ref{eq:e_temp_LTE}}
\tablenotetext{b}{Free-free continuum estimated from \citet{2006ApJ...644..914R}}
\tablerefs{(1) \citet{2006ApJ...644..914R}; (2) \citet{2011ApJ...739L..24K}; (3) \citet{2015MNRAS.450L..80B}}
\end{deluxetable*}

Within the Milky Way, typical electron temperatures in the Galactic Center range from \textasciitilde5400--7000~K \citep[e.g.][]{2001AJ....121.2681L}, significantly higher than those in NGC~253 found in both our and archival calculations. Typical temperatures for other nearby starbursts are also in the above range \citep[e.g. NGC 4945,][]{2016MNRAS.463..252B}, while others such as M82 are even higher at \textasciitilde10000~K \citep{1996ApJ...465..691S}. This suggests that either NGC~253 has an unusually low electron temperature in its nucleus, or other non-LTE effects are in play.

\subsubsection{Non-LTE effects}

As Figure \ref{fig:n253_e_temp} shows, the electron temperature appears to increase with increasing line frequency/decreasing quantum number, both within our data and including the archival data. Equation \ref{eq:e_temp_LTE} is accurate if LTE conditions are present within the galaxy, but if non-LTE conditions predominate then the calculated LTE electron temperature $T_e^*$ is not an accurate representation of the true electron temperature $T_e$. Instead, an additional factor must be added to address stimulated emission, such that \citep{2000tra..book.....R}
\begin{equation}
\label{eq:e_temp}
T_e = T_e^* \left[b_n\left(1-\frac{\beta_n\tau_c}{2}\right)\right]^{0.87}.
\end{equation}
Here, $\tau_c$ is the continuum optical depth, $b_n$ are the LTE departure coefficients defined as $b_n = N_n/N_n^*$ (the ratio of true to LTE-expected number density), and $1 - \beta_n \propto d \ln b_n/dn$. 

The departure coefficients $b_n$ do not vary significantly from unity at our frequencies, and most of the deviations from LTE line strengths arise from the $\beta_n$ term. $b_n$ is generally an increasing function of $n$, and therefore $\beta_n$ is usually negative. We thus expect the values of $T_e^*$ to increasingly deviate from LTE for lower-frequency RRLs assuming optically-thin emission. This trend is noticed in Figure \ref{fig:n253_e_temp}, and it thus seems reasonable that the ionized hydrogen in NGC~253 is not in LTE. In non-LTE conditions, the emission measure of the galaxy cannot be calculated without knowledge of the departure coefficients.

Within Galactic \ion{H}{2} regions, trends of increasing electron temperature are commonly observed with decreasing quantum number between 3~GHz and 10~GHz \citep{1980AaA....90...34S}. Non-LTE effects are usually presumed to be negligible, with low values of $T_e^*$ primarily resulting from other phenomena such as pressure broadening of lines. However, towards the Galactic Center, non-LTE effects do indeed dominate at low frequencies \citep{1980AaA....90...34S}. In extragalactic sources, in contrast, high-frequency RRLs often appear brighter, and \citet{1993ApJ...419..585A} noted that clumping of the gas into high-density \ion{H}{2} regions could explain this. The true nature of stimulated emission and non-LTE effects from extragalactic sources has not been definitively solved, but non-LTE models often better fit extragalactic observational data than LTE models \citep[e.g.][]{2017ApJ...844...73B}. Unfortunately, constrained calculations of $b_n$ and $\beta_n$ are not possible from our observations, so it is not possible to derive an unambiguous quantitative interpretation.

The archival studies listed in Table \ref{table:n253_rrl_hist} have vastly higher angular resolution compared to our data; on the order of 1$^{\prime\prime}$ rather than \textasciitilde100$^{\prime\prime}$. It is thus possible that there is an artificial offset between the archival measurements and our measurements. However, the temperature calculated from the H(92)$\alpha$ line from \citet{2006ApJ...644..914R} is consistent with the temperatures we calculate from our H(93)$\alpha$ and H(91)$\alpha$ lines and consistent with our overall trend. Thus, the higher-frequency measurements from other authors also show agreement with the trend, lending additional support to our observations. Hence, the difference in resolution appears not to affect the derived $T_e^*$ here, and so is not likely the cause of the observed trend when comparing to archival data. We thus conclude that calculations of electron temperature and other quantities derived from single-line cm-wave RRL strengths should be viewed with some caution. Instead, high-resolution multi-line RRL studies are likely important for drawing conclusions regrading the physical properties of \ion{H}{2} regions.

\subsection{Nondetections of RRLs in other galaxies}
In contrast to the bright and numerous NGC~253 detections, the other galaxies with 4cm data (NGC~1266, NGC~1365, NGC~1808) contain no visible RRLs above the RMS noise (Appendix \ref{A1}). If the emitted RRLs from these galaxies are of the same intrinsic luminosity as NGC~253's but at increased distance, then NGC~1808's lines should be on approximately 1.5~mJy at peak strength, NGC~1365's around 0.4 mJy, and NGC~1266's only 0.15 mJy. This calculation assumes the same intrinsic line widths and luminosities, and unresolved emission. Thus, it comes as no surprise that we fail to detect the RRLs in the remaining galaxies, considering that our RMS noises for these spectra (\textasciitilde1~mJy) are greater than the expected RRL strengths. However, taking stacking of RRLs into account, the RRLs within NGC~1808 and NGC~1365 must be intrinsically much weaker than those in NGC~253, consistent with their lower SFRs (Table \ref{table:sources}).

Although all of the galaxies within our sample other than NGC 1068 and NGC 1266 do have previous RRL detections in both the 4cm and 15mm data \citep{1993ApJ...419..585A,1996ApJ...472...54Z,2005AaA...435..831R,2008AaA...483...79R}, their peak intensities lie below our thresholds for detection. The exception to this is NGC~4945, in which both the H(92)$\alpha$ and H(91)$\alpha$ \citep[17.8~mJy;][]{2010AaA...517A..82R} and the H(42)$\alpha$ line \citep[\textasciitilde40~mJy;][]{2016MNRAS.463..252B} have been imaged and detected, but we fail to detect the H(67)$\alpha$, H(66)$\alpha$, and H(64)$\alpha$ lines to thresholds of \textasciitilde4 mJy. The latter is located at the same frequency as the NH$_3$ (5,5) line and is not easily distinguishable (see Section \ref{sec:n4945_NH3} for further discussion), but the remaining two show no evidence of emission and have no possible confounding lines. Thus, the flux densities of these RRLs must be significantly less than those of the higher- and lower-frequency transitions observed at over 15~mJy in previous studies. However, these lines lie at frequency ranges where non-LTE effects are minimal but the lines are high-enough order to be intrinsically weaker than the lower-order lines \citep{2000tra..book.....R}. Thus, it is not entirely surprising that these lines are weaker than both lower-frequency and higher-frequency lines.

%% file: OH.tex
\section{EXCITED OH TRANSITIONS}\label{sec:OH}
\begin{figure}
  \plotone{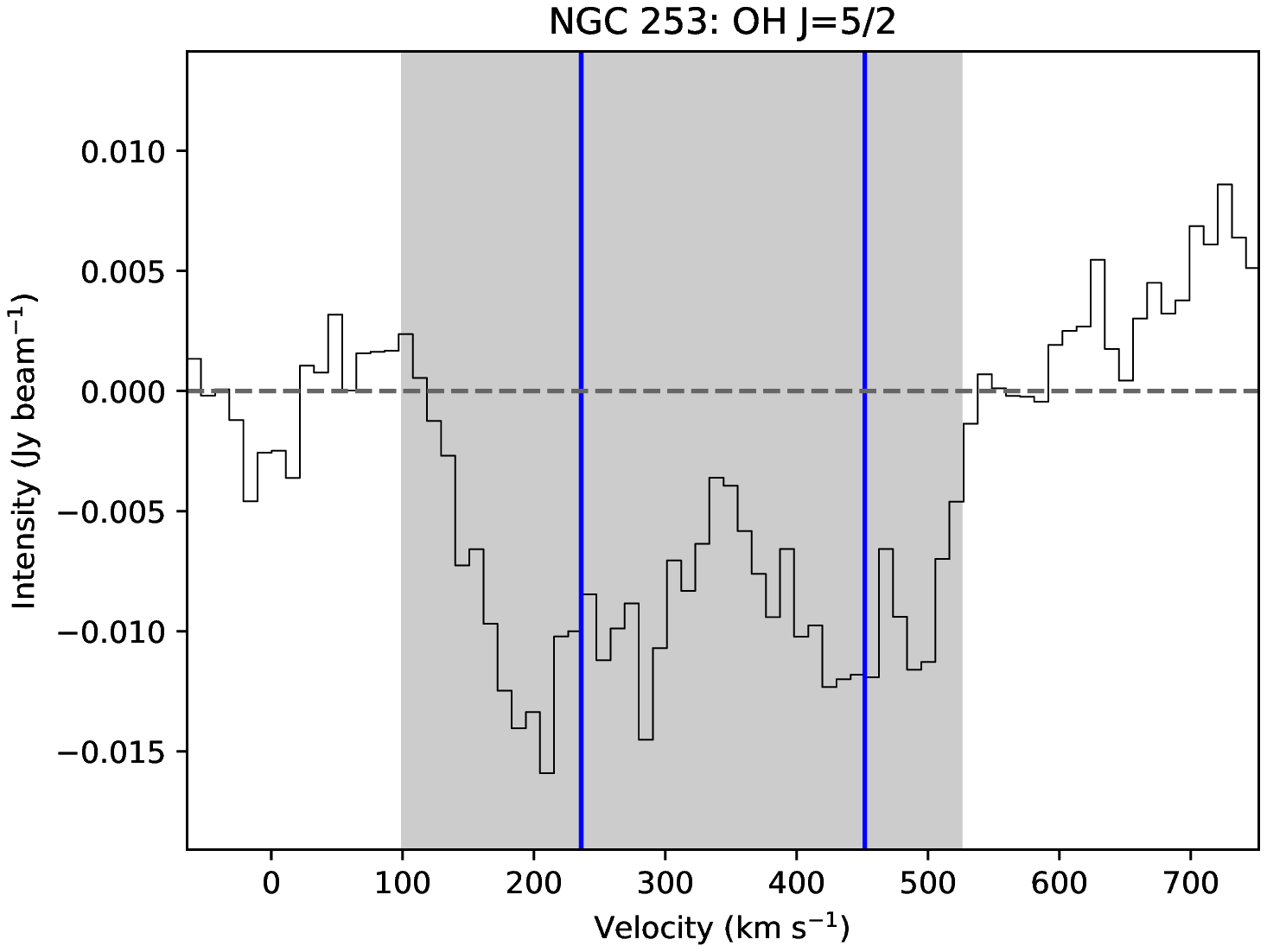}
  \plotone{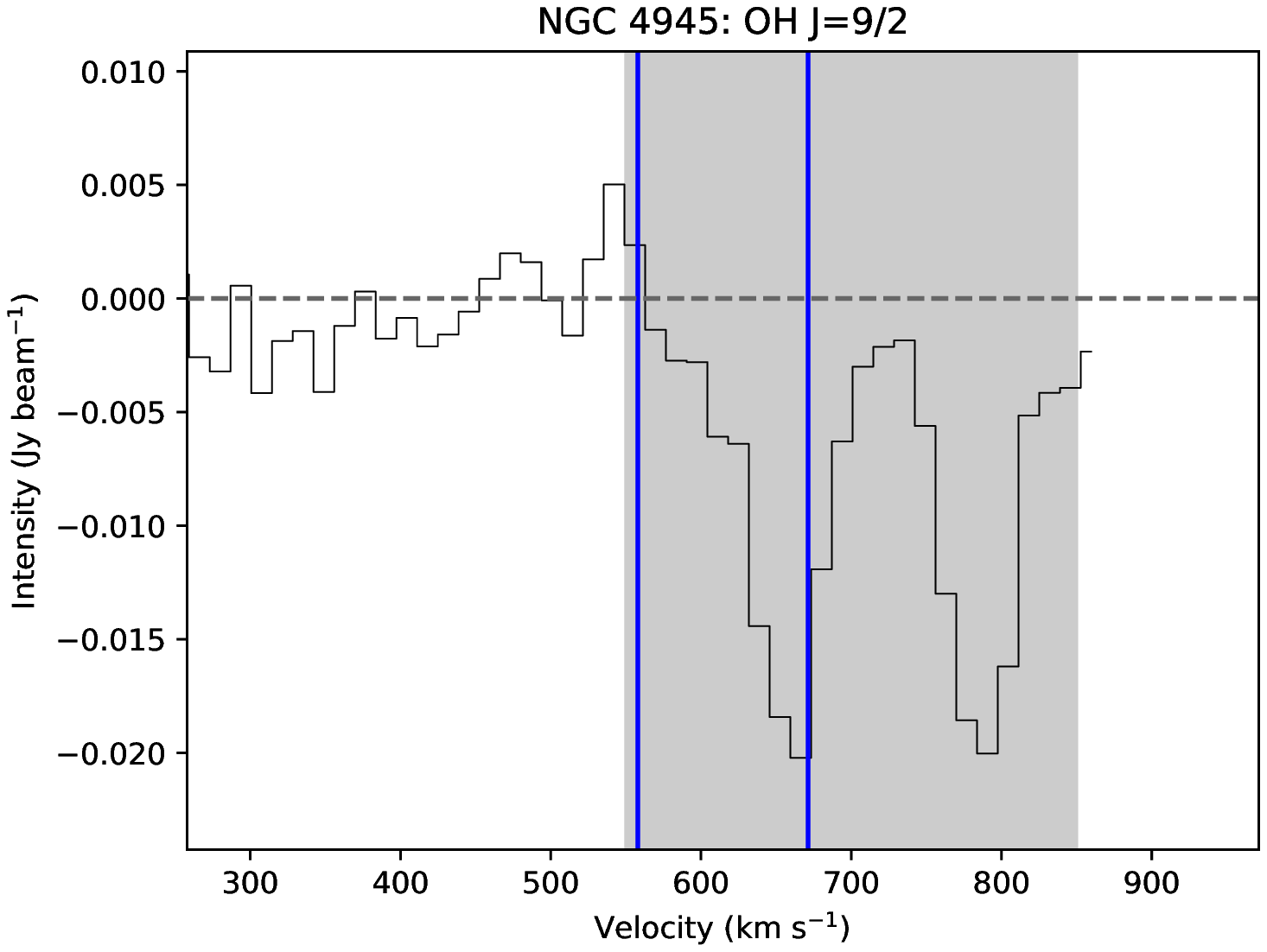}
  \plotone{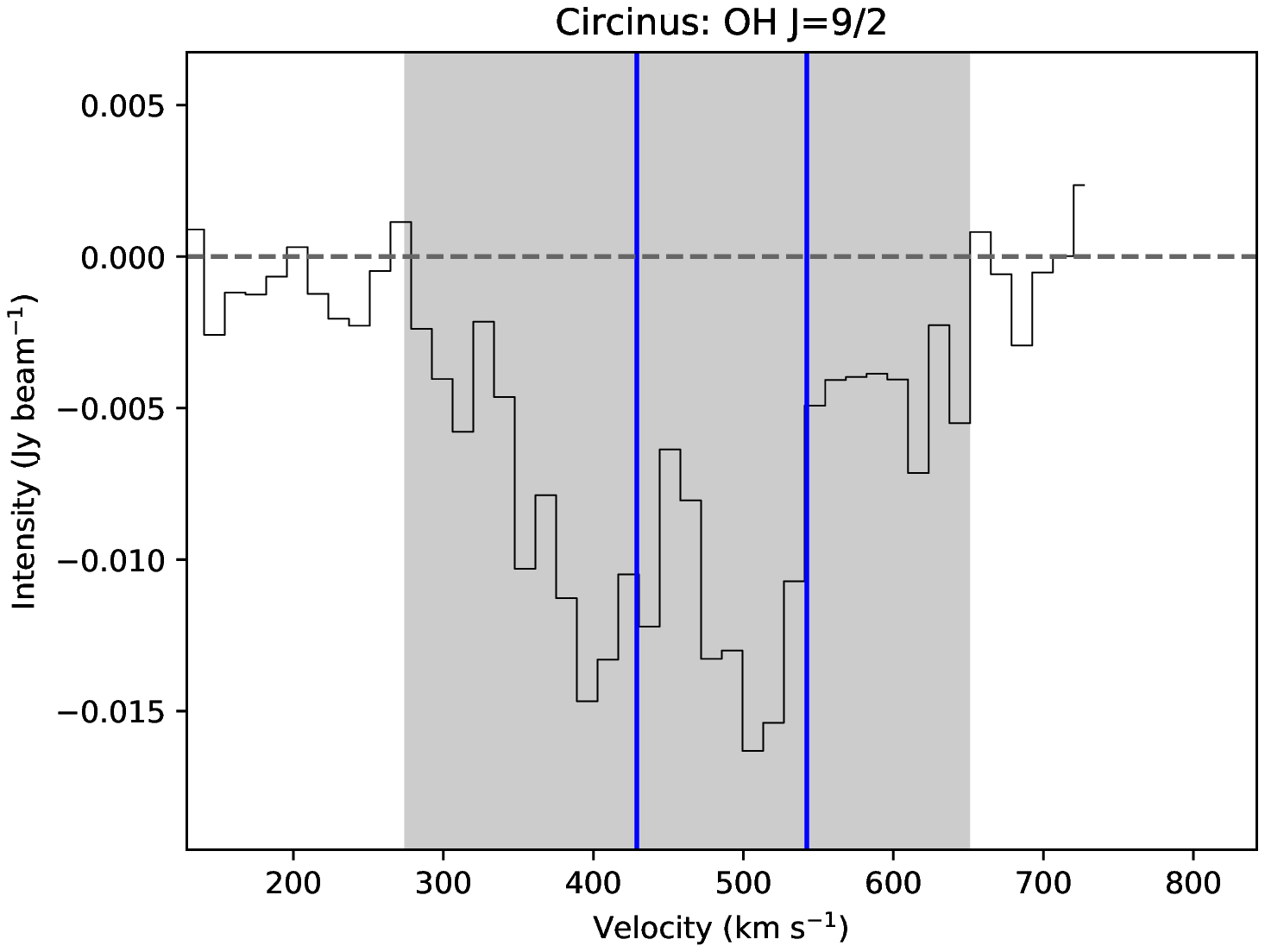}
  \caption{Profiles of detected OH lines within NGC~253, NGC~4945, and Circinus. The blue vertical lines show the systemic velocities from Table \ref{table:sources}, while the shaded regions represent the approximate extents of the lines. Both hyperfine components (Higher F on the left) are shown in each figure; the velocity axis is centered on the F=3 component.}
  \label{fig:OH_lines}
\end{figure}

We detect excited-state transitions of OH in three galaxies. In NGC~253, we observe the J=5/2 transitions at 6~GHz, while in NGC~4945 and Circinus the J=9/2 lines at 24~GHz were detected. We show the lines in Figure \ref{fig:OH_lines} We are unaware of any previous detections of these lines in the three galaxies. The lines are detected in absorption in all three galaxies, and all detected lines display approximately Gaussian profiles.

In NGC~253, we detect both of the main J=5/2 lines. The F=3 and F=2 lines' peak intensities (-13.3~$\pm$~2.2~mJy~bm$^{-1}$ and -12.3~$\pm$~2.3~mJy~bm$^{-1}$) have the ratio expected under local thermodynamic equilibrium (LTE). Both lines are centered on 230~km~s$^{-1}$, a statistically insignificant difference from the systemic velocity of 236~km~s$^{-1}$. Only the lower-frequency F=2--3 satellite line is within our spectral windows, but it is not detected in our data.

Within Circinus, both of the J=9/2 lines are slightly blueshifted from what would be expected based on the systemic velocity of the galaxy. To 1$\sigma$, the two lines' velocities are indistinguishable (392~$\pm$~4~km~s$^{-1}$ for the F=4 line and 402~$\pm$~8~km~s$^{-1}$ for the F=5 line), but the lines are $\geq 3\sigma$ from the systemic velocity of 429~km~s$^{-1}$. In the case of NGC~4945, the two J=9/2 lines are also significantly displaced from the systemic velocity, but unlike Circinus the lines are redshifted by 114~$\pm$~3 and 100~$\pm$~4~km~s$^{-1}$ rather than blueshifted from $v_{sys} = 558$~km~s$^{-1}$. The optical depths can be seen in Table \ref{table:OH_opticaldepths}; all lines are optically thin.

\begin{deluxetable}{llc}
\tablecaption{Peak optical depths of detected OH absorption lines in NGC~253, NGC~4945, and Circinus.\label{table:OH_opticaldepths}}
\tablehead{
\colhead{Galaxy} & \colhead{Transition} & \colhead{$\tau_\nu$} \\
}
\startdata
NGC~253  & J=5/2 F=2 & 0.0081~$\pm$~0.0017 \\
NGC~253  & J=5/2 F=3 & 0.0087~$\pm$~0.0017 \\
NGC~4945 & J=9/2 F=4 & 0.0224~$\pm$~0.0036 \\
NGC~4945 & J=9/2 F=5 & 0.0212~$\pm$~0.0035 \\
Circinus & J=9/2 F=4 & 0.065~$\pm$~0.012 \\
Circinus & J=9/2 F=5 & 0.0534~$\pm$~0.0099
\enddata
\end{deluxetable}

The OH molecule is a free radical, containing an odd number of electrons. Free radicals have much more complex spectra than the purely-rotational spectra of closed-shell molecules such as CO. In all such diatomic radicals, two distinct ladders ($^2\Pi_{3/2}$ and $^2\Pi_{1/2}$ in the case of OH) are present due to fine structure splitting, in which the orbital angular momentum of the lone electron is either parallel or antiparallel to its spin.

Each rotational state in both ladders is split again by $\Lambda$-doubling. $\Lambda$-doubling results from the interaction between the spin of the electron and the total rotational angular momentum of the molecule, splitting the energy levels of each state. The energy splitting is a function of the rotational state of the molecule. Each of these resulting $\Lambda$ states is further split by the interaction of the electron's spin with the spin of the hydrogen nucleus, described by the quantum number $F = J \pm 1/2$. This hyperfine splitting is the origin of the doublet nature of the lines.

Both observed transitions of OH within our spectral windows ($^2\Pi_{3/2}$ J=5/2 and J=9/2) are relatively little-studied transitions, especially outside of the Milky Way. None of our galaxies had been previously detected in J=9/2, while a J=5/2 detection has only been achieved in NGC~4945 \citep{1990MNRAS.245..665W}. The first extragalactic detection of the J=9/2 doublet occurred as recently as 2011 towards Arp 220 by \citet{2011ApJ...742...95O}.

\subsection{Calculations of properties from OH}

From the OH optical depths and an estimate of the gas temperature, it is possible to calculate relative column densities between levels. The total column density, assuming optically thin absorption, for the lower state $N_l$ of any molecular absorption transition is approximately given by \citep[e.g.][]{2015PASJ...67....5M}
\begin{equation}
\label{eq:OH_column_density}
\frac{N_l}{\textrm{cm}^{-2}} = 2.07\times10^3 \frac{\nu^2}{\textrm{GHz}^2} \frac{g_l}{g_u} A_{ul}^{-1} \tau \frac{\Delta \upsilon_{1/2}}{\textrm{km s}^{-1}} \frac{T_{ex}}{\textrm{K}},
\end{equation}
where $\nu$ is the transition frequency, $g_l$ and $g_u$ are the lower and upper state degeneracies, $\tau$ is the line-center optical depth of the line, $\Delta \upsilon_{1/2}$ is the line velocity FWHM, and $T_{ex}$ is the as-of-yet-unknown excitation temperature of the transition. In the case of the OH molecule, $g = 2F_l + 1$, and thus the degeneracies ratio for all of our detected transitions is unity. $A_{ul}$ represents the Einstein emission coefficient. The excitation temperature $T_{ex}$ is much more difficult to determine, and the most often-quoted values are $N_l/T_{ex}$ rather than pure column densities. Table \ref{table:OH_column_densities} gives calculated column densities per excitation temperature for each line.

\begin{deluxetable*}{llccc}
\tablecaption{Einstein Coefficients from Splatalogue, and calculated column densities for all detected OH transitions\label{table:OH_column_densities}}
\tablehead{
\colhead{Galaxy} & \colhead{Transition} & \colhead{$A_{ul}$} & \colhead{$\tau$} & \colhead{$N_l/T_{ex}$ (10$^{13}$ cm$^{-2}$ K$^{-1}$)} \\
}
\startdata
NGC~253  & J=5/2 F=2 & $1.548 \times 10^{-9}$ & 0.0081 $\pm$ 0.0017 & 4.8 $\pm$ 1.5 \\
NGC~253  & J=5/2 F=3 & $1.583 \times 10^{-9}$ & 0.0087 $\pm$ 0.0017 & 6.2 $\pm$ 1.7 \\
NGC~4945 & J=9/2 F=4 & $3.141 \times 10^{-8}$ & 0.0224 $\pm$ 0.0036 & 4.36 $\pm$ 0.78 \\
NGC~4945 & J=9/2 F=5 & $3.158 \times 10^{-8}$ & 0.0212 $\pm$ 0.0035 & 4.42 $\pm$ 0.86 \\
Circinus & J=9/2 F=4 & $3.141 \times 10^{-8}$ & 0.065 $\pm$ 0.012   & 20.1 $\pm$ 4.5 \\
Circinus & J=9/2 F=5 & $3.158 \times 10^{-8}$ & 0.0534 $\pm$ 0.0099 & 23.9 $\pm$ 6.1
\enddata
\end{deluxetable*}

Following the same line of thought as in \citet{1995AaA...298..905B}, the total column density per excitation temperature of a single rotational state of OH can be calculated by 
\begin{equation}
\label{eq:column_density}
\frac{N_J}{\textrm{cm}^{-2}} = 2N_{l,F=J-1/2}+2N_{l,F=J+1/2},
\end{equation}
assuming that $T_{ex}$ is equal for both doublet lines (i.e. LTE conditions). The prefactors of 2 account for both the upper and lower $\Lambda$-doubling state. Due to the similar energies of both OH $\Lambda$-doubling transitions for a given spin state, this is a reasonable assumption. From the J=5/2 state of NGC~253, we calculate a total column density $N_{5/2}/T_{ex} = 22.1 \pm 4.5 \times 10^{13}$ cm$^{-2}$ K$^{-1}$. For the J=9/2 state, we find that $N_{9/2}/T_{ex} = 17.6 \pm 2.3 \times 10^{13}$ cm$^{-2}$ K$^{-1}$ for NGC~4945 and $N_{9/2}/T_{ex} = 88 \pm 15 \times 10^{13}$ cm$^{-2}$ K$^{-1}$ for Circinus.

From the calculated column densities of the different rotational states of OH, it is possible to estimate the rotational temperature $T_{rot}$ of the molecular gas responsible for the excitation, by the Boltzmann Formula
\begin{equation}
\label{eq:OH_Trot}
T_{rot} = \frac{E_u-E_l}{k}/\ln\left(\frac{2J_u+1}{2J_l+1}\frac{N_{J_l}}{N_{J_u}}\right).
\end{equation}
Here, $E_u$ and $E_l$ are the upper and lower state energies (obtained from Splatalogue), $k$ is the Boltzmann constant, $J_u$ and $J_l$ are the upper and lower state total angular momenta, and $N_{J_l}$ and N$_{J_u}$ are the column densities in the lower states of each transition.  This assumes a single excitation temperature (i.e. rotation temperature) is applicable across all states, likely a poorer approximation than for lines of equal $J$, but the quantity we obtain is still of interest. If the gas is emitted from multiple clouds, each with its own intrinsic excitation temperature range, then the emission from different rotational states will not be partitioned across the states in the same way between the clouds. However, this still allows us to understand, in a sense, the average properties of the nuclear region. Using archival line intensities in which the galaxies are unresolved, we are able to calculate the temperatures between the different rotation states. Our calculated rotation temperatures can be seen in Table \ref{table:OH_Trot}.

\begin{deluxetable}{lllcc}
\tablecaption{Rotational temperatures calculated between OH rotation state column densities\label{table:OH_Trot}}
\tablehead{
\colhead{Galaxy} & \colhead{$J_u$} & \colhead{$J_l$} & \colhead{Approx. $T_{rot}$ (K)} & \colhead{Reference}
}
\startdata
NGC~253  & 5/2 & 3/2 & 60   & 1,5 \\
NGC~4945 & 9/2 & 5/2 & 140  & 2,5 \\
NGC~4945 & 9/2 & 3/2 & 80   & 3,5 \\
NGC~4945 & 5/2 & 3/2 & 36   & 2   \\
Circinus & 9/2 & 3/2 & 2000 & 4,5
\enddata
\tablerefs{(1) \citet{1974ApL....15..211W}; (2) \citet{1990MNRAS.245..665W}; (3) \citet{1975ApJ...195L..81W}; (4) \citet{1990MNRAS.244..130H}; (5) This paper}
\tablecomments{None of the previous studies resolve the nucleus, and thus their data are directly comparable to our unresolved measurements. However, we do not quote uncertainties due to the lack of quoted uncertainties in the previous studies of other transitions. In addition, none of the authors calculate or list values for $N_J$ and $\Delta \upsilon$. We estimate the latter from their plots, and calculate the former from their listed optical depth. Due to the logarithmic variation of T$_{rot}$ on these variables (Equation \ref{eq:OH_Trot}), we believe that these $T_{rot}$ values are accurate estimates.}
\end{deluxetable}

\subsubsection{Excited OH in NGC~253}

While the $^2\Pi_{3/2}$ J=3/2 1.6~GHz transitions of NGC~253 have been well-studied and characterized \citep[e.g.][]{1985ApJ...299..312T,1998AJ....115..559F} since their discovery by \citet{1971ApJ...167L..47W}, the presence of maser emission adds uncertainties to the calculation of aspects such as rotational temperatures and optical depths. The maser emission from the NGC~253 J=3/2 lines was ignored for the purposes of calculations, with maximum amplitudes and FWHMs estimated from only the absorption components, assuming a roughly Gaussian absorption morphology. As the J=5/2 transitions that we observe lack any evidence of maser emission and are further in the optically-thin regime, they can provide better estimates of the intrinsic properties of the absorbing OH than the J=3/2 transitions that are contaminated by emission.

The frequencies of the J=9/2 transitions were covered by \citet{2017ApJ...842..124G} in their survey of NGC~253, but neither component of the doublet was detected. While the authors do not provide specific thresholds for the nondetections, their Figure 2 indicates that the lines' intensities must be no stronger than approximately 0.3 mJy~bm$^{-1}$. However, \citet{2017ApJ...842..124G} have a beam size of 6$^{\prime\prime}$ by 4$^{\prime\prime}$, easily resolving the nuclear regions, so line strengths are not directly comparable.

In addition to the $\Lambda$-doubling transitions of OH, several high-frequency rotational transitions have also been observed by \citet{2018ApJ...860...23P}. Of particular note, the transitions solely within the $^2\Pi_{3/2}$ ladder are seen in absorption, while those involving the $^2\Pi_{1/2}$ state are in emission. From the relative strengths of the OH and H$_2$O lines, \citet{2018ApJ...860...23P} favor production of OH from cosmic ray ionization and dissociation of H$_2$O.

Previous interferometric observations of the ground-state J=3/2 lines have revealed maser activity within both the main lines and the satellite lines \citet{1971ApJ...167L..47W,1974ApL....15..211W,1985ApJ...299..312T,1998AJ....115..559F}. In particular, \citet{1985ApJ...299..312T} found that the J=3/2 main line maser emission originates in a large plume of gas to the north of the nucleus. We find no conclusive evidence of any emission in our observations in either the spectral or spatial profile. With our angular resolution of 140$^{\prime\prime}$, the plume should be marginally detectable were its emission strength in the J=5/2 lines on the same order as the J=3/2 emission. The plume is thus not emitting in the J=5/2 lines to a threshold of \textasciitilde1~mJy~bm$^{-1}$ over a 140$^{\prime\prime}$ beam.

\subsubsection{Excited OH in NGC~4945}

Under LTE and optically thin conditions, the F=5 to F=4 ratio should be equal to $11/9$, consistent with our results for both NGC~4945 and Circinus. Based on these observations alone, we find no reason to necessitate deviations from LTE. Within the Galaxy, the J=9/2 OH transitions have only been observed in dense molecular clouds \citep[e.g.][]{1981AaA...102..287B,1986AaA...167..151W}. The transition has only recently been detected in the extragalactic sources Arp~220 \citep{2011ApJ...742...95O,2013ApJ...779...33M} and NGC~3079 \citep{2015PASJ...67....5M}. Arp~220 is a starburst galaxy like NGC~4945, while NGC~3079 is a Seyfert galaxy with a superwind. Unlike the ground-state J=3/2 transitions or the less-excited J=5/2 and J=7/2 transitions, all J=9/2 detections to date have been in absorption; no emission or masing has ever been observed in any source.

\citet{1990MNRAS.245..665W}, using the J=3/2 and J=5/2 lines of NGC~4945, obtained a $T_{rot} = 36$ K, much lower than our values for either pair using the J=9/2 transitions in Table \ref{table:OH_Trot}. In addition, the FWHMs of our J=9/2 lines are smaller by a factor of \textasciitilde3 than those of the J=3/2 and J=5/2 lines. We thus find it unlikely that the source of the absorption in NGC~4945 is the same for all three rotational states. Instead, it is likely that there are multiple molecular clouds with different average rotation temperatures responsible for the absorption seen.

The previous observations of the J=5/2 transitions within NGC~4945 displayed significant overlap between the two lines of the doublet in the velocity domain \citep{1990MNRAS.245..665W}. While the simple structure of the two lines allows for disentangling of each line, it would still be beneficial to be able to trace and characterize the excited gas without any of these effects to allow for higher confidence of results. In our J=9/2 spectrum, the two lines are far enough apart in frequency space and are narrow enough that the overlap between the two is minimal.

\citet{2016ApJ...826..111S} observed the far-IR J=5/2--3/2 transitions at 2510~GHz in NGC~4945 using \textit{Herschel}. Here, the authors again find evidence of an infall to the nucleus at a similar velocity to our transitions. This molecular infall is consistent with our data, where the OH absorption velocity is \textasciitilde100~km~s$^{-1}$ redshifted from the systemic velocity. Of note is that the absorption profile from \citet{2016ApJ...826..111S} has an integrated flux of $>$65000~Jy~km~s$^{-1}$, by far the highest flux of any of their sources, and even taking distance into account, it is one of the intrinsically strongest in their sample. This is again consistent with our detection of the J=9/2 lines, and suggestive either of the presence of larger-than-typical amounts of OH or of higher-than-typical excitation. The gas is likely being excited to high enough excitation temperatures through infrared absorption such that the J=9/2 lines are visible.

The 14 GHz J=7/2 doublet was not included in our bands. However, these transitions lie between the J=5/2 and J=9/2 doublets, which, as mentioned earlier, displayed different relative strengths between the two transitions. An observation of the J=7/2 transitions would be an worthwhile exercise, to add another data point in order to further constrain the excitation and abundance of OH.

\subsubsection{Excited OH in Circinus}

The \textasciitilde2000~K rotation temperature we derive for the OH in Circinus is unphysically high. There thus appears to be an overpopulation of the J=9/2 state relative to the J=3/2 state. No clear evidence of nonthermal processes is seen in the previously-observed OH J=3/2 transitions or in our J=9/2 transitions. It is possible that there is instead a nonthermal mechanism responsible for pumping the molecules to these higher energy levels through an infrared rotational transition. The other option is strong variability of the OH absorption as a function of time. This in turn would imply either a substantial increase in available photons, or an increase in the column density of OH. All of these possibilities are intriguing, and worthy of further study.

A study of the infrared J=5/2--3/2 2510~GHz transition \citep{2016ApJ...826..111S} in Circinus found a strong inverted P-Cygni profile in both components of the doublet, suggesting an infall. It is plausible that the infall is not in LTE and the OH is being pumped to a higher level. However, the OH absorption we observe in the J=9/2 state from the Circinus nucleus is slightly blueshifted, rather than at the 50~km~s$^{-1}$ redshift we would expect from an infall. In addition, the J=3/2 lines observed by \citet{1990MNRAS.244..130H} have two components: one at the systemic velocity and one at a \textasciitilde50~km~s$^{-1}$ redshift. This suggests different origins of the J=9/2, J=5/2, and J=3/2 lines. It is clear that further study is necessary to determine the nature of the OH within this system.

\subsection{An OH and H$_2$O connection?}
An interesting trend is that the two galaxies in which we detect highly excited OH (NGC~4945 and Circinus) are also the two galaxies in our sample that are host to extremely luminous H$_2$O megamasers. While two is not a large enough sample size to draw any definite conclusions, the circumstantial evidence points toward a possible connection. Our velocities of the OH lines in NGC~4945 (672 $\pm$ 2 and 658 $\pm$ 2 km s$^{-1}$ are in agreement with the velocity of the second-strongest H$_2$O peak (668.5 $\pm$ 0.4 km s$^{-1}$), both of which are around 100 km s$^{-1}$ larger than the systemic velocity and NH$_3$ peak absorption velocities, although the linewidths are substantially different. In contrast the brightest H$_2$O peak does not match up with any OH feature in NGC 4945, so the correlation between OH absorption and H$_2$O emission cannot be direct.

Previously, connections between the presence of H$_2$O and OH masers have been noticed. In fact, in NGC~4945 itself, \citet{1982MNRAS.200P..19B} noticed that the ground-state OH maser and H$_2$O maser emission from NGC~4945 also originated at the same velocities. Others have noted connections between OH masers and H$_2$O masers, where both appear correlated with color temperatures of their host galaxies \citep{1986AaA...155..193H}. \citet{2011AaA...525A..91T} found that weaker maser emission from both OH and H$_2$O is commonly found in the same galaxies. \citet{2016ApJ...816...55W} noticed that while H$_2$O masers have been detected in around 4\% of galaxies searched, they have been detected in 25\% of galaxies with known OH masers. However, in galaxies with OH megamasers, H$_2$O kilomasers are actually rarer than would be expected if there were no correlation \citep{2013AaA...560A..12W}. From all these studies, it is thus reasonable that we would find a correlation between H$_2$O maser and thermal OH lines.

\subsection{OH Nondetections}
NGC~1068 is host to a luminous J=3/2 1.667 GHz megamaser \citep{1996ApJ...462..740G}. However, we do not detect any evidence of either emission or absorption at the J=9/2 transitions. The short duration of our observation has resulted in a high RMS (14.2 mJy bm$^{-1}$) noise in the spectrum containing both transitions. If the J=9/2 transitions within NGC~1068 have the same intrinsic luminosity as those within NGC~4945, we would observe them at a peak flux density of approximately 3 mJy bm$^{-1}$, well below the RMS noise of the spectrum. It thus comes as no surprise that we fail to detect any evidence of the doublet.

Of the remaining galaxies, ground-state OH observations have only been published for NGC 1266, and none have been studied in the excited states. The NGC 1266 detection by \citet{1992AJ....103..728B} found the J=3/2 state purely in absorption, with a peak flux density of 110~mJy. Assuming a similar ratio of J=3/2 to J=5/2 and J=9/2 fluxes as in NGC 4945, the two excited transitions would be well below our upper limits.

The J=9/2 lines have only recently been detected beyond the Milky Way. The first detection was by \citet{2011ApJ...742...95O} in the starburst Arp 220, later confirmed by \citet{2013ApJ...779...33M} and \citet{2016ApJ...833...41Z}. Since then, the only other source with detected J=9/2 absorption is NGC~3079 \citep{2015PASJ...67....5M}, and studies in other galaxies are nearly nonexistent. Thus, these two detections represent the third and fourth extragalactic detections of the J=9/2 doublet. As these transitions were found twice in what is meant to be a representative sample of nearby star-forming galaxies, it is likely that the lines are present in other unstudied galaxies as well.

%% file: NH3.tex
\section{NH$_3$ AND BOLTZMANN DIAGRAMS}\label{sec:NH3}
\begin{figure*}
  \plottwo{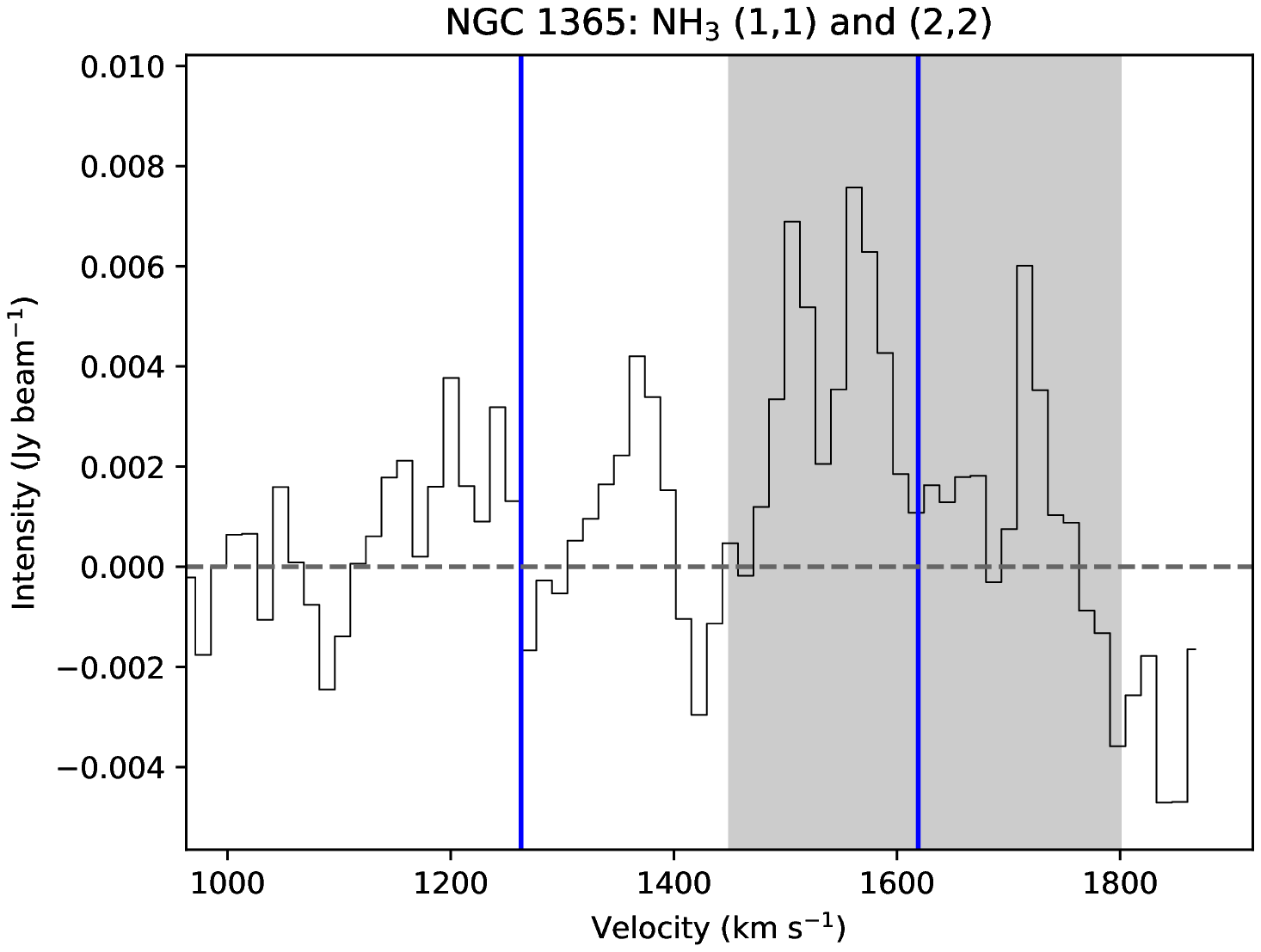}{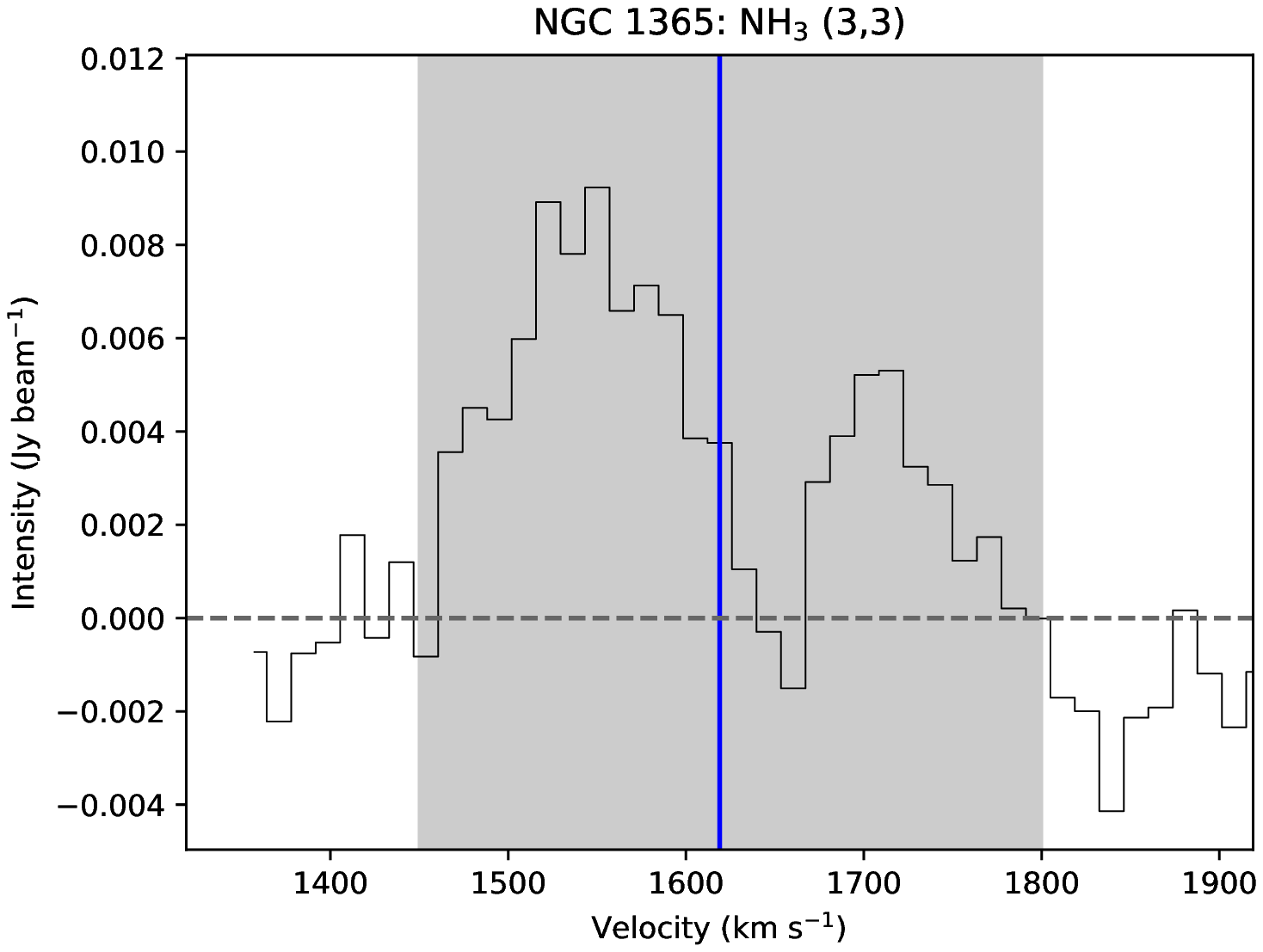}
  \plottwo{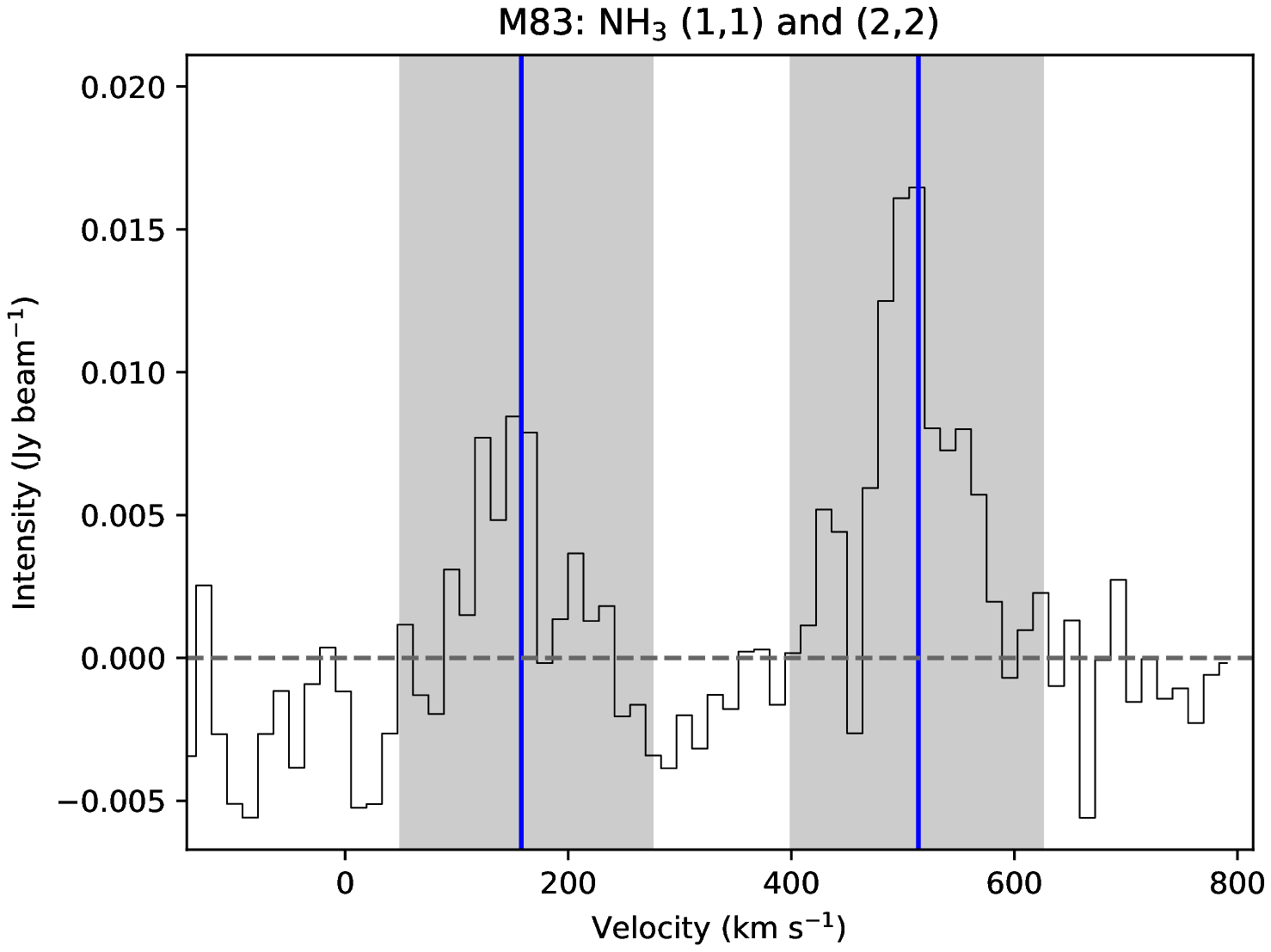}{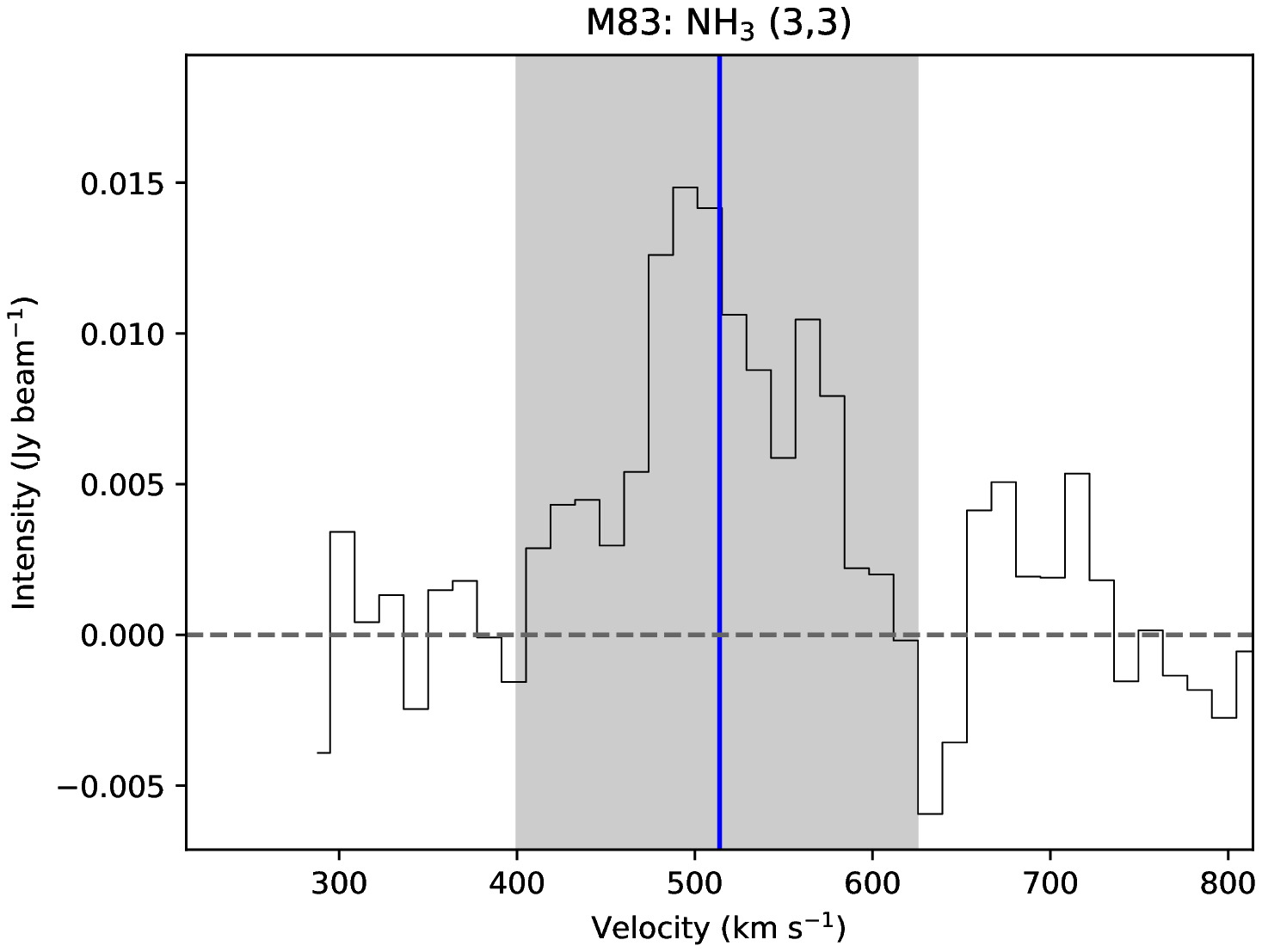}
  \includegraphics[scale=0.5]{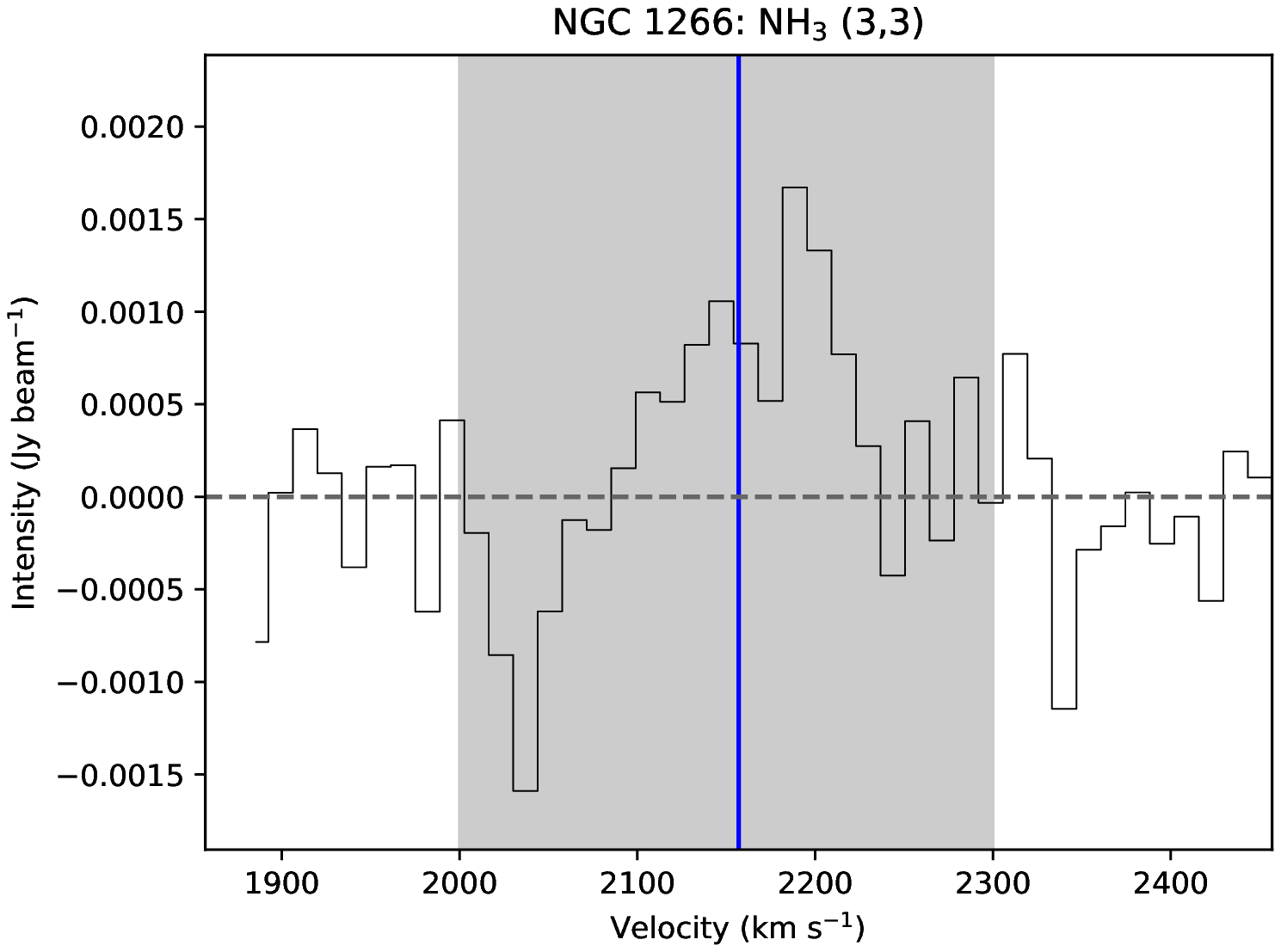}
  \caption{Profiles of detected NH$_3$ and OH lines within NGC~1365, M83, and NGC~1266. The blue vertical lines show the systemic velocities from Table \ref{table:sources}, while the shaded regions represent the approximate extents of the lines. The NH$_3$ (1,1) (right) and (2,2) (left) lines are shown on the same plot, with the velocity axis centered on the (1,1) line. The expected frequency of NGC~1365's (2,2) line if at the systemic velocity is included on the (1,1) line figure for comparison.}
  \label{fig:NH3_lines}
\end{figure*}

\begin{figure*}
  \plottwo{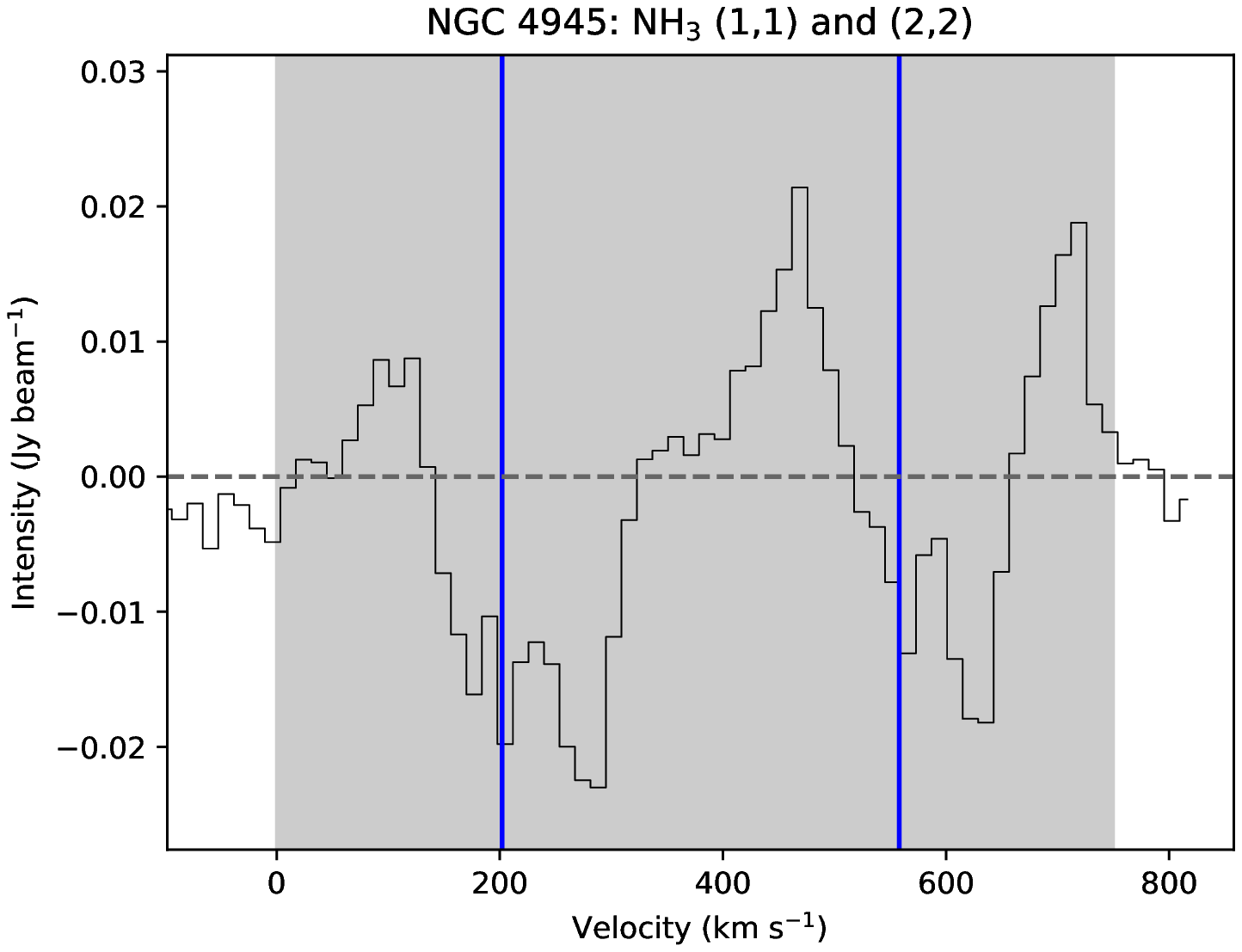}{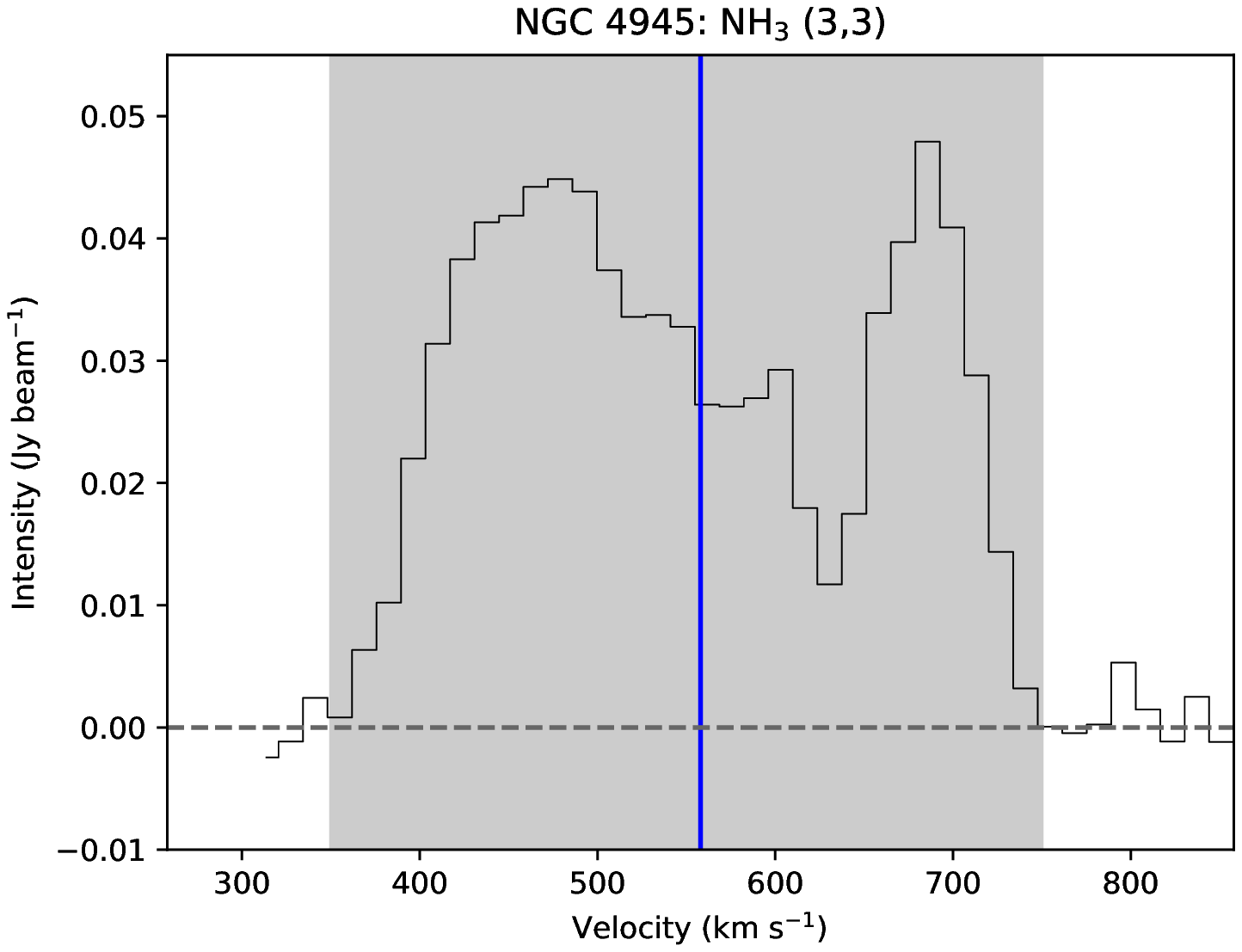}
  \plottwo{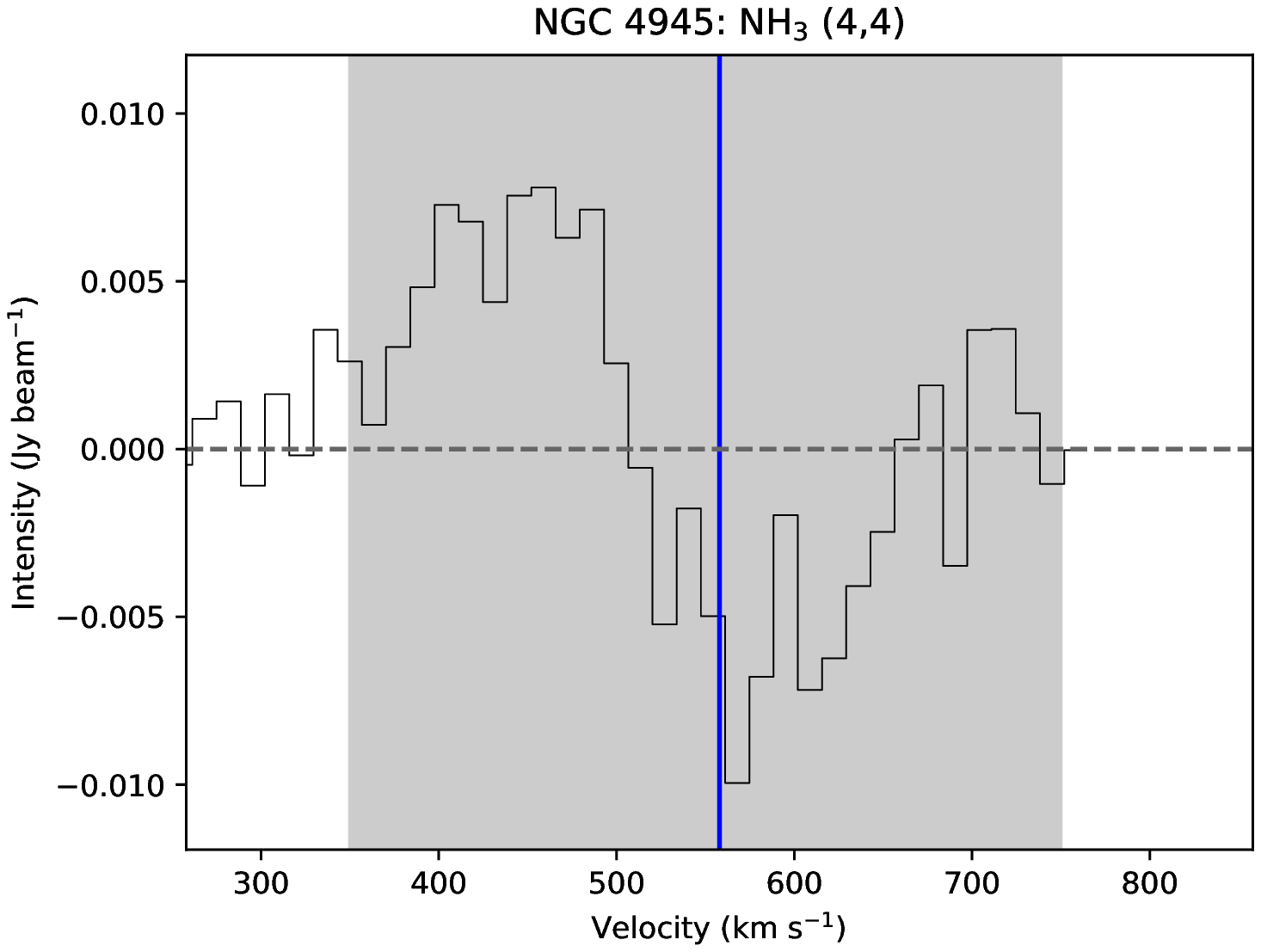}{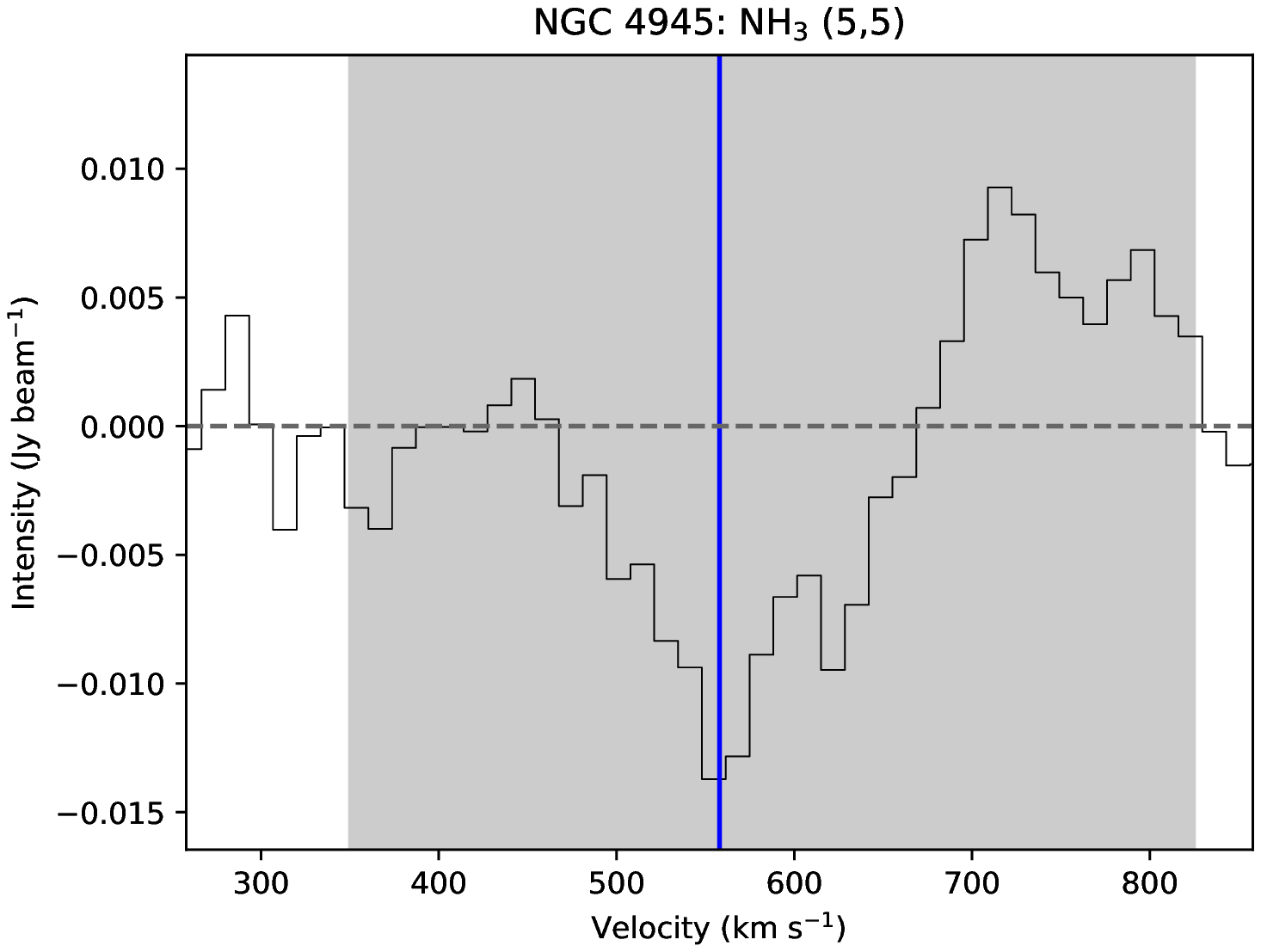}
  \includegraphics[scale=0.5]{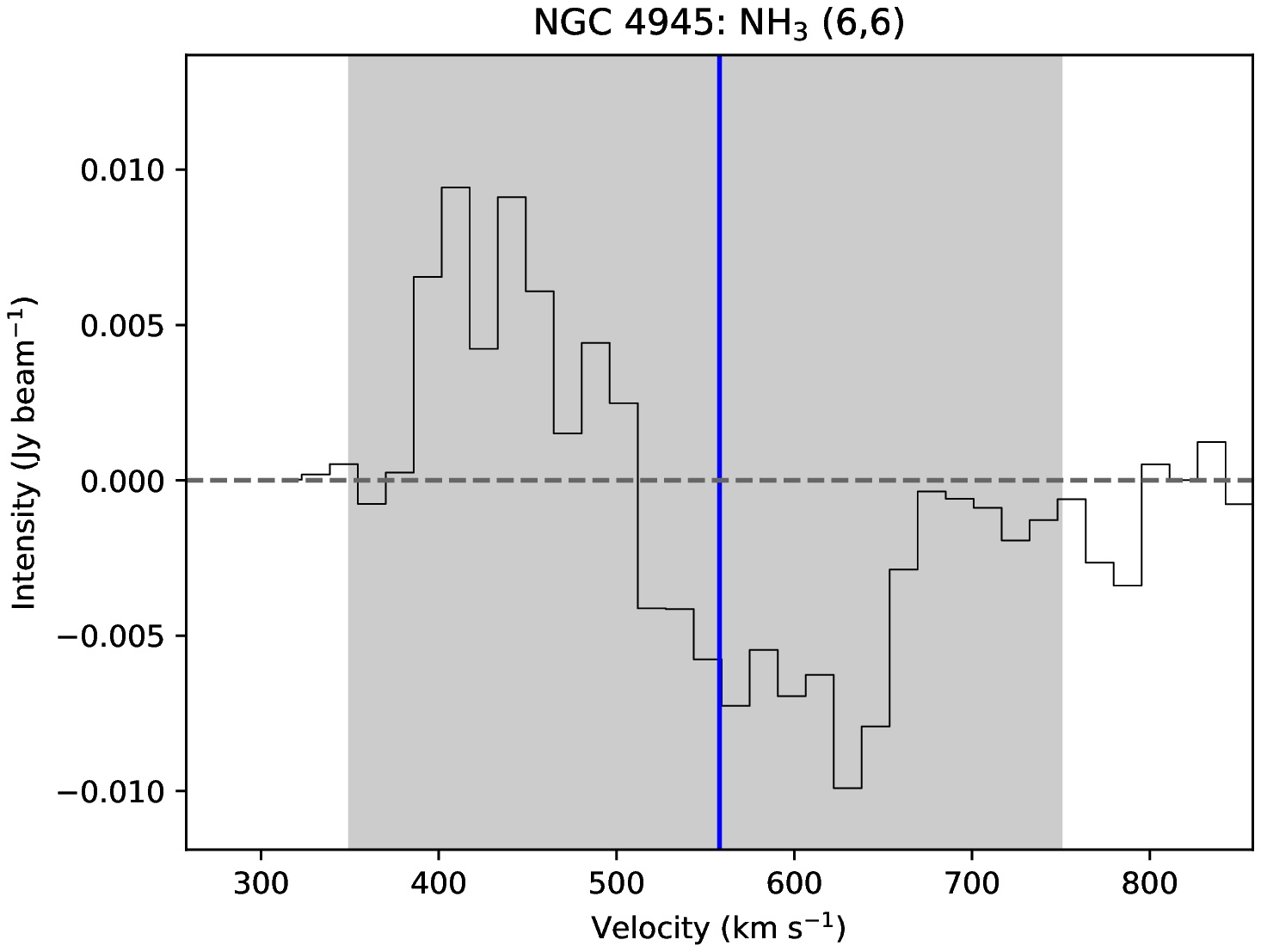}
  \caption{Profiles of detected NH$_3$ and OH lines within NGC~4945. The blue vertical lines show the systemic velocity from Table \ref{table:sources}, while the shaded regions represent the approximate extents of the lines. The NH$_3$ (1,1) (right) and (2,2) (left) lines are shown on the same plot, with the velocity axis centered on the (1,1) line.}
  \label{fig:n4945_NH3_lines}
\end{figure*}

We detect NH$_3$ within four of our galaxies: NGC~1266, NGC~1365, M83, and NGC~4945. The first three exhibit the lines in emission, while the NGC~4945 NH$_3$ spectrum is a complicated superposition of absorption and emission that will be discussed in further detail below. For all galaxies, the linewidths are large enough that the different hyperfine components are unresolved. The lines are shown in Figures \ref{fig:NH3_lines} and \ref{fig:n4945_NH3_lines}.

In NGC~1266, we tentatively detect only the (3,3) line with 3$\sigma$ confidence, with only upper limits on the strengths of the remaining lines. The (3,3) line consists of a single component at 2175.6~$\pm$~1.5~km~s$^{-1}$, close to the systemic LSRK velocity of the system (2157~km~s$^{-1}$). The velocity-integrated main beam brightness temperature is $\int T_{mb} d\upsilon = 0.34 \pm 0.17$~K~km~s$^{-1}$. We do note tentative evidence for the existence of a (2,2) line at the systemic velocity, but the feature bears little to no resemblance to the (3,3) line and is low S/N, so we do not claim it as a detection.

For the remainder of the galaxies with NH$_3$ detections, multiple lines are detected. In NGC~1365, we detect the (1,1) and (3,3) lines. The (3,3) line consists of two components: the primary peak is located at a velocity of 1545~$\pm$~6~km~s$^{-1}$, while a secondary peak is at a larger velocity of 1711~$\pm$~8~km~s$^{-1}$. The (1,1) line displays a peak at a similar velocity to the brighter (3,3) peak of 1552~$\pm$~10~km~s$^{-1}$. The secondary peak potentially also has a counterpart in the (1,1) transition, but the S/N is too low to consider a detection. Within M83, we detect the (2,2) line in addition to the (1,1) and (3,3) lines. All three lines are roughly Gaussian in nature. The (1,1) and (3,3) lines both emanate from near the systemic velocity of 514~km~s$^{-1}$, while the (2,2) line is found at a slightly larger velocity of 546~$\pm$~8~km~s$^{-1}$.

No previous observations of NH$_3$ exist for NGC~1266 or NGC~4945, but both NGC~1365 and M83 were observed in the (1,1), (2,2), and (4,4) transitions by \citet{2013ApJ...779...33M}. NGC~1365 was detected in all three lines, while M83 was only detected in (1,1) and (2,2). \citet{2013ApJ...779...33M} did not target the (3,3) line. The M83 results are thus consistent with the previous results, but we fail to detect two of the NGC~1365 lines previously found. The nondetection of the (4,4) line is expected from the noise within our cubes, but the (2,2) line should have been detected with our observational surface brightness sensitivity assuming the same proportionality constant between their $T_A^*$ and our $T_{mb}$ as is found in the (1,1) line. Our nondetection is disparate by a factor of \textasciitilde2 from the relative strengths found in \citet{2013ApJ...779...33M}.

The final galaxy in which we detect NH$_3$ is NGC~4945. Here, the lines are not easily fit by Gaussians, with complicated entanglements of emission and absorption present within all six (1,1) to (6,6) lines detected. The (1,1) and (2,2) lines have almost identical morphologies, as do the (4,4), (5,5), and (6,6) lines. The (3,3) line displays a much stronger emission base, on top of which we see a profile reminiscent of that of the lower-order lines. All other lines are primarily found in absorption with weaker emission features present.

The NH$_3$ molecule consists of three hydrogen atoms in a trigonal pyramidal geometry with a nitrogen atom. NH$_3$ possesses inversion transitions, where the nitrogen atom tunnels through the plane of the hydrogen atoms. Each state is denoted by ($J$,$K$), where $J$ is the total angular momentum of the NH$_3$ molecule, while $K$ is its projection onto the symmetry axis. The ground (0,0) state is not split as there is no angular momentum along the symmetry axis, so the total NH$_3$ column density cannot be determined from the inversion transitions alone, but those of the higher ($J$,$K \neq 0$) states are calculable. The relative strengths of the inversion transitions are strongly dependent on the temperature of the gas via the Boltzmann equation but nearly independent of density, making the NH$_3$ inversion transitions one of the preferred temperature probes \citep[e.g.][]{1983ARAaA..21..239H,2005ApJ...629..767O,2013ApJ...779...33M,2018ApJ...856..134G}.

There exist two varieties of NH$_3$; those with $K = 3n$ where $n$ is an integer are known as ortho-NH$_3$ (spins of all hydrogen atoms parallel), while the others are known as para-NH$_3$ (some spins antiparallel). For optically thin emission, a one-to-one relationship exists between $T_{mb}$ of an NH$_3$ state transition and the column density given by
\begin{equation}
\label{eq:NH3_col_density}
\frac{N(J,K)}{\textrm{cm}^{-2}} = \frac{1.55 \times 10^{14 }}{\nu} \frac{J(J+1)}{K^2} \int T_{mb} d\upsilon,
\end{equation}
where $N(J,K)$ represents the combined column density of the upper and lower states. Within this formula, $\nu$ is given in GHz and $\int T_{mb} d\upsilon$ in K km s$^{-1}$ \citep[e.g.][]{2003AaA...403..561M,2011AaA...529A.154A,2015PASP..127..266M}. If NH$_3$ is detected in absorption rather than emission, optical depths must be used in place of $T_{mb}$. The formula for column density is thus instead, as per \citet{1995AaA...294..667H,2011ApJ...742...95O}:
\begin{equation}
\label{eq:NH3_col_density_abs}
\frac{N(J,K)}{\textrm{cm}^{-2}} = 1.61 \times 10^{14} \frac{J(J+1)}{K^2 \nu^2} \tau \Delta \upsilon_{FWHM}.
\end{equation}

If two or more NH$_3$ transitions are present, the rotation temperature can then be calculated from the slope of a Boltzmann plot. In a Boltzmann plot, the x-axis is given by the energy above the ground state in units of $E/k$, while on the y-axis the normalized column density $\log_{10}{[N(J,K)/(g_{op}(2J+1))]}$ is plotted. $g_{op}$ represents the ortho-para degeneracy of each state, with $g_{op} = 2$ for ortho-NH$_3$ ($K = 3n$) and $g_{op} = 1$ for para-NH$_3$ ($K \neq 3n$). The rotation temperature is then related the slope, $m$, of a linear fit to the data $T_{rot} = -0.434/m$.

The only galaxy with previous NH$_3$ detections in which we fail to detect the molecule in is NGC~1068. However, the RMS noises of our NGC~1068 spectra are insufficient to detect the lines at the intensities measured by \citet{2011AaA...529A.154A} due to the short duration of that observation. Within M83 and NGC~4945 we detect enough NH$_3$ transitions to construct Boltzmann plots and calculate rotation temperatures, and in NGC~1365 we are able to derive a $T_{rot}$ upper bound. In all three cases, the (3,3) and (6,6) lines are excluded from the fits, as ortho-NH$_3$ ($J = 3n$) is not directly comparable to para-NH$_3$ ($J \neq 3n$) and the potential exists for (3,3) masing \citep{2005ApJ...629..767O,2015PASJ...67....5M,2017ApJ...842..124G}. In all of our Boltzmann plots, we compare the (1,1) and (2,2) transitions to determine $T_{rot}$ for the cool gas, and the (2,2) and (4,4) lines for the warm gas.

For calculations of $T_{rot}$ in NGC~1365, we only consider the velocity components around 1550~km~s$^{-1}$ due to the lack of a (1,1) peak at the larger velocity. From an upper bound to the (2,2) line strength, we calculate a maximum rotation temperature between the (1,1) and (2,2) lines of 26~K. As mentioned in Section \ref{sec:NH3}, \citet{2013ApJ...779...33M} observed the same double-peaked structure but also detected the (2,2) line. Their peak line strength ratio between the (1,1) and (2,2) low-velocity lines of 1.15~$\pm$~0.27 in \citet{2013ApJ...779...33M} is much smaller than our upper limit of 1.73. From these two lines, they calculate a true gas kinetic temperature of $\approx 100$~K, significantly higher than our upper-bound rotation temperature of 26~K. As neither study detected or analyzed the (4,4) or higher para-NH$_3$ transitions, the true temperature range in NGC~1365 is still relatively unconstrained.

\begin{figure}
  \centering
  \plotone{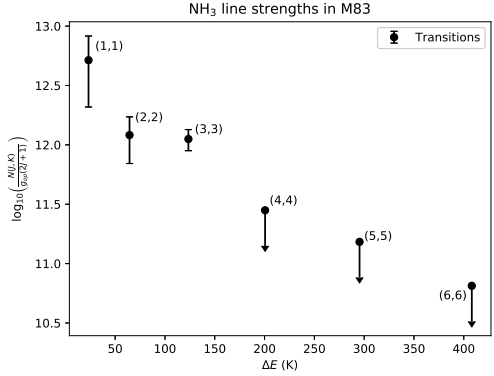}
  \caption{Boltzmann diagram of NH$_3$ in M83}
  \label{fig:m83_Trot}
\end{figure}

Within M83, we conclusively detect the (1,1) and (2,2) lines, allowing for a robust calculation of the cool gas temperature and an upper bound on that of the warm gas. Both lines are fit well by a single-component Gaussian, and the resulting Boltzmann diagram can be seen in Figure \ref{fig:m83_Trot}. We derive a rotation temperature $T_{rot} = 28 \pm 9$~K from the (1,1) and (2,2) line strength ratios, and $T_{rot} \lesssim 89$~K from the (2,2) detection and (4,4) upper bound. Our rotation temperatures serve as lower bounds to the kinetic temperatures of the gas components, with the (2,2) to (4,4) being closer to the true value \citep[e.g.][]{1983AaA...122..164W,1988MNRAS.235..229D,2005ApJ...629..767O}.

\subsection{NH$_3$ in NGC~4945}\label{sec:n4945_NH3}
The most interesting galaxy in the survey from an NH$_3$ perspective is NGC~4945. The identified emission and absorption profiles bear strong resemblance to the HCN (1--0) found in the ALMA survey from \citet{2018AaA...615A.155H}. HCN and other lines show blueshifted emission to the southwest of the nucleus, redshifted emission to the northeast, and slightly redshifted absorption centered on the nucleus. Although we lack the resolving power to see this, our profiles fit this model. We spectrally resolve two distinct absorption components, separated by roughly 75~km~s$^{-1}$. One of these is close to the systemic velocity, while the other is slightly redshifted. This profile is most reminiscent of their H$^{13}$CN (1--0) line, which is similar to our NH$_3$ lines but shows an additional absorption component at the systemic velocity. The H$^{13}$CN line's systemic component is centered on the nucleus, while the redshifted absorption is to the northeast of the nucleus, at the same position as the redshifted emission but at a lower velocity. As is also likely to be true of NH$_3$, H$^{13}$CN is optically thin, and thus both should probe deep into the clouds. Thus, we conclude that the NH$_3$ emission and absorption likely follows a similar morphology to the dense molecular gas traced in H$^{13}$CN.

\begin{figure*}
  \centering
  \plottwo{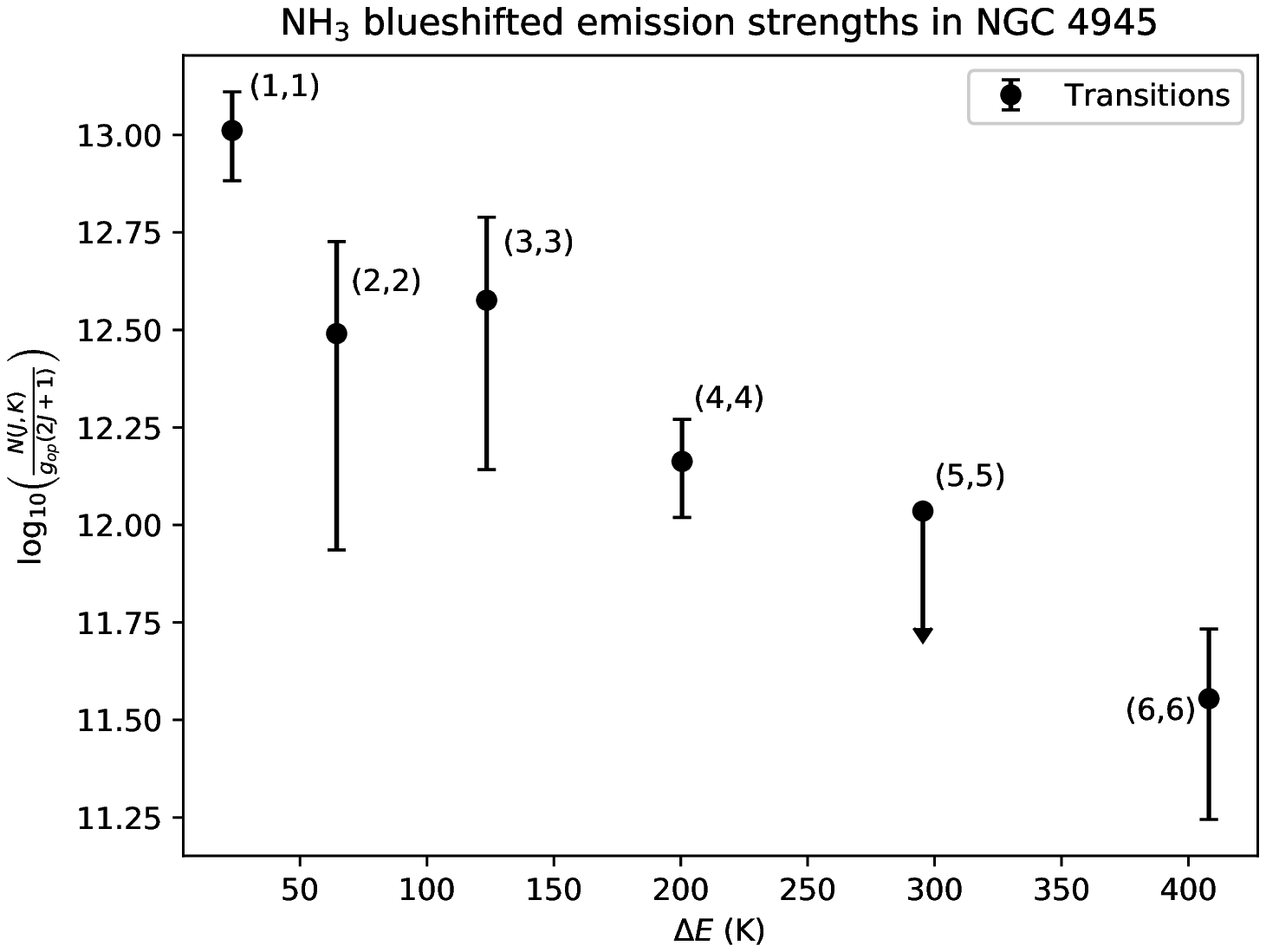}{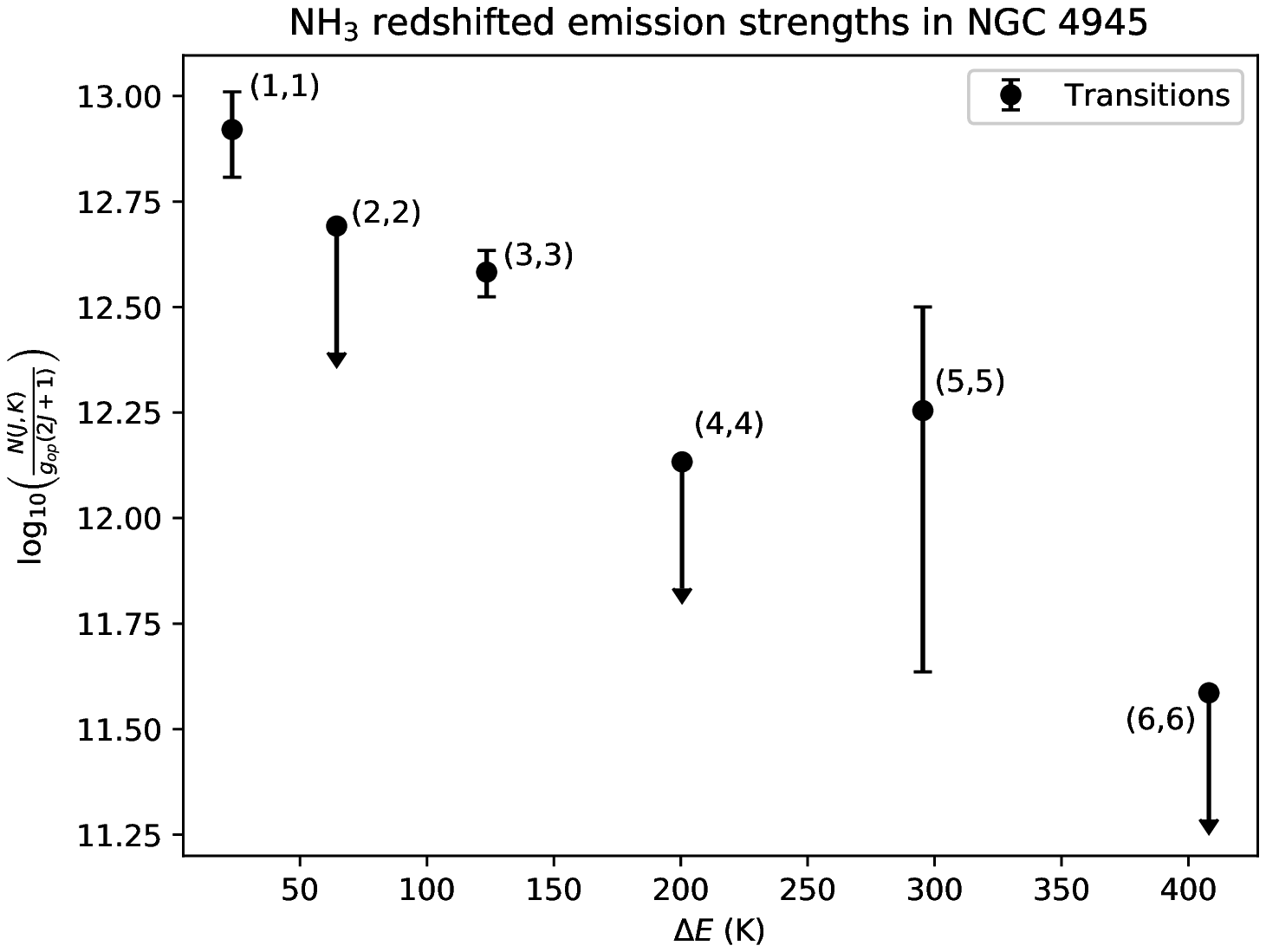}
  \caption{Boltzmann diagrams of blueshifted (left) and redshifted (right) NH$_3$ emission in NGC~4945. The (5,5) point in the redshifted-component diagram is not corrected for RRL contamination since both lines' true strengths are unknown}
  \label{fig:n4945_Trot}
\end{figure*}

Using the two emission components in the line profiles near velocities of 700~km~s$^{-1}$ and 450~km~s$^{-1}$, we created Boltzmann diagrams, seen in Figure \ref{fig:n4945_Trot}. From the Boltzmann diagram and from the line profiles in Figure \ref{fig:n4945_NH3_lines} it appears that the 710~km~s$^{-1}$ (5,5) component appears anomalously strong compared to the other para-NH$_3$ lines. As it is located at a similar velocity to what would be expected of the H(64)$\alpha$ line, contamination is a possibility. If the remaining RRLs surveyed but not detected (H(67)$\alpha$ and H(66)$\alpha$) are just under the noise limit, and assuming similar strengths of all three lines, the H(64)$\alpha$ line alone could have a maximum flux density of $\approx$5~mJy. Subtracting this from the peak flux density of the NH$_3$ (5,5) emission, the transition strength drops significantly into the noise and we would only be able to report an upper limit. Compounding the contamination argument is the very large FWHM of the (5,5) line compared to the (1,1) and (3,3) lines. The other possible explanation, maser emission, should not produce such a wide FWHM.

\begin{figure*}
  \centering
  \plottwo{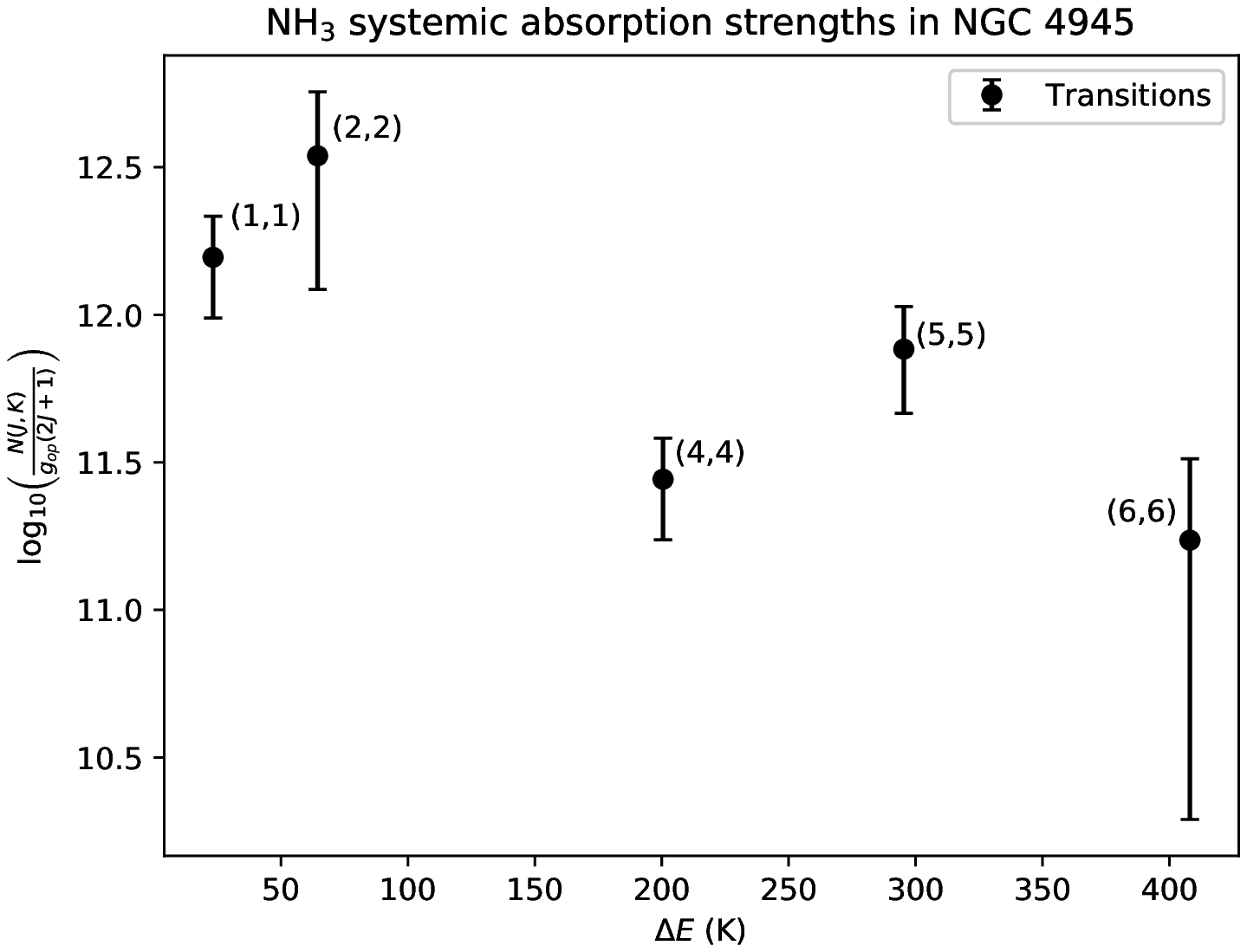}{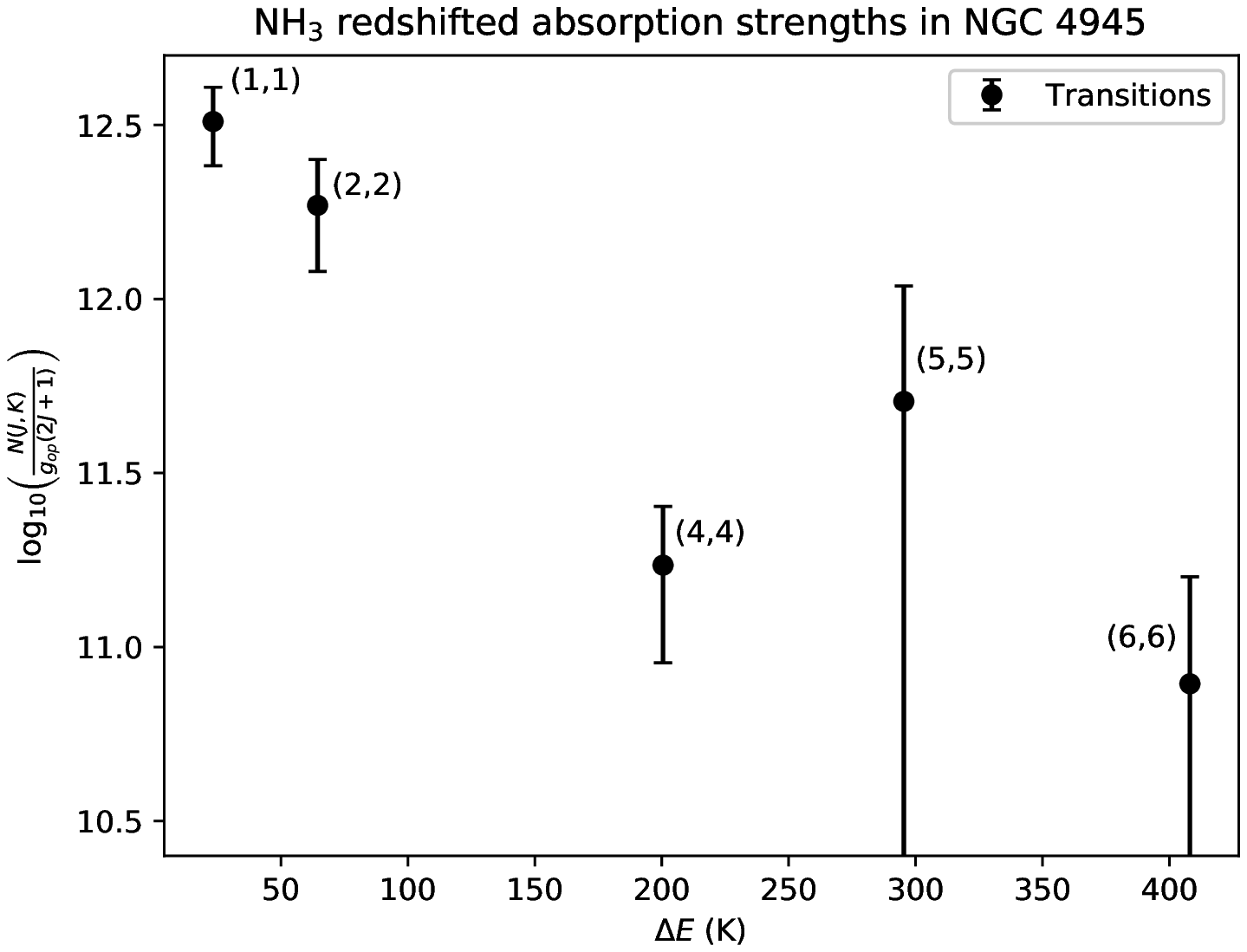}
  \caption{Boltzmann diagrams of systemic (left) and redshifted (right) NH$_3$ absorption in NGC~4945}
  \label{fig:n4945_TrotA}
\end{figure*}

On the other hand, the (3,3) line does appear to be intrinsically strong compared to the other lines, in both the redshifted and blueshifted emission. In addition, the same absorption profile as in the other lines appears superimposed on top of emission. No plausible confounding transitions are located nearby in frequency space. The other ortho-NH$_3$ (6,6) line does not display this same effect, suggesting that the anomalous behavior is not due to an abundance difference between ortho- and para-NH$_3$. The most logical explanation is that the (3,3) line is masing. If confirmed as a maser transition, the (3,3) line of NGC~4945 would be only the third or fourth extragalactic NH3 (3,3) maser known, after NGC~253 \citep{2005ApJ...629..767O,2017ApJ...842..124G}, NGC~3079 \citep{2015PASJ...67....5M}, and a candidate maser in IC~342 \citep{2018ApJ...856..134G}. Interferometric studies could also conclusively determine whether the anomalously high intensity of the (3,3) line is due to maser activity, following the methodology \citet{2017ApJ...842..124G} used for the NGC~253 confirmation. If the (3,3) emission is not cospatial with emission in other lines, then it is almost certainly a maser. Covering the entire velocity range of NGC 4945, the maser must be spatially extended.

From fits to the (1,1) and (2,2) lines, we obtain rotation temperatures of 34~$\pm$~11~K for the blueshifted component's cool gas temperature and $<$78~K for the redshifted component's. Fitting the (2,2) and (4,4) lines of the blueshifted component, we find a warm gas temperature of 180~$\pm$ 61~K. Due to the likelihood of (3,3) masing, we do not report a "hot" gas temperature from the (3,3)--(6,6) line ratio. In fact, the line strength ratio suggests an cooler temperature than calculated from the (2,2) to (4,4) ratio. This should not be seen under LTE conditions, lending further credibility to the (3,3) maser hypothesis.

We also create Boltzmann diagrams utilizing the NH$_3$ absorption in NGC 4945 (Figure \ref{fig:n4945_TrotA}). We exclude the (3,3) line, as determining the emission baseline on which the absorption components are superimposed is not possible. For the remaining lines, it should be emphasized that the absorption optical depths are estimates assuming no emission, and thus may be underestimates.

In an ideal Boltzmann diagram, the curve connecting lines of the same ammonia variety (ortho or para) will be a straight line, unless high opacity is present. Then it will be decreasing concave-up function, with steeper slopes between the (1,1) and (2,2) than between the (2,2) and (4,4), etc. However, the absorption Boltzmann diagrams of NGC 4945 display atypical properties. In particular, the (5,5) line is higher than the (4,4) line, and the (1,1) to (2,2) ratio is shallower than the (2,2) to (4,4) ratio. There is no evidence of (5,5) contamination in the other lower-velocity components of NGC~4945 used for this absorption analysis. Thus, any rotation temperatures calculated from these diagrams would be nonsensical. Most notable is that in the absorption at the systemic velocity, the (1,1) is actually lower than the (2,2) in the Boltzmann plot, and as such a calculated temperatue would be negative.

While these are unexpected and unexplained effects, similar patterns in NH$_3$ absorption have been noted before. In Arp 220, \citet{2011ApJ...742...95O} and \citet{2016ApJ...833...41Z} found a similar profile using the VLA, with depressed (1,1) and (4,4) lines, and high (2,2) and (5,5) lines. The NH$_3$ Boltzmann diagrams of both galaxies are remarkably similar \citep[See figure 5 of][]{2016ApJ...833...41Z}. They suggest the possible hypothesis that the NH$_3$ nonmetastable states are considerably overpopulated, as Arp 220 displays a large number of nonmetastable inversion lines. Although the nonmetastable (2,1), (3,2), and (4,3) transitions are within our NGC~4945 spectral windows, we fail to detect them to 1$\sigma$ RMS noises of \textasciitilde4~mJy. Thus, the non-metastable level population explanation does not appear likely in NGC~4945.

Having been observed in two galaxies with two different instruments, we rule out that this pattern is simply due to miscalibration, instrumental artifacts, or other nonphysical reasons. Evidently, either column densities of the $K=3n+2$ states are systematically higher than those of the $K=3n+1$ states or as-of-yet unknown molecular physics is responsible for the difference in intensities. Clearly, NH$_3$ (1,1) through (6,6) absorption observations of other galaxies are needed, to determine if this effect is present in more galaxies, and constrain the physics behind the anomalous ratios.

%% file: UnidentifiedLine.tex
\section{OTHER LINES DETECTED}

\begin{figure*}
  \plottwo{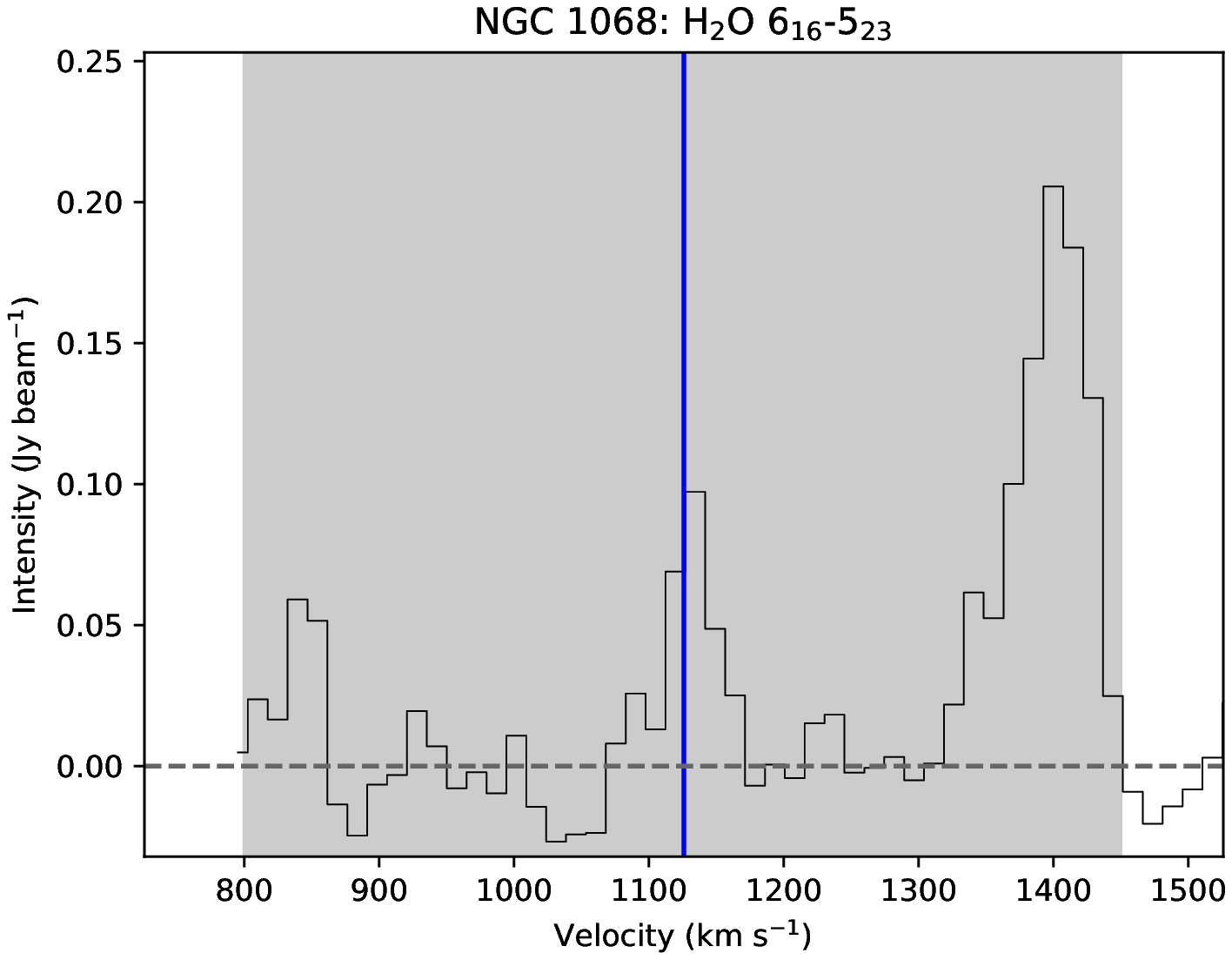}{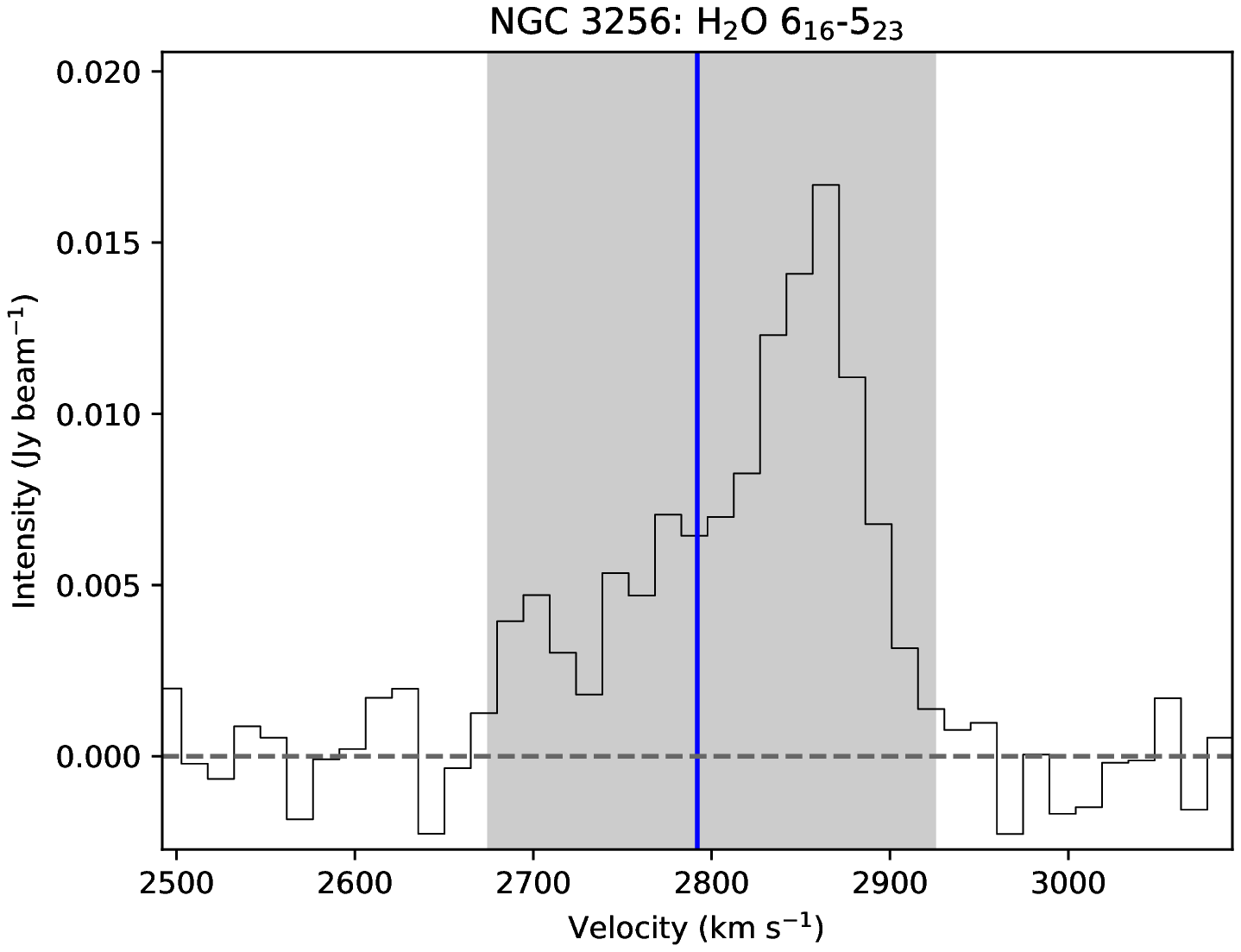}
  \plottwo{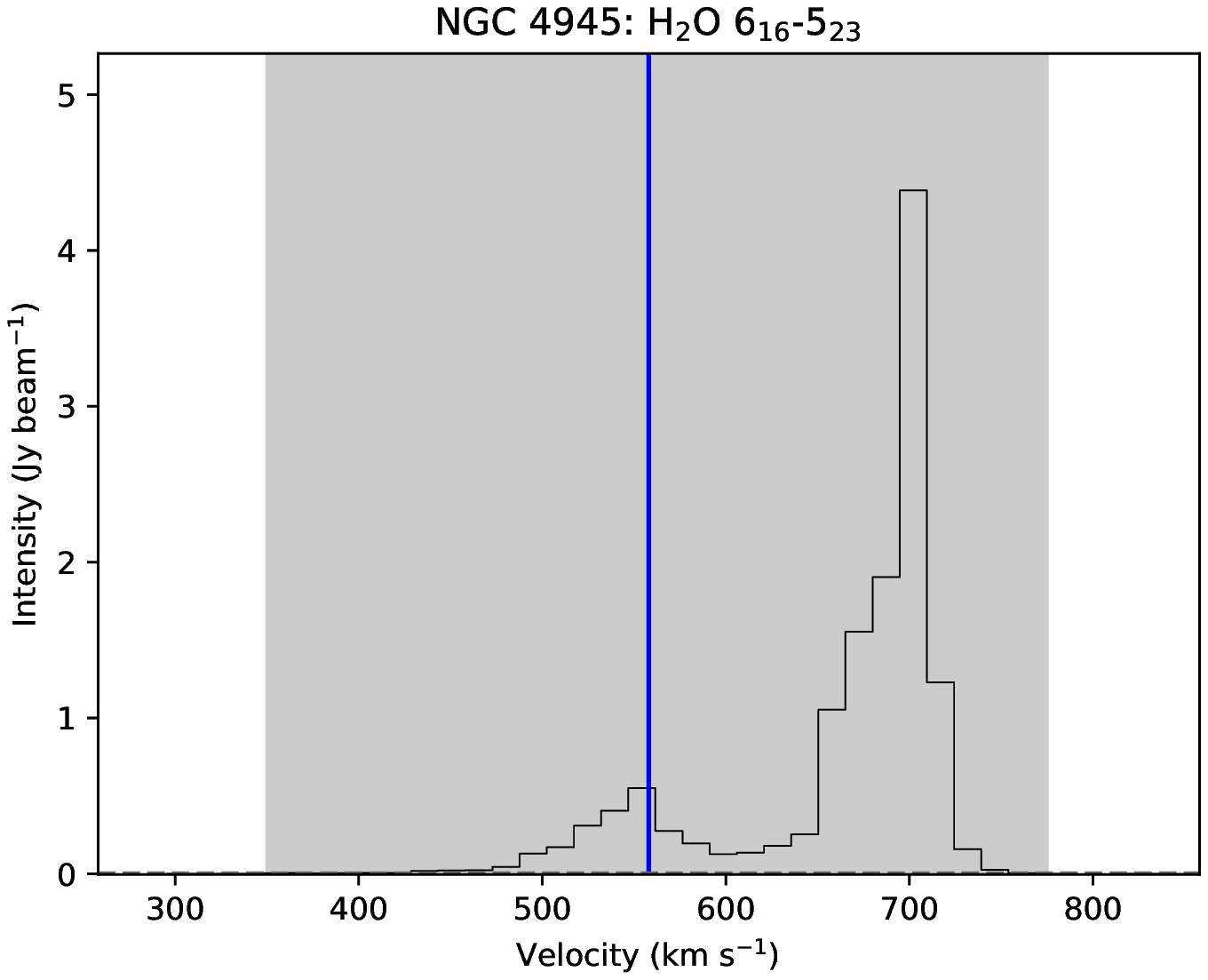}{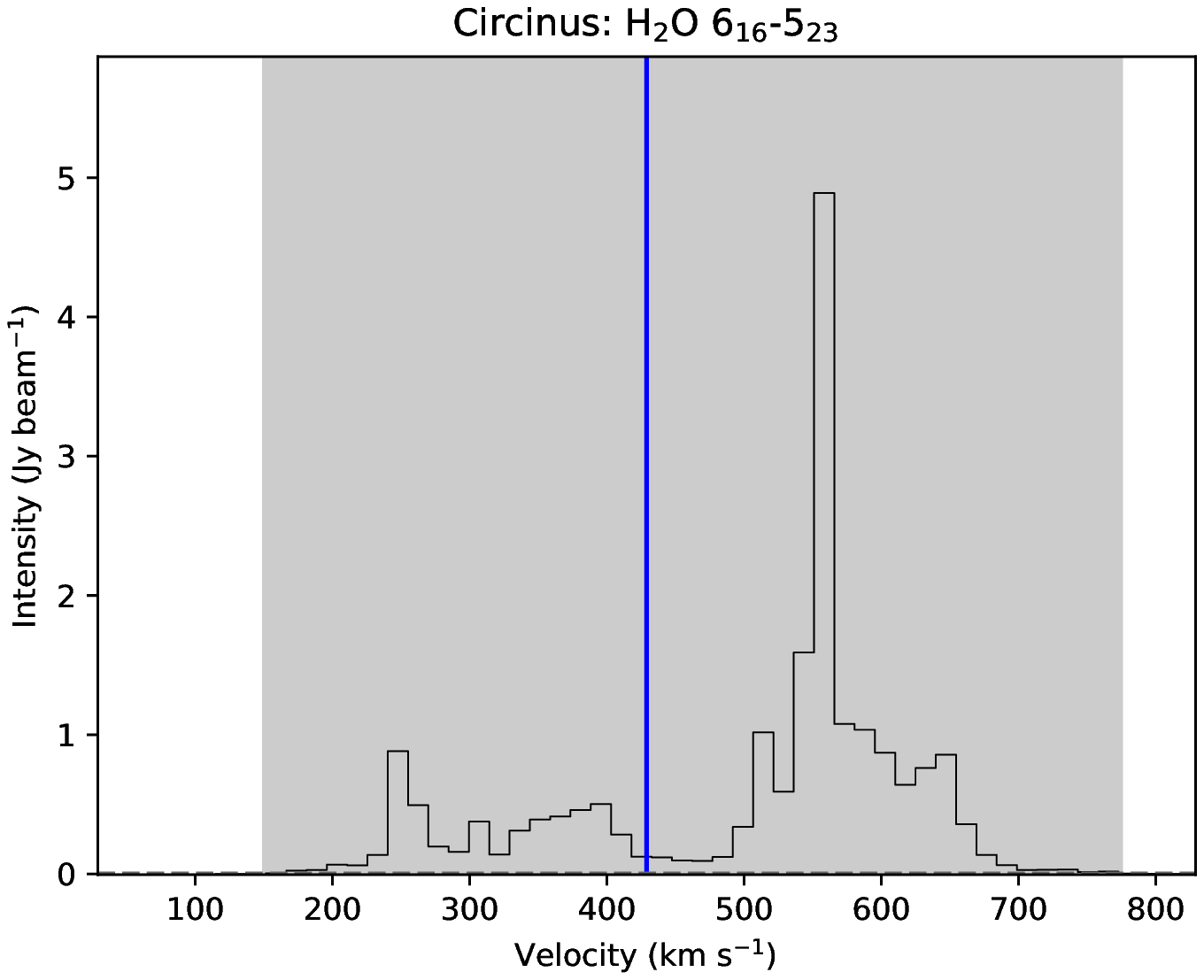}
  \caption{Profiles of detected H$_2$O lines within the sample. The blue vertical lines show the systemic velocities from Table \ref{table:sources}, while the shaded regions represent the approximate extents of the lines.}
  \label{fig:H2O_lines}
\end{figure*}

\begin{figure}
  \plotone{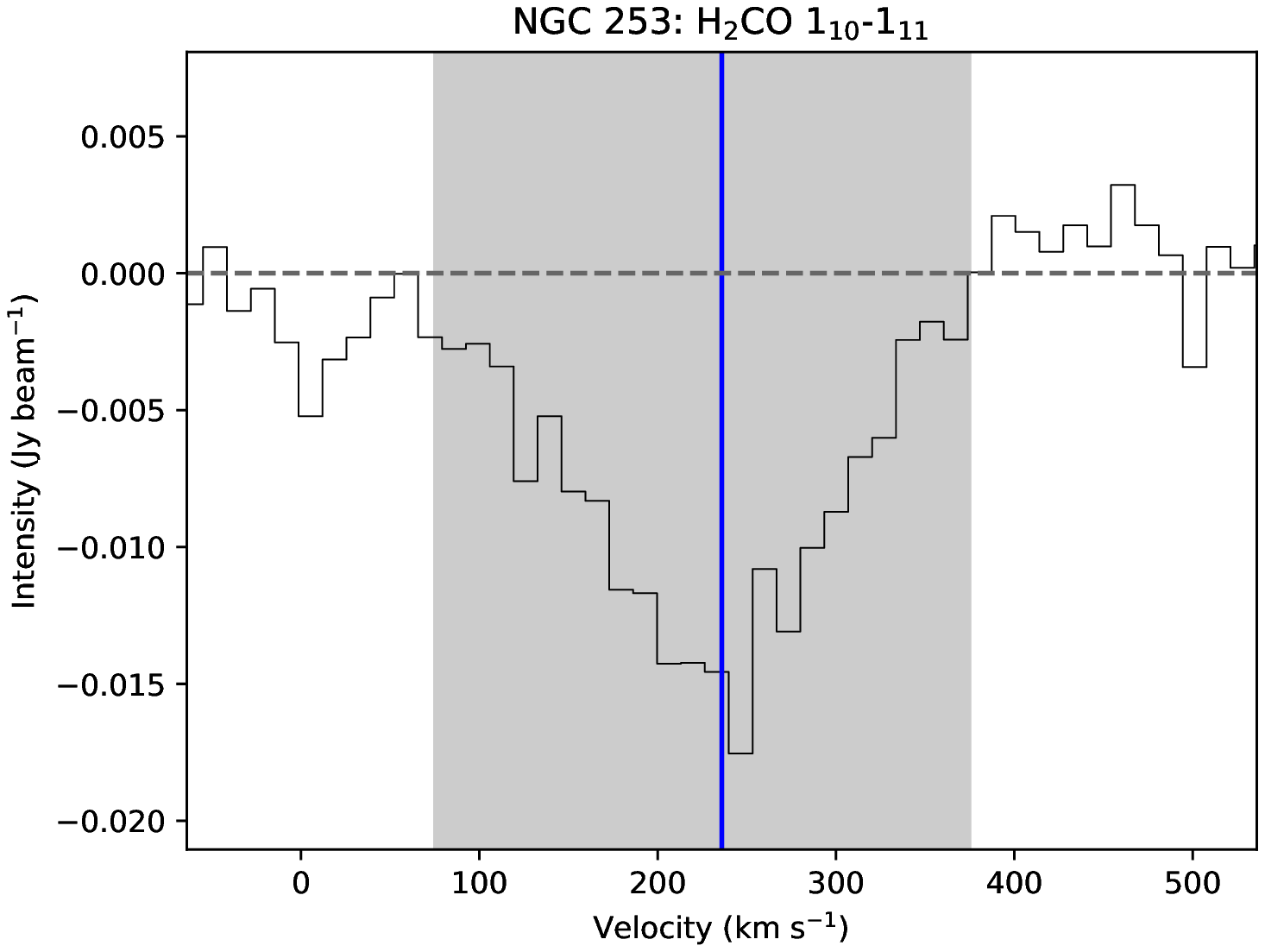}
  \plotone{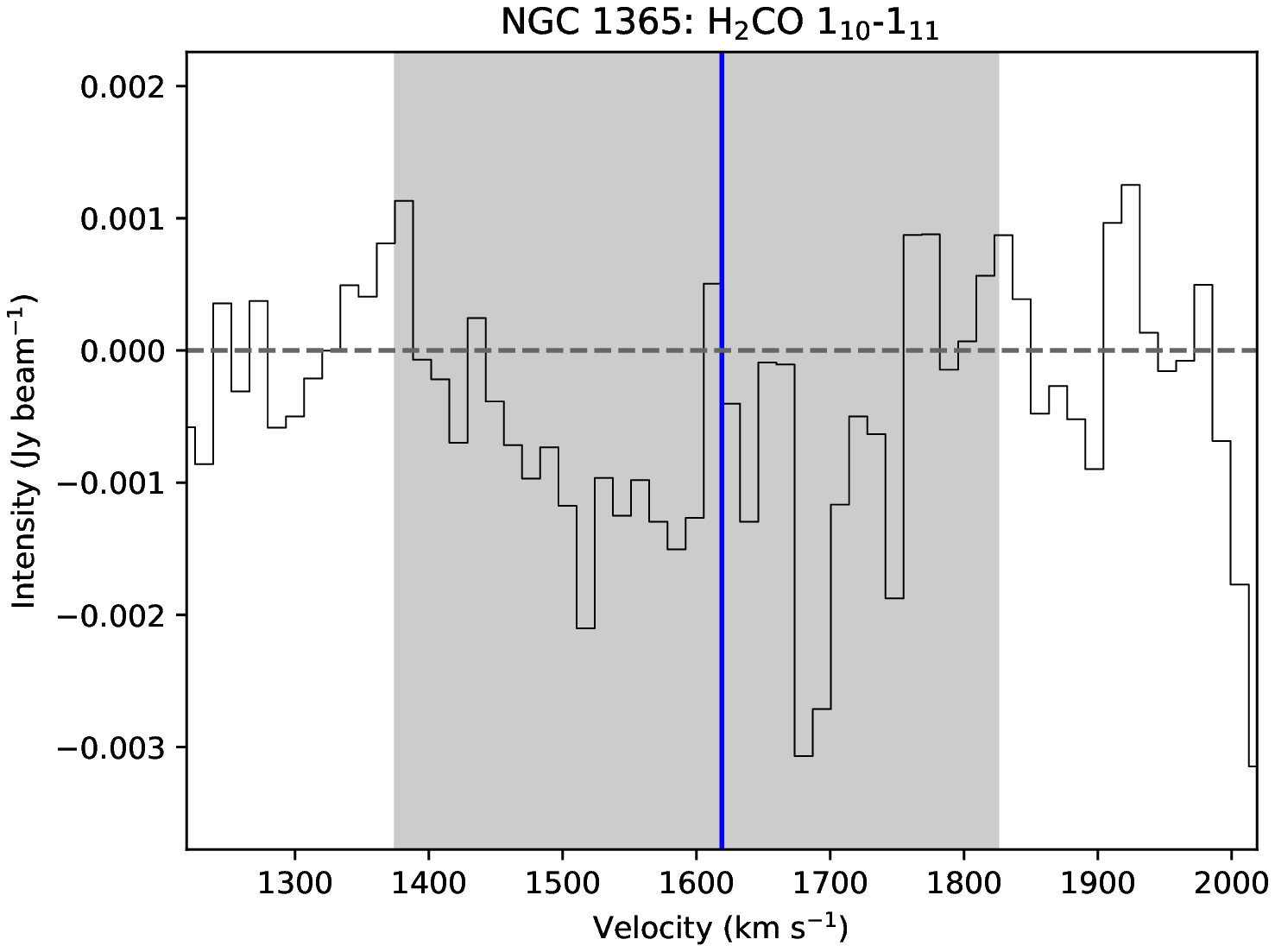}
  \plotone{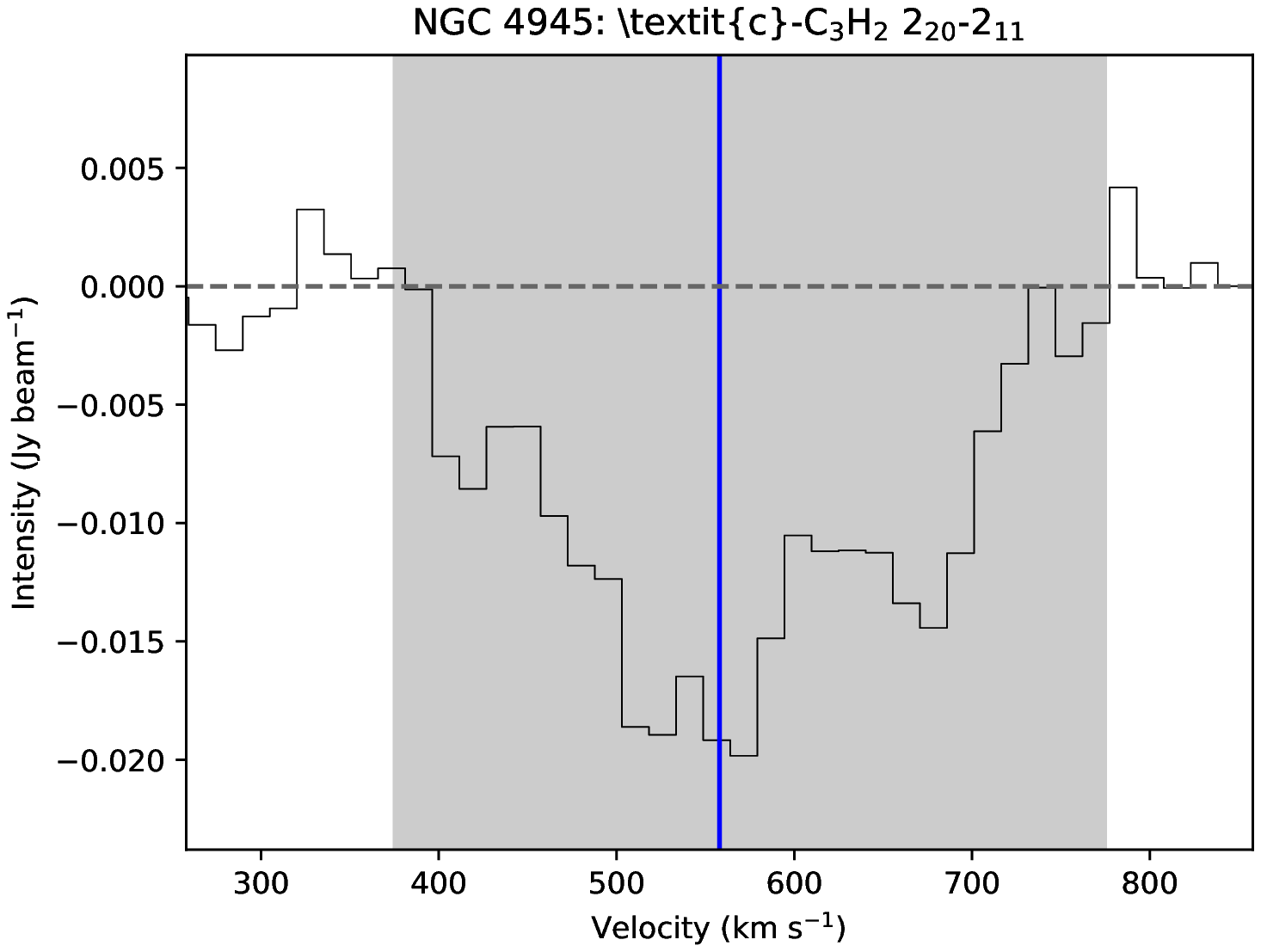}
  \caption{Profiles of other detected lines within the sample. The blue vertical lines show the systemic velocities from Table \ref{table:sources}, while the shaded regions represent the approximate extents of the lines.}
  \label{fig:other_lines}
\end{figure}

We also detect transitions from H$_2$O, H$_2$CO, and \textit{c}-C$_3$H$_2$. However, H$_2$CO has been extensively studied by \citet{2008ApJ...673..832M,2013ApJ...766..108M} and H$_2$O has been subject to numerous studies in each detection in our sample \citep[e.g.][]{2001ApJ...556..694G,2009AaA...502..529S,1997ApJ...481L..23G,2003ApJ...590..162G}. The \textit{c}-C$_3$H$_2$ molecule is only detected from a single transition in a single galaxy in our sample. We thus do not analyze the data from these lines, but do show their spectra in Figures \ref{fig:H2O_lines} and \ref{fig:other_lines}.

\subsection{Unidentified Line in NGC 253}
\begin{figure}
  \plotone{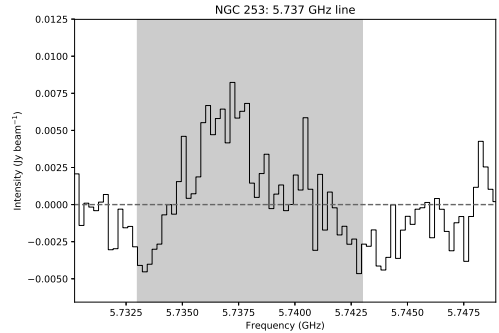}
  \caption{Profile of unidentified line within NGC~253. The shaded region represents the approximate extent of the line.}
  \label{fig:n253_lines_unidentified}
\end{figure}

An emission feature within NGC~253 at \textasciitilde5.737~GHz (Figure \ref{fig:n253_lines_unidentified}) could not be matched with any previously-observed transitions, nor with known RFI sources located nearby in frequency space. The feature is not present in any of the other eight galaxies. Moreover, the ringing effect caused by antenna CA05 is not obviously present in this spectral window. According to Splatalogue \citep{2007AAS...21113211R}, the only potential spectral transition is the SiH $^2\Pi_{1/2}$ J=3/2 $\Lambda$-doubling F=1 transition at 5.742~$\pm$~0.007~GHz redshifted by \textasciitilde40~km~s$^{-1}$ from the systemic velocity. All other possible spectral transitions are either of improbably high energies or from molecules of implausible complexity. This molecule has never been conclusively identified in space, apart from a single tentative detection of a submillimeter transition in Orion-KL \citep{2001ApJS..132..281S}. We therefore conclude this identification is unlikely.
 
%% file: Nondetections.tex
\section{NONDETECTIONS}
None of our observed galaxies show the 6.7~GHz CH$_3$OH, the 4.8~GHz OH $^2\Pi_{1/2}$, the 9.1~GHz HC$_3$N, or the 22~GHz HNCO transitions (RMS values between \textasciitilde1 and 5~mJy~bm$^{-1}$; see Table \ref{table:nondetections}). CH$_3$OH at 6.7~GHz is known to be a common maser transition within the Milky Way Galaxy, but is usually very weak in extragalactic sources compared to the higher-frequency CH$_3$OH transitions. The only published extragalactic detections are from M31 \citep{2010ApJ...724L.158S} and the LMC \citep{2008MNRAS.385..948G}. HC$_3$N \citep[e.g.][]{1988AaA...201L..23H} and HNCO \citep[e.g.][]{2005ApJ...618..259M} have been studied extensively in extragalactic sources through transitions at higher frequencies, and their transitions at lower frequencies have been detected in Galactic sources. No previous studies of these transitions within our observed bands exist from extragalactic sources, and even the HC$_3$N transitions at 18~GHz and 45~GHz has only recently been detected in several nearby galaxies \citep{2011AJ....142...32M,2017AaA...600A..15J}.

The parallel-ladder $^2\Pi_{1/2}$ J=1/2 OH transition at 4.7~GHz is commonly seen in galaxies hosting ground-state OH megamasers \citep{1986AaA...155..193H}, which are not known from any of our galaxies. The transition has an upper state energy between that of the $^2\Pi_{3/2}$ J=5/2 and J=9/2 upper states, but a lower Einstein coefficient than both. Thus, our nondetections are unsurprising.

The lack of an HC$_3$N 1--0 detection in NGC~253 is somewhat notable, given the strength of the multitude of mm-wave lines observed in the past \citep[e.g.][]{2015AaA...579A.101A}. However, this nondetection is expected; the energy of the first excited state is low (0.44~K) with a small Einstein coefficient, so the transition is expected to be significantly weaker than the higher $J$ transitions for the reasonably warm temperatures and high densities in NGC~253. Our nondetections are consistent the weakness of the 2--1 transitions observed in other galaxies \citep{2017AaA...600A..15J}.

%% file: Conclusions.tex
\section{CONCLUSIONS}\label{Ch6}

We have detected spectral lines from six different species (\ion{H}{2}, OH, H$_2$O, H$_2$CO, NH$_3$, and \textit{c}-C$_3$H$_2$) within the nine galaxies in our sample utilizing the ATCA in its H75 configuration within the 4cm and 15mm bands. Eight out of the nine surveyed galaxies have at least one line detection, while ten different transitions are detected in NGC 4945. Primary conclusions include:

1) Within NGC 253, we detect all 18 H(n)$\alpha$ recombination lines within our spectral windows, the largest number of RRLs observed in an extragalactic source to date. When calculating electron temperatures, a clear trend towards increasing $T_e$ with an increasing line frequency presents itself. Comparison to literature RRL observations serves to enhance this conclusion. Seen before in Galactic sources, this trend is likely indicative of non-LTE conditions within the nuclear regions of NGC 253. An unidentified line at 5.737~GHz is also detected, showcasing the need for further extragalactic cm-wave surveys.

2) We detect highly excited OH transitions from the $^2\Pi_{3/2}$ ladder within three galaxies: NGC 253 (J=5/2), NGC 4945 (J=9/2), and Circinus (J=9/2). While the J=5/2 state has been detected in several extragalactic sources in the past, it has never before been observed in NGC 253. In contrast, we do not detect any transitions from the $^2\Pi_{1/2}$ ladder. Neither do we detect the OH plume to the north of the nucleus, suggesting that the material within it has a low excitation temperature. No evidence of masing is found, in contrast to the ground-state J=3/2 transitions.

3) Within NGC 4945 and Circinus, we detect highly-excited OH in the $^2\Pi_{3/2}$ J=9/2 state, over 500~K above the ground state. The J=9/2 transitions are only found in regions of dense gas in the Galaxy, and these two galaxies represent only the third and fourth extragalactic detections. Within NGC 4945, the absorption is significantly redshifted, suggesting inflow towards the nucleus, consistent with previous infrared observations. In contrast, previous infrared observations of Circinus have found inflow evidence, but our absorption lines are slightly blueshifted, more consistent with outflow. Comparison of the Circinus J=9/2 data to previous J=3/2 data also yields unphysically high rotation temperatures, suggesting that nonthermal processes are involved. In NGC 4945, OH rotation temperatures between transitions of higher $J$ yield higher rotation temperatures, similar to and consistent with NH$_3$ temperatures. As our sample was constructed mainly on ease of observation and not likelihood of OH detection, it is likely that highly-excited OH is more common in other galaxies than previously thought.

4) A potential connection between the appearance of highly-excited OH absorption and H$_2$O masers is identified. The J=9/2 OH state is detected in and only in the two galaxies displaying the strongest H$_2$O maser emission. In one of these galaxies (NGC 4945), the absorption and emission are at the same velocity, redshifted by \textasciitilde100~km~s$^{-1}$ from the systemic velocity. Connections between extragalactic ground-state OH masers and H$_2$O maser emission have been noticed in the past, but the little-studied nature of excited OH absorption makes our study a first.

5) NH$_3$ is detected in four galaxies: M83, NGC 1266, NGC 1365, and NGC 4945. Within M83 and NGC 4945, enough transitions are detected to create Boltzmann diagrams and calculate rotational temperatures. For M83, we calculate a rotation temperature of 28~$\pm$~9~K for the cool gas from the (1,1) to (2,2) line strength ratio. Using an upper limit to the (4,4) line strength, we constrain the warm gas temperature to $\lesssim$89 K. In NGC 1365, we observe a (1,1) to (2,2) temperature of $\leq$26~K, at odds with previous data.

6) The NH$_3$ in NGC 4945, detected in the (1,1) through (6,6) lines, shows a complicated superposition of emission and absorption. The profile is similar to that of several mm-wave tracers of dense gas. The (3,3) line displays the same general profile as the others, but superimposed on top of a much stronger emission base. This anomalous behavior is not seen in the other ortho-NH$_3$ (6,6) line, suggesting that spatially-extended maser activity may be the culprit. Spatially-resolved studies of the (3,3) line should be able to confirm this hypothesis by noting if the emission comes from regions where the other lines are detected in absorption.

7) Boltzmann diagrams created using the NH$_3$ absorption in NGC 4945 display depressed (1,1) and (4,4) lines and/or enhanced (2,2) and (5,5) lines. The same effect has previously been observed in the ULIRG Arp 220. There is no clear physical reason for why these anomalous ratios should be present in either galaxy. Further observations are needed of the NH$_3$ inversion spectra of other galaxies where NH$_3$ is in absorption, to determine how common this trend is, and what the physical cause is. Further observations of NH$_3$ inversion lines in other extragalactic sources should lead to further insight as to the true nature of the (1,1) weakening, and potentially new physics in the process.

Our works shows that the diagnostic power of such studies is large. Factors such as LTE departure that are difficult to show from mm-wave lines are more evident in cm-wave studies. We find unexpected phenomena, such as the NH$_3$ line strength anomalies and OH J=9/2 detections, in the centimeter. The centimeter wavelengths thus appear to harbor significant untapped potential to assist in characterization of star-forming regions.

%% file: NondetectionsAppendix.tex
\section{THRESHOLDS FOR NONDETECTIONS}\label{A1}
The appendix contains the detection thresholds for all target lines that were nondetections in our galaxies (Table \ref{table:nondetections}).
\startlongtable
\begin{deluxetable}{lllc}
\tablecaption{Nondetections and 1$\sigma$ thresholds\label{table:nondetections}}
\tablehead{
\colhead{Galaxy} & \colhead{Species} & \colhead{Transition} & \colhead{1$\sigma$ RMS noise\tablenotemark{a} (mJy bm$^{-1}$)}
}

\startdata
NGC 253 & OH $^2\Pi_{1/2}$ & J=1/2 F=1$^-$--1$^+$ & 2.56\tablenotemark{b} \\
 & CH$_3$OH & 5$_{15}$--6$_{06}$ & 3.42 \\
 & HC$_3$N & J=1--0 & 3.74 \\
NGC 1068 & \textit{c}-C$_3$H$_2$ & 2$_{20}$--2$_{11}$ & 13.8 \\
 & HNCO & 1$_{01}$--0$_{00}$ & 15.2 \\
 & NH$_3$ & (1,1) & 14.5 \\
 & NH$_3$ & (2,2) & 14.5 \\
 & OH $^2\Pi_{3/2}$ & J=9/2 F=4$^+$--4$^-$ & 14.2 \\
 & OH $^2\Pi_{3/2}$ & J=9/2 F=5$^+$--5$^-$ & 14.2 \\
 & NH$_3$ & (3,3) & 14.5 \\
 & NH$_3$ & (4,4) & 14.4 \\
 & NH$_3$ & (5,5) & 13.0 \\
 & NH$_3$ & (6,6) & 12.5 \\
NGC 1266 & RRL & H(110)$\alpha$ & 0.66 \\
 & RRL & H(108)$\alpha$ & 0.69 \\
 & RRL & H(107)$\alpha$ & 0.93 \\
 & RRL & H(106)$\alpha$ & 0.78 \\
 & RRL & H(105)$\alpha$ & 1.21 \\
 & RRL & H(104)$\alpha$ & 0.84 \\
 & RRL & H(103)$\alpha$ & 0.89 \\
 & OH $^2\Pi_{3/2}$ & J=5/2 F=2$^+$--2$^-$ & 0.78 \\
 & OH $^2\Pi_{3/2}$ & J=5/2 F=3$^+$--3$^-$ & 0.78 \\
 & RRL & H(102)$\alpha$ & 0.79 \\
 & RRL & H(101)$\alpha$ & 0.77 \\
 & RRL & H(100)$\alpha$ & 0.72 \\
 & CH$_3$OH & 5$_{15}$--6$_{06}$ & 0.76 \\
 & RRL & H(099)$\alpha$ & 0.76 \\
 & RRL & H(094)$\alpha$ & 3.15 \\
 & RRL & H(093)$\alpha$ & 1.05 \\
 & RRL & H(091)$\alpha$ & 1.24 \\
 & HC$_3$N & J=1--0 & 1.44 \\
 & RRL & H(089)$\alpha$ & 1.60 \\
 & RRL & H(088)$\alpha$ & 1.41 \\
 & \textit{c}-C$_3$H$_2$ & 2$_{20}$--2$_{11}$ & 0.98 \\
 & HNCO & 1$_{01}$--0$_{00}$ & 1.08 \\
 & H$_2$O & 6$_{16}$--5$_{23}$ & 1.12 \\
 & NH$_3$ & (1,1) & 1.18 \\
 & NH$_3$ & (2,2) & 1.18 \\
 & OH $^2\Pi_{3/2}$ & J=9/2 F=4$^+$--4$^-$ & 1.22 \\
 & OH $^2\Pi_{3/2}$ & J=9/2 F=5$^+$--5$^-$ & 1.22 \\
 & NH$_3$ & (4,4) & 1.14 \\
 & NH$_3$ & (5,5) & 1.14 \\
 & NH$_3$ & (6,6) & 1.23 \\
NGC 1365 & RRL & H(111)$\alpha$ & 0.59 \\
 & OH $^2\Pi_{1/2}$ & J=1/2 F=1$^-$--1$^+$ & 0.59 \\
 & RRL & H(110)$\alpha$ & 0.77 \\
 & RRL & H(109)$\alpha$ & 0.72 \\
 & RRL & H(108)$\alpha$ & 0.62 \\
 & RRL & H(107)$\alpha$ & 0.62 \\
 & RRL & H(106)$\alpha$ & 0.65 \\
 & RRL & H(105)$\alpha$ & 0.69 \\
 & RRL & H(104)$\alpha$ & 0.78 \\
 & RRL & H(103)$\alpha$ & 0.73 \\
 & OH $^2\Pi_{3/2}$ & J=5/2 F=2$^+$--2$^-$ & 0.74 \\
 & OH $^2\Pi_{3/2}$ & J=5/2 F=3$^+$--3$^-$ & 0.74 \\
 & RRL & H(102)$\alpha$ & 0.81 \\
 & RRL & H(101)$\alpha$ & 0.83 \\
 & CH$_3$OH & 5$_{15}$--6$_{06}$ & 1.67 \\
 & RRL & H(099)$\alpha$ & 1.67 \\
 & RRL & H(093)$\alpha$ & 1.41 \\
 & RRL & H(091)$\alpha$ & 2.11 \\
 & HC$_3$N & J=1--0 & 1.97 \\
 & RRL & H(089)$\alpha$ & 1.87 \\
 & RRL & H(088)$\alpha$ & 1.87 \\
 & \textit{c}-C$_3$H$_2$ & 2$_{20}$--2$_{11}$ & 2.72 \\
 & HNCO & 1$_{01}$--0$_{00}$ & 3.17 \\
 & H$_2$O & 6$_{16}$--5$_{23}$ & 3.14 \\
 & NH$_3$ & (2,2) & 3.45 \\
 & OH $^2\Pi_{3/2}$ & J=9/2 F=4$^+$--4$^-$ & 3.03 \\
 & OH $^2\Pi_{3/2}$ & J=9/2 F=5$^+$--5$^-$ & 3.03 \\
 & NH$_3$ & (4,4) & 2.84 \\
 & NH$_3$ & (5,5) & 3.02 \\
 & NH$_3$ & (6,6) & 2.37 \\
NGC 1808 & RRL & H(111)$\alpha$ & 1.01 \\
 & OH $^2\Pi_{1/2}$ & J=1/2 F=1$^-$--1$^+$ & 1.01 \\
 & H$_2$CO & 1$_{10}$--1$_{11}$ & 1.01 \\
 & RRL & H(110)$\alpha$ & 0.98 \\
 & RRL & H(108)$\alpha$ & 1.09 \\
 & RRL & H(107)$\alpha$ & 1.13 \\
 & RRL & H(106)$\alpha$ & 1.19 \\
 & RRL & H(105)$\alpha$ & 15.5 \\
 & RRL & H(104)$\alpha$ & 1.25 \\
 & RRL & H(103)$\alpha$ & 1.26 \\
 & OH $^2\Pi_{3/2}$ & J=5/2 F=2$^+$--2$^-$ & 1.27 \\
 & OH $^2\Pi_{3/2}$ & J=5/2 F=3$^+$--3$^-$ & 1.27 \\
 & RRL & H(102)$\alpha$ & 1.49 \\
 & RRL & H(101)$\alpha$ & 1.27 \\
 & RRL & H(100)$\alpha$ & 1.69 \\
 & CH$_3$OH & 5$_{15}$--6$_{06}$ & 1.41 \\
 & RRL & H(099)$\alpha$ & 1.41 \\
 & RRL & H(094)$\alpha$ & 2.03 \\
 & RRL & H(093)$\alpha$ & 1.66 \\
 & RRL & H(091)$\alpha$ & 1.95 \\
 & HC$_3$N & J=1--0 & 2.06 \\
 & RRL & H(089)$\alpha$ & 1.82 \\
 & RRL & H(088)$\alpha$ & 1.91 \\
NGC 3256 & \textit{c}-C$_3$H$_2$ & 2$_{20}$--2$_{11}$ & 4.00 \\
 & HNCO & 1$_{01}$--0$_{00}$ & 3.43 \\
 & NH$_3$ & (1,1) & 5.51 \\
 & NH$_3$ & (2,2) & 5.51 \\
 & OH $^2\Pi_{3/2}$ & J=9/2 F=4$^+$--4$^-$ & 4.32 \\
 & OH $^2\Pi_{3/2}$ & J=9/2 F=5$^+$--5$^-$ & 4.32 \\
 & NH$_3$ & (3,3) & 4.56 \\
 & NH$_3$ & (4,4) & 3.32 \\
 & NH$_3$ & (5,5) & 2.90 \\
 & NH$_3$ & (6,6) & 2.56 \\
NGC 4945 & HNCO & 1$_{01}$--0$_{00}$ & 4.48 \\
M83 & \textit{c}-C$_3$H$_2$ & 2$_{20}$--2$_{11}$ & 4.91 \\
 & HNCO & 1$_{01}$--0$_{00}$ & 3.87 \\
 & H$_2$O & 6$_{16}$--5$_{23}$ & 3.11 \\
 & OH $^2\Pi_{3/2}$ & J=9/2 F=4$^+$--4$^-$ & 4.52 \\
 & OH $^2\Pi_{3/2}$ & J=9/2 F=5$^+$--5$^-$ & 4.52 \\
 & NH$_3$ & (4,4) & 4.66 \\
 & NH$_3$ & (5,5) & 3.24 \\
 & NH$_3$ & (6,6) & 3.48 \\
Circinus & \textit{c}-C$_3$H$_2$ & 2$_{20}$--2$_{11}$ & 4.36 \\
 & HNCO & 1$_{01}$--0$_{00}$ & 3.61 \\
 & NH$_3$ & (1,1) & 4.35 \\
 & NH$_3$ & (2,2) & 4.35 \\
 & NH$_3$ & (3,3) & 4.24 \\
 & NH$_3$ & (4,4) & 3.95 \\
 & NH$_3$ & (5,5) & 2.93 \\
 & NH$_3$ & (6,6) & 3.06
\enddata

\tablenotetext{a}{All channel widths are 224 kHz}
\tablenotetext{b}{This transition suffers from the antenna 5 problems in the NGC 253 reduction, and the RMS value may not be representative of the detection threshold}
\end{deluxetable}